%% file: main.tex
\documentclass[journal,a4paper,onecolumn]{IEEEtran}

\input{defs}
\input{acronyms}

\begin{document}
    \title{Syndrome-Based Error-Erasure Decoding of Interleaved Linearized Reed--Solomon Codes}
    \author{%
        Felicitas Hörmann\,\orcidlink{0000-0003-2217-9753},~\IEEEmembership{Graduate Student Member,~IEEE},
        \and
        Hannes Bartz\,\orcidlink{0000-0001-7767-1513},~\IEEEmembership{Member,~IEEE}
        \thanks{F.~Hörmann and H.~Bartz are with the Institute of Communications and Navigation, German Aerospace Center (DLR), Germany (e-mail: \{felicitas.hoermann, hannes.bartz\}@dlr.de). Moreover, F. Hörmann is affiliated with the School of Computer Science, University of St.\,Gallen, Switzerland. F. Hörmann and H. Bartz acknowledge the financial support by the Federal Ministry of Education and Research of Germany in the programme of ``Souverän. Digital. Vernetzt.'' Joint project 6G-RIC, project identification number: 16KISK022.}
    }

\maketitle

\begin{abstract}
    \Ac{LRS} codes are sum-rank-metric codes that generalize both Reed--Solomon and Gabidulin codes.
    We study vertically and horizontally \emph{interleaved} \ac{LRS} (VILRS and HILRS) codes whose codewords consist of a fixed number of stacked or concatenated codewords of a chosen \ac{LRS} code, respectively.
    Our unified presentation of results for horizontal \emph{and} vertical interleaving is novel and simplifies the recognition of resembling patterns.
    
    This paper's main results are syndrome-based decoders for both \acs{VILRS} and \acs{HILRS} codes.
    We first consider an error-only setting and then present more general error-erasure decoders, which can handle full errors, row erasures, and column erasures simultaneously.
    Here, an erasure means that parts of either the row space or the column space of the error are already known before decoding.
    We incorporate this knowledge directly into Berlekamp--Massey-like key equations and thus decode all error types jointly.
    The presented error-only and error-erasure decoders have an average complexity in $O(\intOrder n^2)$ and $\softoh{\intOrder n^2}$ in most scenarios, respectively, where $\intOrder$ is the interleaving order and $n$ denotes the length of the component code.
    
    Errors of sum-rank weight $\numbErrors = \numbFullErrors + \numbRowErasures + \numbColErasures$ consist of $\numbFullErrors$ full errors, $\numbRowErasures$ row erasures, and $\numbColErasures$ column erasures.
    Their successful decoding can be guaranteed for $\numbFullErrors \leq \tfrac{1}{2} (n - k - \numbRowErasures - \numbColErasures)$, where $n$ and $k$ represent the length and the dimension of the component \acs{LRS} code.
    Moreover, probabilistic decoding beyond the unique-decoding radius is possible with high probability when $\numbFullErrors \leq \tfrac{\intOrder}{\intOrder + 1}(n - k - \numbRowErasures - \numbColErasures)$ holds for interleaving order $\intOrder$.
    We give an upper bound on the failure probability for probabilistic unique decoding and showcase its tightness via Monte Carlo simulations.
\end{abstract}


\IEEEpeerreviewmaketitle
\acresetall

\input{introduction}
\acresetall

\input{preliminaries}

\input{vilrs-codes}

\input{hilrs-codes}

\input{simulations}

\input{efficient_subroutines}

\acresetall
\input{conclusion}

\section*{Acknowledgments}

The authors want to thank Sven Puchinger for fruitful brainstorming sessions and stimulating discussions that helped to develop and improve this work.

\bibliographystyle{IEEEtran}
\bibliography{references}

\end{document}

%% file: defs.tex
\usepackage[utf8]{inputenc}
\usepackage{amsmath,amsfonts,amssymb,bm}
\usepackage{mathrsfs}
\usepackage{cite}
\usepackage[dvipsnames]{xcolor}
\usepackage[nolist]{acronym}
\usepackage{orcidlink}
\usepackage{setspace}
\usepackage{pgfplots}
\usepackage[normalem]{ulem}
\usepackage{multirow}
\usepackage{adjustbox}
\usepackage[caption=false,font=footnotesize]{subfig}

\usepackage{tikz}
\usetikzlibrary{plotmarks,patterns,positioning,decorations.pathreplacing,math,matrix}

\pgfplotsset{
	compat=1.15,
    every axis/.append style={
        scale only axis,
	width=0.8\textwidth,
	height=0.8\textwidth,
	label style={inner sep=0, font=\normalsize},
	tick label style={font=\scriptsize},
	legend style={font=\scriptsize},
	mark size=3,
	major grid style={dashed},
	line width=0.8pt,
	axis line style = thin}
}

\pgfplotsset{compat=1.8}
\pgfplotsset{
mystyle/.style={
    scale only axis,
    width=0.7\columnwidth,
    height=0.55\columnwidth,
    label style={inner sep=0, font=\normalsize},
    tick label style={font=\scriptsize},
    legend style={font=\scriptsize},
    mark size=3,
    major grid style={dashed},
    line width=0.8pt,
    axis line style = thin}
}

\usepackage[ruled,vlined,titlenumbered,linesnumbered]{algorithm2e}
\SetKwInOut{Input}{Input}\SetKwInOut{Output}{Output}
\DontPrintSemicolon

\usepackage{mathtools}
\mathtoolsset{showonlyrefs}

\allowdisplaybreaks

\usepackage{hyperref}

\def\ve#1{{\mathchoice{\mbox{\boldmath$\displaystyle #1$}}%
		{\mbox{\boldmath$\textstyle #1$}}%
		{\mbox{\boldmath$\scriptstyle #1$}}%
		{\mbox{\boldmath$\scriptscriptstyle #1$}}}}

\newtheorem{theorem}{Theorem}
\newtheorem{definition}{Definition}
\newtheorem{lemma}{Lemma}

\newtheorem{remark}{Remark}

\newtheorem{example}{Example}

\newcommand{\Fqm}{\ensuremath{\mathbb F_{q^m}}}
\newcommand{\Fqms}{\ensuremath{\mathbb F_{q^{ms}}}}

\newcommand{\Fq}{\ensuremath{\mathbb F_{q}}}

\newcommand{\F}{\ensuremath{\mathbb F}}

\newcommand{\NN}{\ensuremath{\mathbb{N}}}

\newcommand{\aut}{\ensuremath{\theta}}
\newcommand{\autinv}{\ensuremath{\aut^{-1}}}

\newcommand{\SkewPolyringZeroDer}{\ensuremath{\Fqm[x;\aut]}}
\newcommand{\SkewPolyringZeroDerInv}{\ensuremath{\Fqm[x;\autinv]}}

\newcommand{\opev}[3]{\ensuremath{{#1}(#2)_{#3}}}
\newcommand{\opfull}[2]{\ensuremath{\mathcal{D}_{\aut}(#2)_{#1}}}
\newcommand{\opfullinv}[2]{\ensuremath{\mathcal{D}_{\autinv}(#2)_{#1}}}
\newcommand{\opfullexp}[3]{\ensuremath{\mathcal{D}_{\aut}^{#3}(#2)_{#1}}}
\newcommand{\opfullexpinv}[3]{\ensuremath{\mathcal{D}_{\autinv}^{#3}(#2)_{#1}}}

\newcommand{\opexp}[3]{\ensuremath{\mathcal{D}^{#3}(#2)_{#1}}}

\DeclareMathOperator{\mpol}{mpol}

\newcommand{\mpolArgs}[2]{\ensuremath{\mpol_{(#1)_{#2}}}}

\newcommand{\OComplNoArg}{\ensuremath{O}}
\newcommand{\softoh}[1]{\bnd{\softO}{#1}}
\newcommand{\softO}{\widetilde{\OComplNoArg}} %

\newcommand{\algoname}[1]{{\normalfont\textsc{#1}}} %

\newcommand{\defeq}{:=}
\newcommand{\eqdef}{=:}

\renewcommand{\bar}{\overline}

\newcommand{\modl}{\; \mathrm{mod}_\mathrm{l} \;}
\newcommand{\modr}{\; \mathrm{mod}_\mathrm{r} \;}

\DeclareMathOperator{\wt}{wt}
\DeclareMathOperator{\rk}{rk}

\DeclareMathOperator{\diag}{diag}

\DeclareMathOperator{\Id}{\textrm{Id}}

\newcommand{\rkq}{\ensuremath{\rk_q}}
\newcommand{\rkqm}{\ensuremath{\rk_{q^m}}}

\renewcommand{\vec}[1]{\ve{#1}} %
\newcommand{\mat}[1]{\ensuremath{\bm{#1}}}

\newcommand{\opMoore}[3]{\ensuremath{\mathfrak{M}_{\aut}^{#1}(#2)_{#3}}}
\newcommand{\opMooreInv}[3]{\ensuremath{\mathfrak{M}_{\autinv}^{#1}(#2)_{#3}}}
\newcommand{\genNorm}[2]{\ensuremath{\mathcal{N}_{\aut}^{#1}(#2)}}
\newcommand{\genNormInv}[2]{\ensuremath{\mathcal{N}_{\autinv}^{#1}(#2)}}
\newcommand{\gcrd}{\ensuremath{\mathrm{gcrd}}}
\newcommand{\lclm}{\ensuremath{\mathrm{lclm}}}
\newcommand{\ldivNoArg}{\ensuremath{\mathrm{ldiv}}}
\newcommand{\ldiv}[1]{\ensuremath{\ldivNoArg\left(#1\right)}}
\newcommand{\rdivNoArg}{\ensuremath{\mathrm{rdiv}}}
\newcommand{\rdiv}[1]{\ensuremath{\rdivNoArg\left(#1\right)}}

\renewcommand{\a}{\vec{a}}
\renewcommand{\b}{\vec{b}}
\renewcommand{\c}{\vec{c}}
\newcommand{\e}{\vec{e}}

\newcommand{\s}{\vec{s}}
\renewcommand{\t}{\vec{t}}

\renewcommand{\v}{\vec{v}}

\newcommand{\x}{\vec{x}}
\newcommand{\y}{\vec{y}}

\newcommand{\A}{\mat{A}}
\newcommand{\B}{\mat{B}}
\newcommand{\C}{\mat{C}}

\newcommand{\E}{\mat{E}}
\newcommand{\G}{\mat{G}}
\newcommand{\I}{\mat{I}}
\renewcommand{\H}{\mat{H}}

\newcommand{\Q}{\mat{Q}}

\renewcommand{\S}{\mat{S}}

\newcommand{\X}{\mat{X}}
\newcommand{\Y}{\mat{Y}}
\newcommand{\0}{\ensuremath{\mathbf 0}}

\newcommand{\vecbeta}{\ensuremath{\boldsymbol{\beta}}}
\newcommand{\vecgamma}{\ensuremath{\boldsymbol{\gamma}}}

\newcommand{\vecxi}{\ensuremath{\boldsymbol{\xi}}}

\newcommand{\vecsigma}{\ensuremath{\boldsymbol{\sigma}}}
\newcommand{\vectau}{\ensuremath{\boldsymbol{\tau}}}
\newcommand{\veclambda}{\ensuremath{\boldsymbol{\lambda}}}
\newcommand{\nVec}{\ensuremath{\vec{n}}}
\newcommand{\vecntilde}{\ensuremath{\vec{\widetilde{n}}}}

\newcommand{\mycode}[1]{\ensuremath{\mathcal{#1}}}

\newcommand{\linRS}[4]{\ensuremath{\mathrm{LRS}[#1, #2; #3, #4]}} %
\newcommand{\linRSWithAut}[5]{\ensuremath{\mathrm{LRS}_{#1}[#2, #3; #4, #5]}} %

\newcommand{\vertIntLinRS}[5]{\ensuremath{\mathrm{VI}\mathrm{LRS}[#1, #2, #3; #4, #5]}} %
\newcommand{\horIntLinRS}[5]{\ensuremath{\mathrm{HI}\mathrm{LRS}[#1, #2, #3; #4, #5]}} %

\newcommand{\SumRankWeight}{\ensuremath{\wt_{\Sigma R}}}

\newcommand{\SumRankDist}{d_{\ensuremath{\Sigma}R}}

\newcommand{\SumRankWeightWPartition}[1]{\SumRankWeight^{#1}}
\newcommand{\SumRankDistWPartition}[1]{\SumRankDist^{#1}}

\newcommand{\numbFullErrors}{{t_{\indFullErrors}}}

\newcommand{\numbRowErasures}{{t_{\indRowErasures}}}
\newcommand{\numbRowErasuresBar}{{\bar{t}_{\indRowErasures}}}
\newcommand{\numbRowErasuresWIndex}[1]{{t_{\indRowErasures, #1}}}
\newcommand{\numbColErasures}{{t_{\indColErasures}}}
\newcommand{\numbColErasuresBar}{{\bar{t}_{\indColErasures}}}
\newcommand{\numbColErasuresWIndex}[1]{{t_{\indColErasures, #1}}}
\newcommand{\numbErrors}{\tau}
\newcommand{\numbErrorsWIndex}[1]{\tau_{j}}
\newcommand{\numbErrorsMax}{{\numbErrors_{\mathsf{max}}}}
\newcommand{\numbErrorsWeightedVertical}{{{\numbErrors}_{\mathsf{vert}}^{\ast}}}
\newcommand{\numbErrorsWeightedHorizontal}{{{\numbErrors}_{\mathsf{hor}}^{\ast}}}

\newcommand{\numbErrorsVec}{\vectau}
\newcommand{\numbFullErrorsVec}{\t_{\indFullErrors}}
\newcommand{\numbRowErasuresVec}{\t_{\indRowErasures}}
\newcommand{\numbColErasuresVec}{\t_{\indColErasures}}

\newcommand{\numbFullErrorsInBlock}[1]{{t_{\indFullErrors}^{(#1)}}}
\newcommand{\numbRowErasuresInBlock}[1]{{t_{\indRowErasures}^{(#1)}}}
\newcommand{\numbColErasuresInBlock}[1]{{t_{\indColErasures}^{(#1)}}}
\newcommand{\numbErrorsInBlock}[1]{{\numbErrors^{(#1)}}}

\newcommand{\indFullErrors}{\mathcal{F}}
\newcommand{\indRowErasures}{\mathcal{R}}
\newcommand{\indColErasures}{\mathcal{C}}

\newcommand{\indErrorType}{\mathcal{T}}
\newcommand{\numbErrorType}{t_{\indErrorType}}
\newcommand{\numbErrorTypeVec}{\t_{\indErrorType}}
\newcommand{\numbErrorTypeInBlock}[1]{\numbErrorType^{(#1)}}

\newcommand{\eFullErrors}{\e_\indFullErrors}
\newcommand{\eRowErasures}{\e_\indRowErasures}
\newcommand{\eColErasures}{\e_\indColErasures}

\newcommand{\matGivenSRWeight}[4]{\ensuremath{\mathcal{M}_{#1}(#2, #3; #4)}}
\newcommand{\vecGivenSRWeight}[3]{\ensuremath{\mathcal{V}_{#1}(#2; #3)}}

\newcommand{\autInvXiWIndex}[1]{\widetilde{\xi}_{#1}} %
\newcommand{\autInvXiInvWIndex}[1]{\widehat{\xi}_{#1}} %
\newcommand{\autInvXiVec}{\widetilde{\vecxi}}
\newcommand{\autInvXiInvVec}{\widehat{\vecxi}}

\newcommand{\gammaq}{\ensuremath{\kappa_q}}

\newcommand{\oh}[1]{\bnd{O}{#1}}
\newcommand{\bnd}[2]{\ensuremath{#1\mathopen{}\left(#2\right)\mathclose{}}}

\newcommand{\shot}[2]{\ensuremath{{#1}^{(#2)}}}
\newcommand{\subShot}[3]{\ensuremath{{#1}^{(#3)}}_{#2}}

\newcommand{\pe}{\ensuremath{\gamma}}

\newcommand{\intOrder}{\ensuremath{s}}

\newcommand{\shots}{\ensuremath{\ell}}

\newcommand{\skewrev}[1]{\ensuremath{\bar{#1}}}

\newcommand{\ELP}{\ensuremath{\lambda}}
\newcommand{\ELPrev}{\ensuremath{\skewrev{\ELP}}}
\newcommand{\ELPfull}{\ensuremath{\ELP_{\indFullErrors}}}
\newcommand{\ELPfullWIndex}[1]{\ensuremath{\ELP_{\indFullErrors, #1}}}
\newcommand{\ELProw}{\ensuremath{\ELP_{\indRowErasures}}}

\newcommand{\ELPcol}{\ensuremath{\ELP_{\indColErasures}}}
\newcommand{\ELPcolWIndex}[1]{\ensuremath{\ELP_{\indColErasures, #1}}}

\newcommand{\ELPcolRevWIndex}[1]{\ensuremath{\ELPrev_{\indColErasures, #1}}}

\newcommand{\ELPfullcol}{\ensuremath{\ELP_{\indFullErrors \indColErasures}}}
\newcommand{\ELPfullcolWIndex}[1]{\ensuremath{\ELP_{\indFullErrors \indColErasures, #1}}}

\newcommand{\ESP}{\ensuremath{\sigma}}
\newcommand{\ESPfull}{\ensuremath{\ESP_\indFullErrors}}
\newcommand{\ESPfullWIndex}[1]{\ensuremath{\ESP_{\indFullErrors, #1}}}
\newcommand{\ESProw}{\ensuremath{\ESP_\indRowErasures}}
\newcommand{\ESProwWIndex}[1]{\ensuremath{\ESP_{\indRowErasures, #1}}}
\newcommand{\ESPcol}{\ensuremath{\ESP_\indColErasures}}

\newcommand{\ESPrev}{\ensuremath{\skewrev{\ESP}}}

\newcommand{\ESProwRevWIndex}[1]{\ensuremath{\ESPrev_{\indRowErasures, #1}}}

\newcommand{\ESPfullrow}{\ensuremath{\ESP_{\indFullErrors \indRowErasures}}}
\newcommand{\ESPfullrowWIndex}[1]{\ensuremath{\ESP_{\indFullErrors \indRowErasures, #1}}}

\newcommand{\ESPcomponentSyndrome}[1]{s_{RC, #1}}

\newcommand{\ELPcomponentSyndrome}[1]{s_{CR, #1}}
\newcommand{\coeffpower}[2]{\Theta^{#2}(#1)}

\newcommand{\componentSyndrome}[1]{s_{#1}}
\newcommand{\componentSyndromeWIndex}[2]{s_{#1, #2}}

\newcommand{\revComponentSyndromeWIndex}[2]{\skewrev{s}_{#1, #2}}

\newcommand{\errEvalPoly}{\ensuremath{\omega}}
\newcommand{\erasureEvalPoly}{\ensuremath{\psi}}

\newcommand{\h}{\vec{h}}

\newcommand{\M}{\ve{M}}

\newcommand{\sample}{\overset{\$}{\gets}}

\newcommand{\q}{\vec{q}}

\newcommand{\n}{\vec{n}}

\DeclareMathOperator{\VInt}{VInt}
\DeclareMathOperator{\HInt}{HInt}

\newcommand{\xvert}{\ensuremath{\x_{\mathsf{vert}}}}
\newcommand{\xhor}{\ensuremath{\x_{\mathsf{hor}}}}

\tikzmath{
	\frameDist = .1cm;
	\leftDotsDist = -.12;
	\rightDotsDist = -.17;
	\braceDist = .1cm;
	\braceTextDist = .3cm;
	\tikzPictureDist = .15cm;
}

\tikzset{
	box/.style={minimum width=1cm, node distance=.1cm, inner sep=.1cm},
	fillblue/.style={fill=blue!20!white},
	fillred/.style={fill=red!20!white},
	fillgreen/.style={fill=green!20!white},
	brace/.style={decorate, decoration={brace, mirror, amplitude=7pt}},
	bracetext/.style={below, yshift=-\braceTextDist},
}

\newcommand{\myfmargbox}[1]{\marginbox{0pt 3pt}{\fbox{\ensuremath{#1}}}}

\newcommand{\colorblock}[3]{
	\begin{tikzpicture}[baseline=(xblock.base)]
		\node[box,#1] (xblock) {$\subShot{\x}{#2}{#3}$};
	\end{tikzpicture}
}

\newcommand{\raisedLeftPar}{\raisebox{1pt}{$\Biggl($}}
\newcommand{\raisedRightPar}{\raisebox{1pt}{$\Biggr)$}}

\newcommand{\Prob}{\ensuremath{\mathrm{Pr}}}
\newcommand{\failureProb}{\ensuremath{\Prob_{\mathsf{fail}}}}

\DeclareMathOperator{\ext}{ext}
\newcommand{\coeffq}[1]{\ext_q(#1)}

%% file: acronyms.tex
\begin{acronym}
	\acro{BCH}{Bose--Chaudhuri--Hocquenghem}
	\acro{BMD}{bounded minimum distance}
	\acro{ESP}{error-span polynomial}
	\acro{ELP}{error-locator polynomial}
	\acro{GMD}{generalized minimum distance}
	\acro{SRS}{skew Reed--Solomon}
	\acro{ISRS}{interleaved skew Reed--Solomon}
	\acro{lclm}{least common left multiple}
	\acro{LRS}{linearized Reed--Solomon}
	\acro{LLRS}{lifted linearized Reed--Solomon}
	\acro{ILRS}{interleaved linearized Reed--Solomon}
	\acro{HILRS}{horizontally interleaved linearized Reed--Solomon}
	\acro{VILRS}{vertically interleaved linearized Reed--Solomon}
	\acro{LILRS}{lifted interleaved linearized Reed--Solomon}
	\acro{MDS}{maximum distance separable}
	\acro{MSRD}{maximum sum-rank distance}
	\acro{MSD}{maximum skew distance}
	\acro{MS-SSRS}{multi-sequence skew shift-register synthesis}
	\acro{RS}{Reed--Solomon}
	\acro{SkaRo}{Skachek--Roth}
	\acro{SkaAndRo}{Skachek and Roth}
	\acro{LEEA}{linearized extended Euclidean algorithm}
	\acro{SEEA}{skew extended Euclidean algorithm}
	\acro{SkaRo}{Skachek--Roth}
	\acro{SkaAndRo}{Skachek and Roth}
	\acro{gcrd}{greatest common right divisor}
	\acro{PIR}{private information retrieval}
	\acro{PQC}{post-quantum cryptography}
	\acro{LRPC}{low-rank parity-check}
	\acro{REF}{row-echelon form}
\end{acronym}

%% file: introduction.tex
\section{Introduction}\label{sec:introduction}

The sum-rank metric is a rather modern alternative metric in coding theory and covers the Hamming metric as well as the rank metric as special cases.
A prominent family of sum-rank-metric codes are \ac{LRS} codes~\cite{martinez2018skew} which generalize \acl{RS} codes in the Hamming metric and Gabidulin codes in the rank metric, respectively.
\ac{LRS} codes attain the Singleton-like bound in the sum-rank metric with equality and have been studied from many different perspectives.
While the sum-rank metric was first proposed in the context of space-time coding~\cite[Sec.~III]{lu2005unified}, it now has a variety of applications in e.g.\ coherent and non-coherent multishot network coding~\cite{nobrega2009multishot,nobrega2010multishot,martinez2019reliable,BartzPuchinger2024Fastdecodinglifted}, \acl{PIR}~\cite{martinez2019private}, and distributed storage~\cite{martinez2019Universal}.
Another emerging use case for the sum-rank metric is \ac{PQC}, which deals with the design and analysis of cryptosystems that remain secure in the realm of powerful quantum computers.
First results for \ac{PQC} include the study of generic decoding in the sum-rank metric~\cite{puchinger2020generic} as well as a work on distinguishers for disguised \ac{LRS} codes~\cite{HoermannBartzEtAl2023DistinguishingRecoveringGeneralized}.

Code-based cryptography mostly uses channels with an additive error of fixed weight in a certain metric.
When side channels are accessible on top of the encrypted data, additional information about the secret error might be leaked and facilitate attacks.
For example, the knowledge of an entry of the error, of an erroneous position, or of an error-free position in the Hamming metric decreases the complexity of information-set decoding~\cite{HorlemannPuchingerEtAl2022InformationSetDecoding}.
Note that the knowledge of an erroneous position is precisely what is classically called an \emph{erasure} in the Hamming metric.
As each erasure influences exactly one position of the error vector, it can be thought of as a weight-one error with side information.
The notion of erasures in the rank metric has to be adapted to the different way of measuring the gravity of errors.
For erasures in the rank metric, the focus is on an error decomposition whose two parts share the column space and the row space with the error, respectively.
A rank error is called a \emph{column erasure}, if only the first part is unknown, and a \emph{row erasure}, if only the second part is unknown~\cite{gabidulin2008errorAndErasure,Wachter-ZehZeh2014ListUniqueError,silva_rank_metric_approach,richter2004ErrorAndErasure}.
The same strategy can be applied blockwise to the sum-rank metric and~\autoref{fig:error_decomposition} shows how an error is composed of full errors, row erasures, and column erasures.
Moreover, it is highlighted which parts of the respective decompositions are known to the receiver.

\begin{figure*}[ht!]
    \centering
        \input{./error_decomposition.tikz}
    \caption{Visualization of the sum-rank error decomposition into parts containing full errors, row erasures, and column
    erasures, respectively. The filled green parts are known to the receiver, the unknown parts are hatched in red.}
    \label{fig:error_decomposition}
\end{figure*}

The simultaneous decoding of errors and erasures for Gabidulin codes in the rank metric was e.g.\ considered in~\cite{richter2004ErrorAndErasure} and the signature scheme RankSign is based on error-erasure decoding of \ac{LRPC} codes~\cite{GaboritRuattaEtAl2014RankSignEfficientSignature}.
Further, a series of work deals with the natural occurrence of error-erasure decoding of rank-metric codes in linear network coding, where some packets are erroneous and others are lost~\cite{silvakschischang2007,silva_rank_metric_approach,silva2009error}.
There is also a syndrome-based error-erasure decoder for \ac{LRS} codes in the sum-rank metric~\cite{hoermann2022error_erasure} which generalizes the error-only decoder from~\cite{Martinez-PenasPuchinger2024MaximumSumRank} and can be applied to randomized decoding of \ac{LRS} codes~\cite{JerkovitsBartzEtAl2023RandomizedDecodingLinearized,JerkovitsBartzEtAl2024SupportGuessing}, for example.

A typical coding-theoretical approach to handle burst errors is \emph{code interleaving}.
When a code is interleaved \emph{vertically} with an interleaving order $\intOrder$, each codeword of the new code is a matrix consisting of $\intOrder$ stacked codewords of the original code.
In contrast, any codeword of a \emph{horizontally} $\intOrder$-interleaved code is a concatenation of $\intOrder$ codewords of the original code.
While communication channels endowed with the Hamming metric mostly benefit from vertical interleaving and the horizontal analog was barely considered, both notions occur quite naturally in the rank metric.
In fact, interleaved rank-error channels lead to a common row space for the component errors in the vertical case and to a shared column space in the horizontal setting.
The decoding of Gabidulin codes was studied for vertical interleaving~\cite{LoidreauOverbeck2006DecodingRankErrors,Wachter-ZehZeh2014ListUniqueError,wachter2013decoding,BartzJerkovitsEtAl2021FastDecodingCodes} as well as for horizontal interleaving~\cite{li2014transform,SidorenkoBossert2010DecodingInterleavedGabidulin,Sidorenko2011SkewFeedback,PuchingerRosenkildeneNielsenEtAl2016Rowreductionapplied,PuchingerMueelichEtAl2017DecodingInterleavedGabidulin}.
Moreover, interleaving of rank-metric codes and similar ideas were used in cryptographic proposals such as Durandal~\cite{AragonBlazyEtAl2019DurandalRankMetric}, LIGA~\cite{RennerPuchingerEtAl2021LigaCryptosystemBased}, LowMS~\cite{AragonDyserynEtAl2023LowMSnewrank}, and a variant of Loidreau's system~\cite{RennerPuchingerEtAl2019InterleavingLoidreausRank}.

In the sum-rank-metric regime, the decoding of vertically interleaved \acs{LRS} codes and its applications were studied in~\cite{bartz2022fast,bartz2021decoding,BartzPuchinger2024Fastdecodinglifted}.
Further, a Metzner--Kapturowski-like decoder was presented in~\cite{JerkovitsHoermannEtAl2023DecodingHighOrder,JerkovitsHoermannEtAl2024ErrorCodePerspective} and allows to probabilistically decode \emph{any} vertically interleaved linear code with high interleaving order, regardless of the structure of the underlying component code.
In an earlier version of this work, the authors derived a definition of horizontally interleaved \ac{LRS} codes and based their proposal of a Gao-like decoding algorithm in~\cite{HoermannBartz2023FastGaoDecoding} on it.
Note that the Gao-like decoder can correct only errors and a generalization to the error-erasure setting in the spirit of~\cite[Sec.~3.2.3]{wachter2013decoding} will likely introduce restrictions on the block sizes and thus only apply to a smaller family of codes.
We use~\autoref{fig:sum-rank-for-interleaved-vectors} to give an overview on how the definition of the sum-rank weight differs for the vertical and the horizontal setting.
The notions themselves will be discussed in more detail later.

\begin{figure*}[ht]
	\centering
	\subfloat[Vertical interleaving.]{%
			$\begin{gathered}
			\scalebox{.88}{$
				\xvert =
				\begin{pmatrix}
					\x_1 \\
					\x_2 \\
					\vdots \\
					\x_{\intOrder}
				\end{pmatrix}
				=
				\begin{pmatrix}
					\\[-8pt]
					\input{./x1-vector.tikz} \\[8pt]
					\input{./x2-vector.tikz} \\
					\vdots \\[3pt]
					\input{./xs-vector.tikz} \\[7pt]
				\end{pmatrix}
				\qquad \implies
				\SumRankWeight(\xvert) =
				\rkq \left( \myfmargbox{
					\begin{matrix}
						\colorblock{fillblue}{1}{1} \\[4pt]
						\colorblock{fillblue}{2}{1} \\[-1pt]
						\vdots \\
						\colorblock{fillblue}{\intOrder}{1}
					\end{matrix}
				} \right)
				+ \rkq \left( \myfmargbox{
					\begin{matrix}
						\colorblock{fillred}{1}{2} \\[4pt]
						\colorblock{fillred}{2}{2} \\[-1pt]
						\vdots \\
						\colorblock{fillred}{\intOrder}{2}
					\end{matrix}
				} \right)
				+ \dots + \rkq \left( \myfmargbox{
					\begin{matrix}
						\colorblock{fillgreen}{1}{\shots} \\[4pt]
						\colorblock{fillgreen}{2}{\shots} \\[-1pt]
						\vdots \\
						\colorblock{fillgreen}{\intOrder}{\shots}
					\end{matrix}
				} \right)
			$}
		\end{gathered}$}
		\\[15pt]
	\subfloat[Horizontal interleaving.]{%
		$\begin{gathered}
			\scalebox{.88}{$
				\xhor = \left( \x_1 \mid \x_2 \mid \dots \mid \x_{\intOrder} \right) =
				\raisedLeftPar \input{./x1-vector.tikz} ~ \input{./x2-vector.tikz} ~ \cdots ~ \input{./xs-vector.tikz} \raisedRightPar
			$}
			\\[7pt]
			\scalebox{.88}{$
				\implies \SumRankWeight(\xhor) =
				\rkq \raisedLeftPar \input{./x-shot1-vector.tikz} \raisedRightPar
				+ \rkq \raisedLeftPar \input{./x-shot2-vector.tikz} \raisedRightPar
				+ \dots + \rkq \raisedLeftPar \input{./x-shotell-vector.tikz} \raisedRightPar
			$}
		\end{gathered}$}

	\caption{Illustration of the sum-rank weight for vertically and horizontally interleaved vectors.}
	\label{fig:sum-rank-for-interleaved-vectors}
\end{figure*}

\acresetall

\subsubsection*{Contributions}
We present syndrome-based error-only and error-erasure decoders for \ac{VILRS} codes as well as for \ac{HILRS} codes.
Up to our knowledge, this is the first publication giving a unified presentation of \emph{both} types of interleaving, which allows us to highlight the similarities and differences between the two concepts.

The presented decoders are syndrome-based, i.e., they follow a Berlekamp--Massey-like approach and rely on the recovery of the \ac{ELP} or the \ac{ESP}, respectively.
While both approaches are equivalent for non-interleaved \ac{LRS} codes~\cite{hoermann2022error_erasure}, each of the two approaches is tailored to one interleaving type.
More precisely, the decoding of \acs{VILRS} codes relies on an \ac{ELP} key equation and \acs{HILRS} codes are decoded by means of an \ac{ESP} key equation.

Our decoding algorithms are the first syndrome-based \emph{and} the first error-erasure decoders for \acs{HILRS} and \acs{VILRS} codes.
Since we provide fast subroutines for certain steps in the decoding process, the algorithms require on average only $\softoh{\intOrder n^2}$ operations in the ambient field $\Fqm$ for errors and erasures, and $O(\intOrder n^2)$ operations in the error-only setting given that the interleaving order $\intOrder$ grows at most linearly in $m$ and $n$ denotes the length of the component code.

The error-only decoders for \acs{VILRS} and \acs{HILRS} codes guarantee successful decoding up to the unique-decoding radius $\tfrac{1}{2}(n-k)$ and probabilistic unique decoding succeeds with high probability for errors of weight at most $\tfrac{\intOrder}{\intOrder + 1}(n - k)$.
Here, $n$ and $k$ denote the length and the dimension of the component code, respectively, and $\intOrder$ is the interleaving order.

In the error-erasure setting, we focus on errors of sum-rank weight $\numbErrors = \numbFullErrors + \numbRowErasures + \numbColErasures$ that can be decomposed into $\numbFullErrors$ full errors, $\numbRowErasures$ row erasures, and $\numbColErasures$ column erasures.
Our decoders guarantee unique decoding for both code classes when the error satisfies $\numbFullErrors \leq \tfrac{1}{2} (n - k - \numbRowErasures - \numbColErasures)$.
Further, errors of larger weight can be decoded with high probability when $\numbFullErrors \leq \tfrac{\intOrder}{\intOrder + 1}(n - k - \numbRowErasures - \numbColErasures)$ applies.
In fact, the exact decoding radii for guaranteed and probabilistic decoding in the error-erasure regime are slightly better but differ for \acs{VILRS} and \acs{HILRS} codes.

\subsubsection*{Outline}

After the preliminaries, this paper contains two main parts:
\autoref{sec:vilrs} and~\autoref{sec:hilrs} are devoted to \emph{vertically} and \emph{horizontally} interleaved \ac{LRS} codes and their syndrome-based decoding, respectively.
We start each of these sections with describing the corresponding concept and the considered error and channel models in the sum-rank metric, and then present an error-only decoder as well as its generalization to the error-erasure setting.

In~\autoref{sec:simulations}, we then showcase experimental results to verify the tightness of the obtained upper bounds on the decoding-failure probabilities.
\autoref{sec:fast-dec} contains short descriptions of fast subroutines that we use to speed up the derived decoders.
Finally, we conclude in~\autoref{sec:conclusion} and give an outlook on future work and applications of the presented material.

%% file: error_decomposition.tikz
\begin{tikzpicture}[node distance=0cm, ampersand replacement=\&]
	\matrix [draw] (e)
	{
	\node [draw, minimum width=.3cm, minimum height=.3cm] (e1) {};
	\&
	\node (edots) {$\dots$};
	\&
	\node [draw, minimum width=.3cm, minimum height=.3cm] (el) {};
	\\
	};

	\node [right=of e] (equal) {$=$};

	\matrix [draw, pattern color=red!50!white, pattern=north east lines, right=of equal] (aF)
	{
	\node [draw, minimum width=.3cm, minimum height=.3cm] (aF1) {};
	\&
	\node (aFdots) {$\dots$};
	\&
	\node [draw, minimum width=.3cm, minimum height=.3cm] (aFl) {};
	\\
	};

	\node [right=of aF] (firstTimes) {$\cdot$};

	\matrix [draw, pattern color=red!50!white, pattern=north east lines, right=of firstTimes] (BF)
	{
	\node [draw, minimum width=.3cm, minimum height=.3cm] (BF1) {};
	\& \& \\[-7pt]
	\&
	\node (BFdots) {$\ddots$};
	\& \\[-3pt]
	\& \&
	\node [draw, minimum width=.3cm, minimum height=.3cm] (BFl) {};
	\\
	};

	\node [right=of BF] (firstPlus) {$+$};

	\matrix [draw, fill=Green!30!white, right=of firstPlus] (aR)
	{
	\node [draw, minimum width=.3cm, minimum height=.3cm] (aR1) {};
	\&
	\node (aRdots) {$\dots$};
	\&
	\node [draw, minimum width=.3cm, minimum height=.3cm] (aRl) {};
	\\
	};

	\node [right=of aR] (secondTimes) {$\cdot$};

	\matrix [draw, pattern color=red!50!white, pattern=north east lines, right=of secondTimes] (BR)
	{
	\node [draw, minimum width=.3cm, minimum height=.3cm] (BR1) {};
	\& \& \\[-7pt]
	\&
	\node (BRdots) {$\ddots$};
	\& \\[-3pt]
	\& \&
	\node [draw, minimum width=.3cm, minimum height=.3cm] (BRl) {};
	\\
	};

	\node [right=of BR] (secondPlus) {$+$};

	\matrix [draw, pattern color=red!50!white, pattern=north east lines, right=of secondPlus] (aC)
	{
	\node [draw, minimum width=.3cm, minimum height=.3cm] (aC1) {};
	\&
	\node (aCdots) {$\dots$};
	\&
	\node [draw, minimum width=.3cm, minimum height=.3cm] (aCl) {};
	\\
	};

	\node [right=of aC] (thirdTimes) {$\cdot$};

	\matrix [draw, fill=Green!30!white, right=of thirdTimes] (BC)
	{
	\node [draw, minimum width=.3cm, minimum height=.3cm] (BC1) {};
	\& \& \\[-7pt]
	\&
	\node (BCdots) {$\ddots$};
	\& \\[-3pt]
	\& \&
	\node [draw, minimum width=.3cm, minimum height=.3cm] (BCl) {};
	\\
	};

	\node (etext) at ([yshift=-10pt] e.center |- BF.south east) {$\e$};

	\node (equalText) at ([yshift=-10pt] equal.center |- BF.south east) {$=$};

	\node (aFtext) at ([yshift=-10pt] aF.center |- BF.south east) {$\a_{\indFullErrors}$};
	\node (firstTimesText) at ([yshift=-10pt] firstTimes.center |- BF.south east) {$\cdot$};
	\node (BFtext) at ([yshift=-10pt] BF.center |- BF.south east) {$\B_{\indFullErrors}$};

	\node (firstPlusText) at ([yshift=-10pt] firstPlus.center |- BF.south east) {$+$};

	\node (aRtext) at ([yshift=-10pt] aR.center |- BF.south east) {$\a_{\indRowErasures}$};
	\node (secondTimesText) at ([yshift=-10pt] secondTimes.center |- BF.south east) {$\cdot$};
	\node (BRtext) at ([yshift=-10pt] BR.center |- BF.south east) {$\B_{\indRowErasures}$};

	\node (secondPlusText) at ([yshift=-10pt] secondPlus.center |- BF.south east) {$+$};

	\node (aCtext) at ([yshift=-10pt] aC.center |- BF.south east) {$\a_{\indColErasures}$};
	\node (thirdTimesText) at ([yshift=-10pt] thirdTimes.center |- BF.south east) {$\cdot$};
	\node (BCtext) at ([yshift=-10pt] BC.center |- BF.south east) {$\B_{\indColErasures}$};

	\draw [decorate, decoration = {brace, mirror, amplitude=8pt}]
			([yshift=-18pt] aF.south west |- BF.south east) --  ([yshift=-18pt] BF.south east)
			node (full) [pos=0.5, below=10pt, align=center] {full errors};
	\draw [decorate, decoration = {brace, mirror, amplitude=8pt}]
			([yshift=-18pt] aR.south west |- BR.south east) --  ([yshift=-18pt] BR.south east)
			node (row) [pos=0.5, below=10pt, align=center] {row erasures};
	\draw [decorate, decoration = {brace, mirror, amplitude=8pt}]
			([yshift=-18pt] aC.south west |- BC.south east) --  ([yshift=-18pt] BC.south east)
			node (column) [pos=0.5, below=10pt, align=center] {column erasures};
\end{tikzpicture}%

%% file: x1-vector.tikz
\begin{tikzpicture}[baseline=(x11.base)]
    \node[box,fillblue] (x11) {$\subShot{\x}{1}{1}$};
    \node[box,fillred,right=of x11] (x12) {$\subShot{\x}{1}{2}$};
    \node[box,right=\leftDotsDist of x12] (x12dots) {$\cdots$};
    \node[box,fillgreen,right=\rightDotsDist of x12dots] (x1ell) {$\subShot{\x}{1}{\shots}$};

    \draw ([xshift=-\frameDist, yshift=-\frameDist] x11.south west) rectangle ([xshift=\frameDist, yshift=\frameDist] x1ell.north east);

\end{tikzpicture}

%% file: x2-vector.tikz
\begin{tikzpicture}[baseline=(x21.base)]
    \node[box,fillblue] (x21) {$\subShot{\x}{2}{1}$};
    \node[box,fillred,right=of x21] (x22) {$\subShot{\x}{2}{2}$};
    \node[box,right=\leftDotsDist of x22] (x22dots) {$\cdots$};
    \node[box,fillgreen,right=\rightDotsDist of x22dots] (x2ell) {$\subShot{\x}{2}{\shots}$};

    \draw ([xshift=-\frameDist, yshift=-\frameDist] x21.south west) rectangle ([xshift=\frameDist, yshift=\frameDist] x2ell.north east);
\end{tikzpicture}

%% file: xs-vector.tikz
\begin{tikzpicture}[baseline=(xs1.base)]
    \node[box,fillblue] (xs1) {$\subShot{\x}{\intOrder}{1}$};
    \node[box,fillred,right=of xs1] (xs2) {$\subShot{\x}{\intOrder}{2}$};
    \node[box,right=\leftDotsDist of xs2] (xs2dots) {$\cdots$};
    \node[box,fillgreen,right=\rightDotsDist of xs2dots] (xsell) {$\subShot{\x}{\intOrder}{\shots}$};

    \draw ([xshift=-\frameDist, yshift=-\frameDist] xs1.south west) rectangle ([xshift=\frameDist, yshift=\frameDist] xsell.north east);
\end{tikzpicture}

%% file: x-shot1-vector.tikz
\begin{tikzpicture}[baseline=(x11.base)]
    \node[box,fillblue] (x11) {$\subShot{\x}{1}{1}$};
    \node[box,fillblue,right=of x11] (x21) {$\subShot{\x}{2}{1}$};
    \node[box,right=\leftDotsDist of x21] (x21dots) {$\cdots$};
    \node[box,fillblue,right=\rightDotsDist of x21dots] (xs1) {$\subShot{\x}{\intOrder}{1}$};

    \draw ([xshift=-\frameDist, yshift=-\frameDist] x11.south west) rectangle ([xshift=\frameDist, yshift=\frameDist] xs1.north east);
\end{tikzpicture}

%% file: x-shot2-vector.tikz
\begin{tikzpicture}[baseline=(x12.base)]
    \node[box,fillred] (x12) {$\subShot{\x}{1}{2}$};
    \node[box,fillred,right=of x12] (x22) {$\subShot{\x}{2}{2}$};
    \node[box,right=\leftDotsDist of x22] (x22dots) {$\cdots$};
    \node[box,fillred,right=\rightDotsDist of x22dots] (xs2) {$\subShot{\x}{\intOrder}{2}$};

    \draw ([xshift=-\frameDist, yshift=-\frameDist] x12.south west) rectangle ([xshift=\frameDist, yshift=\frameDist] xs2.north east);
\end{tikzpicture}

%% file: x-shotell-vector.tikz
\begin{tikzpicture}[baseline=(x1ell.base)]
    \node[box,fillgreen] (x1ell) {$\subShot{\x}{1}{\shots}$};
    \node[box,fillgreen,right=of x1ell] (x2ell) {$\subShot{\x}{2}{\shots}$};
    \node[box,right=\leftDotsDist of x2ell] (x2elldots) {$\cdots$};
    \node[box,fillgreen,right=\rightDotsDist of x2elldots] (xsell) {$\subShot{\x}{\intOrder}{\shots}$};

    \draw ([xshift=-\frameDist, yshift=-\frameDist] x1ell.south west) rectangle ([xshift=\frameDist, yshift=\frameDist] xsell.north east);
\end{tikzpicture}

%% file: preliminaries.tex
\section{Preliminaries}\label{sec:preliminaries}

We fix a prime power $q$ and denote the finite field of order $q$ by $\Fq$.
Like~\cite{bartz2022fast}, we use the following constant depending on the field size $q$ in some expressions:
\begin{equation}\label{eq:def_kappa_q}
    \kappa_q \defeq \prod_{i=1}^{\infty} \frac{1}{1-q^{-i}} \quad \text{for any } q \in \NN \text{ with } q \geq 2.
\end{equation}
As a sequence, $\kappa_q$ is monotonically decreasing for growing $q$ and converges to $1$ quickly.
In fact, the first values are $\kappa_2 \approx 3.463$, $\kappa_3 \approx 1.785$, and $\kappa_4 \approx 1.452$.
The bound $\kappa_q < 3.5$ is conservative and even $\kappa_q < 2$ applies for non-binary fields with $q > 2$.

We further let $\Fqm$ be an extension field of $\Fq$ and choose a field automorphism $\aut$ on $\Fqm$ whose fixed field is precisely $\Fq$.
We write e.g.\ $\aut(\a)$ to denote the {elementwise} application of $\aut$ to a vector $\a$.
More generally, we use this notation to evaluate unary functions and operators on vectors and matrices elementwise.

We work with \emph{row} vectors and usually denote vectors and matrices by bold lowercase and uppercase letters, respectively.
Moreover, we often divide length-$n$ vectors into $\shots$ blocks and call the vector $\n = (n_1, \dots, n_{\shots})$ of the block lengths a \emph{length partition} of $n$.
A length partition contains only positive integers and satisfies $\sum_{i=1}^{\shots} n_i = n$.
We denote the corresponding block representation of a vector $\x \in \Fqm^n$ by $\x = \left( \shot{\x}{1} \mid \dots \mid \shot{\x}{\shots} \right)$ with $\shot{\x}{i} \in \Fqm^{n_i}$ for all $i = 1, \dots, \shots$ and extend this notation to matrices.

At some points, it will be useful to represent elements of $\Fqm$ over $\Fq$.
Let us therefore fix an arbitrary ordered $\Fq$-basis $\vecgamma = (\gamma_1, \dots, \gamma_m)$ of $\Fqm$ and consider the induced vectorspace isomorphism $\ext: \Fqm \to \Fq^{m}$ which \emph{extends} an $\Fqm$-element to its $\Fq$-representation.
Namely, an element $x \in \Fqm$ is mapped to its coefficient vector $\coeffq{x} \defeq (x_1, \dots, x_m) \in \Fq^{m}$ with respect to the basis $\vecgamma$, which means that $x_1, \dots, x_m \in \Fq$ are chosen such that $x = \sum_{i=1}^{m} x_i \gamma_i$ holds.
When we consider a vector $\x \in \Fqm^{n}$, we represent its entries over $\Fq$ and collect the $\Fq$-representations $\coeffq{x_1}, \dots, \coeffq{x_n}$ as columns in the matrix $\coeffq{\x} \in \Fq^{m \times n}$.
Similarly, the notation $\coeffq{\X} \in \Fq^{s m \times n}$ for a matrix $\X \in \Fqm^{s \times n}$ with rows $\x_1, \dots, \x_s \in \Fqm^{n}$ denotes the $\Fq$-matrix obtained by stacking the $\Fq$-representations $\coeffq{\x_1}, \dots, \coeffq{\x_s}$ of its rows.

We use the common $O(\cdot)$-notation to state the asymptotic time complexity of discussed algorithms.
Moreover, we adopt $\softO(\cdot)$ to indicate that we neglect logarithmic factors.
We usually consider operations in $\Fqm$ and assume that an operation in an extension field $\Fqms$ of $\Fqm$ can be executed in $\softO(\intOrder)$ operations in $\Fqm$ as explained in~\cite[Sec.~II.A]{BartzJerkovitsEtAl2021FastDecodingCodes}.
This is reasonable because a suitable $\Fqm$-basis of $\Fqms$ can be chosen in all cases according to~\cite{CouveignesLercier2009EllipticPeriodsFinite}.

\subsection{Sum-Rank-Metric Codes}

The \emph{sum-rank weight} of a vector $\x \in \Fqm^n$ with respect to the length partition $\nVec \in \NN^{\shots}$ is
\begin{equation}
    \label{eq:sum-rank_weight}
    \SumRankWeightWPartition{\nVec}(\x) = \sum_{i=1}^{\shots} \rkq(\x^{(i)}),
\end{equation}
where $\rkq(\shot{\x}{i})$ denotes the number of $\Fq$-linearly independent entries of $\shot{\x}{i}$ for every $i = 1, \dots, \shots$~\cite[Def.~25]{martinez2018skew}.
Equivalently, $\rkq(\shot{\x}{i})$ with $i = 1, \dots, \shots$ can be interpreted as the $\Fq$-rank of the matrix $\coeffq{\shot{\x}{i}} \in \Fq^{m \times n_i}$.
We further define the \emph{rank partition} of $\x$ as the vector $\numbErrorsVec = (\numbErrorsInBlock{1}, \dots, \numbErrorsInBlock{\shots}) \in \NN^{\shots}$ with $\numbErrorsInBlock{i} \defeq \rkq(\shot{\x}{i})$ for all $i = 1, \dots, \shots$ to emphasize how the sum-rank weight $\numbErrors = \sum_{i=1}^{\shots} \numbErrorsInBlock{i}$ of $\x$ splits across the $\shots$ blocks.
The sum-rank weight induces the \emph{sum-rank metric} on $\Fqm^n$, which is explicitly given as $\SumRankDistWPartition{\nVec}(\x, \y) \defeq \SumRankWeightWPartition{\nVec}(\x - \y)$ for all $\x, \y \in \Fqm^n$.
When $\nVec$ is clear from the context, we simply write $\SumRankWeight$ and $\SumRankDist$, respectively.
Note that the sum-rank metric covers the Hamming metric and the rank metric as special cases for $\shots = n$ and $\shots = 1$, respectively.

This paper considers \emph{$\Fqm$-linear} sum-rank-metric codes, that is, $\Fqm$-linear subspaces of $\Fqm^n$ endowed with the sum-rank metric.
A \emph{linear code} $\mycode{C} \subseteq \Fqm^n$ has dimension $k \defeq \dim_{q^m}(\mycode{C})$ and length $n$.
It can be represented as the $\Fqm$-row space of a full-rank \emph{generator matrix} $\G \in \Fqm^{k \times n}$ or, equivalently, as the kernel of a full-rank \emph{parity-check matrix} $\H \in \Fqm^{(n-k) \times n}$.
Its \emph{minimum sum-rank distance} is
\begin{align}
    \SumRankDist
    &\defeq \min_{{\x, \y \in \mycode{C}, \x \neq \y}} \{ \SumRankDist(\x, \y)\}
    = \min_{{\x \in \mycode{C}, \x \neq \0}} \{ \SumRankWeight(\x)\}.
\end{align}
The sum-rank analog of the Singleton bound reads $\SumRankDist(\mycode{C}) \leq n-k+1$~\cite[Prop.~34]{martinez2018skew} and codes attaining it with equality are called \emph{\acf{MSRD} codes}.

\subsection{Skew Polynomials}

The non-commutative \emph{skew-polynomial ring} $\SkewPolyringZeroDer$ was first studied by Ore~\cite{ore1933special,ore1933theory} and contains all formal polynomials $\sum_i f_i x^{i-1}$ with finitely many nonzero coefficients $f_i \in \Fqm$.
It is equipped with ordinary polynomial addition and a non-commutative multiplication determined by the rule $x f_i = \aut(f_i) x$ for all $f_i \in \Fqm$.
The notion of \emph{degree} naturally carries over from classical polynomials and we denote the set of all skew polynomials of degree at most $k - 1$ by $\SkewPolyringZeroDer_{<k} \defeq \{f \in \SkewPolyringZeroDer: \deg(f) < k\}$.

$\SkewPolyringZeroDer$ is a right Euclidean ring which ensures for every pair $f,g \in \SkewPolyringZeroDer$ the existence of skew polynomials $q_r$ and $r_r$ such that $f(x) = q_r(x) g(x) + r_r(x)$ and $\deg(r_r) < \deg(g)$ holds~\cite{ore1933theory}.
We write $(f \modr g)(x) \defeq r_r(x)$ and if $(f \modr g)(x) = 0$ applies, we call $g$ a \emph{right divisor} of $f$ and write $\rdiv{f, g}(x) \defeq q_r(x)$.
Since $\SkewPolyringZeroDer$ is also a \emph{left} Euclidean ring, we similarly obtain skew polynomials $q_l$ and $r_r$ for every $f, g \in \SkewPolyringZeroDer$ with $f(x) = g(x) q_l(x) + r_l(x)$ and $\deg(r_l) < \deg(g)$~\cite{ore1933theory}.
Again, we use the notation $(f \modl g)(x) \defeq r_l(x)$ and call $g$ a \emph{left divisor} of $f$ if $(f \modl g)(x) = 0$ holds.
In the latter case, we write $\ldiv{f, g}(x) \defeq q_l(x)$.
The \emph{\ac{gcrd}} and the \emph{\ac{lclm}} of two skew polynomials $f,g \in\SkewPolyringZeroDer$ are denoted by $\gcrd(f,g)$ and $\lclm(f,g)$, respectively.
We assume that both \ac{gcrd} and \ac{lclm} are monic to ensure their uniqueness.

The product $p = f \cdot g \in \SkewPolyringZeroDer$ of two skew polynomials $f, g \in \SkewPolyringZeroDer$ with $d_f \defeq \deg(f)$ and $d_g \defeq \deg(g)$ has degree $d_f + d_g$.
The coefficients $p_l$ of $p$ with $l = \min(d_f, d_g) + 1, \dots, \max(d_f, d_g) + 1$ can be computed as follows~\cite{li2014transform}:
\begin{equation}
    p_l =
    \begin{cases}
        \sum\limits_{i=1}^{d_f+1}f_i\aut^{i-1}(g_{l-i+1})
        & \text{if } d_f \leq d_g
        \\
        \sum\limits_{i=1}^{d_g+1}f_{l-i+1}\aut^{l-i}(g_{i})
        &\text{if } d_f > d_g.
    \end{cases}
    \label{eq:skew_product_coeffs}
\end{equation}

Another useful transformation for skew polynomials is the \emph{skew reverse}.
The {$\aut$-reverse} of an $f \in \SkewPolyringZeroDer$ with respect to a fixed integer $t \geq \deg(f)$ is defined as $\skewrev{f}(x) = \sum_{i=1}^{t+1} \skewrev{f}_i x^{i-1} \in \SkewPolyringZeroDer$ with coefficients $\skewrev{f}_i = \aut^{i-t-1}(f_{t-i+2})$ for all $i=1,\dots,t+1$~\cite[p.~574]{li2014transform},~\cite[Sec.~2.4]{silva2009error}.

Next, we define the generalized operator evaluation for skew polynomials, which requires the definition of $\genNorm{i}{a}$ for $i \geq 0$ and $a \in \Fqm$.
We set $\genNorm{0}{a} = 1$ and proceed iteratively with $\genNorm{i}{a}=\aut^{i-1}(a) \cdot \genNorm{i-1}{a} = \prod_{j=0}^{i-1}\aut^{j}(a)$ for all $i > 0$~\cite{lam1988vandermonde}.
Since certain expressions containing this function will appear in later proofs, we anticipate technical auxiliary results in the next lemma:
\begin{lemma}\label{lem:aut_and_norms}
    The following equalities hold for two integers $\alpha, \beta \geq 0$ and an element $x \in \Fqm$:
    \begin{enumerate}
        \item
            $\aut^{\alpha}\bigl( \genNormInv{\beta}{\autinv(x)} \bigr) =
            \begin{cases}
                \genNormInv{\beta-\alpha}{\autinv(x)} \cdot \genNorm{\alpha}{x} &
                \text{if } \beta \geq \alpha
                \\
                \genNorm{\alpha-\beta+2}{\autinv(x)} \cdot \genNorm{\alpha}{\autinv(x^{-1})} &
                \text{if } \beta < \alpha
            \end{cases}
            $ \hfill \refstepcounter{equation} (\theequation) \label{eq:normlemma-1}
        \item
            $\aut^{-\alpha}\bigl( \genNormInv{\beta}{\autinv(x)} \bigr)
            = \genNormInv{\alpha+\beta}{\autinv(x)} \cdot \genNormInv{\alpha}{\autinv(x^{-1})}$
            \hfill \refstepcounter{equation} (\theequation) \label{eq:normlemma-2}
        \item
            $\aut^{-\alpha}\bigl( \genNorm{\beta}{x} \bigr)
            = \genNormInv{\alpha}{\autinv(x)} \cdot \genNormInv{\alpha-\beta}{\autinv(x^{-1})}$
            \hfill \refstepcounter{equation} (\theequation) \label{eq:normlemma-3}
    \end{enumerate}
\end{lemma}

\begin{IEEEproof}
    We start with the proof of 1) and 2) by assuming that $\alpha \geq 1$ holds since the case $\alpha = 0$ is straightforward to check.
    We obtain
    \begin{align*}
        \aut^{\pm\alpha}\bigl( \genNormInv{\beta}{\autinv(x)}) \bigr)
        &= \aut^{\pm\alpha}\bigl( \aut^{-(\beta-1)}(\autinv(x)) \dots \autinv(x) \bigr)
        = \aut^{-(\beta\mp\alpha)}(x) \dots \aut^{\pm\alpha-1}(x)
    \end{align*}
    by definition and the first statement follows for $\beta \geq \alpha$ from
    \begin{align*}
        \aut^{\alpha}\bigl( \genNormInv{\beta}{\autinv(x)}) \bigr)
        &= \aut^{-(\beta-\alpha-1)}(\autinv(x)) \dots \autinv(x) \cdot x \dots \aut^{\alpha}(\autinv(x))
        = \genNormInv{\beta-\alpha}{\autinv(x)} \cdot \genNorm{\alpha}{x}
    \end{align*}
    and for $\beta < \alpha$ from
    \begin{align*}
        \aut^{\alpha}\bigl( \genNormInv{\beta}{\autinv(x)}) \bigr)
        &= \aut^{\alpha-\beta+1}(\autinv(x)) \dots \autinv(x) \cdot \aut^{\alpha-1}(\autinv(x^{-1})) \dots \autinv(x^{-1})
        = \genNorm{\alpha-\beta+2}{\autinv(x)} \cdot \genNorm{\alpha}{\autinv(x^{-1})}.
    \end{align*}
    As $\alpha + \beta \geq 0$ is always satisfied, there is no need for a case distinction for the second statement and we obtain
    \begin{align*}
        \aut^{-\alpha}\bigl( \genNormInv{\beta}{\autinv(x)}) \bigr)
        &= \aut^{-(\alpha+\beta-1)}(\autinv(x)) \dots \autinv(x) \cdot \aut^{-(\alpha-1)}(\autinv(x^{-1})) \dots (\autinv(x^{-1}))
        \\
        &= \genNormInv{\alpha+\beta}{\autinv(x)} \cdot \genNormInv{\alpha}{\autinv(x^{-1})}.
    \end{align*}
    
    We now prove 3) and focus on the nontrivial case $\beta \geq 1$, which yields
    \begin{align}
        \aut^{-\alpha}\bigl( \genNorm{\beta}{x} \bigr)
        &= \aut^{-(\alpha-\beta)}(\autinv(x)) \dots \aut^{-(\alpha-1)}(\autinv(x))
        = \aut^{-(\alpha-1)}(\autinv(x)) \dots \autinv(x) \cdot \aut^{-(\alpha-\beta-1)}(\autinv(x^{-1})) \dots \autinv(x^{-1})
        \\
        &= \genNormInv{\alpha}{\autinv(x)} \cdot \genNormInv{\alpha-\beta}{\autinv(x^{-1})}.
    \end{align}
\end{IEEEproof}

Let us define the operator $\opfull{a}{b} \defeq \aut(b)a$ for all $a,b \in \Fqm$ and set $\opfullexp{a}{b}{0} \defeq b$.
Further, consider its powers $\opfullexp{a}{b}{i} \defeq \opfull{a}{\opfullexp{a}{b}{i-1}} = \aut^i(b) \cdot \genNorm{i}{a}$ for all $i>0$ and any $a, b \in \Fqm$~\cite[Prop.~32]{martinez2018skew}.
The \emph{generalized operator evaluation} of a skew polynomial $f(x) = \sum_i f_i x^{i-1} \in \SkewPolyringZeroDer$ at an element $b \in \Fqm$ with respect to an evaluation parameter $a \in \Fqm$ was studied in e.g.\ \cite{leroy1995pseudolinear,martinez2018skew} and is defined as
\begin{equation}\label{eq:def_gen_op_eval}
  \opev{f}{b}{a} \defeq \sum_{i}f_i\opfullexp{a}{b}{i-1}.
\end{equation}
Note that there is another meaningful way to evaluate skew polynomials which is called \emph{remainder evaluation}~\cite{lam1988vandermonde,martinez2018skew}.
We do not consider remainder evaluation in this paper but it has interesting connections to the generalized operator evaluation.

\begin{remark}
    Generalized operator evaluation collapses to the usual evaluation of classical polynomials for the identity automorphism $\aut = \Id$ and evaluation parameter $a = 1$ and to the evaluation of linearized polynomials~\cite{ore1933special} for the Frobenius automorphism $\aut = \cdot^q$ and evaluation parameter $a = 1$.
\end{remark}

When the product of two skew polynomials $f,g\in\SkewPolyringZeroDer$ is evaluated at $b \in \Fqm$ with respect to $a \in \Fqm$, the generalized operator evaluation follows the product rule $\opev{(f\cdot g)}{b}{a} = \opev{f}{\opev{g}{b}{a}}{a}$~\cite{martinez2019private}.
Moreover, it is useful to note that generalized operator evaluation $\opev{f}{\cdot}{a}$ with a fixed evaluation parameter $a$ is an $\Fq$-linear map because $\aut$ fixes $\Fq$~\cite{leroy1995pseudolinear}.
We write $\opev{f}{\b}{a}$ to denote the vector of the evaluations of $f$ at all entries of a vector $\b$ and use the same notation also for matrices.
When we consider a vector $\b$ with respect to a length partition $\n$ and we choose $\shots$ evaluation parameters $a_1, \dots, a_\shots \in \Fqm$, we use
\begin{align*}
    \opev{f}{\b}{\a} \defeq
    \bigl(  \opev{f}{\shot{\b}{1}}{a_1} \bigm\vert \dots \bigm\vert   \opev{f}{\shot{\b}{\shots}}{a_\shots} \bigr) \in \Fqm^n
\end{align*}
to denote the elementwise evaluation of $f$ at the entries of $\b$, where the $i$-th evaluation parameter $a_i$ is used for the entries of the $i$-th block $\shot{\b}{i}$ for $i = 1, \dots, \shots$.
We use the notation $\opfull{\a}{\b}$ for the operator $\opfull{\cdot}{\cdot}$ and its powers in a similar fashion.

The generalized operator evaluation of a skew polynomial $f(x) = \sum_{i=1}^{d} f_i x^{i-1}$ can also be expressed as a vector-vector product.
In particular, the evaluation vector $\opev{f}{\b}{\a} \in \Fqm^{n}$ for evaluation points $\b \in \Fqm^{n}$ and evaluation parameters $\a \in \Fqm^{\shots}$ is the product of the coefficient vector $(f_1, \dots, f_{d})$ of $f$ and the \emph{generalized Moore matrix} $\opMoore{d}{\b}{\a}$ which was studied in e.g.\ \cite{lam1988vandermonde} and is given by
\begin{equation}\label{eq:def_gen_moore_mat}
    \opMoore{d}{\b}{\a} \defeq
    \begin{pmatrix}
        \b
        \\ 
        \opfull{\a}{\b}
        \\ 
        \vdots 
        \\ 
        \opfullexp{\a}{\b}{d-1}
    \end{pmatrix} \in \Fqm^{d \times n}.
\end{equation}
We further use the notation
\begin{equation*}
    \opMoore{d}{\B}{\a} \defeq
    \begin{pmatrix}
        \opMoore{d}{\b_1}{\a}
        \\ 
        \vdots 
        \\ 
        \opMoore{d}{\b_\intOrder}{\a}
    \end{pmatrix} \in \Fqm^{\intOrder d \times n}
    \quad \text{for matrices }
    \B =
    \begin{pmatrix}
        \b_1
        \\
        \vdots
        \\
        \b_\intOrder
    \end{pmatrix}
    \in \Fqm^{\intOrder \times n}.
\end{equation*}

We will see that generalized Moore matrices can be used to generate \acl{LRS} codes and we are thus naturally interested in their $\Fqm$-rank.
In fact,~\cite[Thm.~2]{martinez2018skew} and~\cite[Thm.~4.5]{lam1988vandermonde} imply that $\rkqm(\opMoore{d}{\b}{\a}) = \min(d, n)$ holds if each block $\shot{\b}{1}, \dots, \shot{\b}{\shots}$ contains $\Fq$-linearly independent entries and if the entries of $\a$ belong to pairwise distinct nontrivial conjugacy classes of $\Fqm$.
Here, the \emph{conjugacy class} of an element $a \in \Fqm$ is the set $\mathfrak{C}(a)\defeq\left\{\aut(c)ac^{-1} :c\in\Fqm^*\right\}$ and this notion depends on the choice of $\aut$~\cite{lam1988vandermonde}.
The set of all conjugacy classes with respect to a fixed $\Fqm$-automorphism $\aut$ is a partition of $\Fqm$ and since we picked an automorphism with fixed field precisely $\Fq$, there are $q-1$ nontrivial conjugacy classes next to the trivial one $\mathfrak{C}(0)$.
In particular, one choice of representatives for all nontrivial conjugacy classes are the powers $1,\pe,\dots,\pe^{q-2}$ of a primitive element $\pe \in \Fqm$, i.e., of an element that generates the multiplicative group $\Fqm^{\ast}$.

The roots of a skew polynomial $f \in \SkewPolyringZeroDer$ with respect to the generalized operator evaluation $\opev{f}{\cdot}{a}$ with fixed evaluation parameter $a \in \Fqm$ form an $\Fq$-linear subspace of $\Fqm$.
The dimension of this subspace is upper-bounded by $\deg(f)$ and equality holds if and only if $f$ divides $p(x) = x^m-\genNorm{m}{a}$, i.e., the polynomial for which $\opev{p}{x}{a} = 0$ holds for all elements $x \in \Fqm$~\cite[Prop.~1.3.4]{caruso2019residues}.
When we consider root spaces of $f$ with respect to different evaluation parameters $a_1, \dots, a_\shots$ and these parameters belong to pairwise distinct nontrivial conjugacy classes of $\Fqm$, the dimensions of the $\shots$ root spaces sum up to at most $\deg(f)$~\cite[Prop.~1.3.7]{caruso2019residues}.
In this setting, equality holds when $f$ divides $\prod_{i=1}^{\shots}(x^m-\genNorm{m}{a_i})$~\cite[Prop.~1.3.7]{caruso2019residues}.

The monic skew polynomial $\mpolArgs{\b}{\a}$ characterized by $\opev{\mpolArgs{\b}{\a}}{\b}{\a} = \0$ is called the \emph{minimal skew polynomial} of the vector $\b \in \Fqm^{n}$ with respect to generalized operator evaluation and evaluation parameters $\a \in \Fqm^{\shots}$.
Recall that every entry of the $i$-th block $\shot{\b}{i}$ is evaluated with respect to the $i$-th evaluation parameter $a_i$ for $i = 1, \dots, \shots$.
The degree of $\mpolArgs{\b}{\a}$ is at most $n$ and equality holds if and only if the entries of each block $\shot{\b}{1}, \dots, \shot{\b}{\shots}$ are $\Fq$-linearly independent and $a_1, \dots, a_{\shots}$ belong to pairwise distinct nontrivial conjugacy classes of $\Fqm$~\cite[Sec.~1.3.1]{caruso2019residues}.
When all entries of $\b$ are nonzero, the minimal skew polynomial can be computed as follows~\cite[Rem.~1.3.6]{caruso2019residues}:
\begin{equation}
    \label{eq:min_poly}
    \mpolArgs{\b}{\a}(x) = \lclm\left(x-\tfrac{\aut(b_{\kappa}^{(i)})a_i}{b_{\kappa}^{(i)}}\right)_{{1 \leq \kappa \leq n_i}, {1 \leq i \leq \shots}}.
\end{equation}

\subsection{Linearized Reed--Solomon Codes}

\Ac{LRS} codes are a prominent family of sum-rank-metric codes and can be described as evaluation codes with respect to the generalized operator evaluation of skew polynomials.
We make use of the concepts and notations from the previous section for their definition.

\begin{definition}[{Linearized Reed--Solomon Codes~\cite[Def.~31]{martinez2018skew}}]\label{def:LRS_codes}
    Let $\vecbeta \in \Fqm^{n}$ be a vector whose blocks $\shot{\vecbeta}{1}, \dots, \shot{\vecbeta}{\shots}$ contain $\Fq$-linearly independent elements and choose $\vecxi \in \Fqm^{\shots}$ with entries belonging to pairwise distinct nontrivial conjugacy classes of $\Fqm$.
    Then, the code
    \begin{equation}
        \linRSWithAut{\aut}{\vecbeta}{\vecxi}{\nVec}{k} \defeq
        \left\{
        \opev{f}{\vecbeta}{\vecxi}
        : f\in\SkewPolyringZeroDer_{<k}\right\} \subseteq \Fqm^{n}
    \end{equation}
    is a \emph{\acf{LRS} code} of length $n$ and dimension $k$.
\end{definition}

Note that the definition of an \ac{LRS} code depends on the choice of the $\Fqm$-automorphism $\aut$.
If $\aut$ is clear from the context, we often omit the index and write $\linRS{\vecbeta}{\vecxi}{\nVec}{k}$ to refer to the respective code.
We further introduce the following shorthand notations for vectors of evaluation parameters because they will recur throughout the remainder of this paper:
\begin{equation}\label{eq:def_autInvXiVec}
    \vecxi, \quad \vecxi^{-1}, \quad \autInvXiVec \defeq \autinv(\vecxi), \quad \text{and } \quad \autInvXiInvVec \defeq \autinv(\vecxi^{-1}).
\end{equation}
Remark that each of these vectors contains elements of pairwise distinct nontrivial conjugacy classes.

\ac{LRS} codes have minimum sum-rank distance $n-k+1$ and are thus~\ac{MSRD} codes~\cite[Thm. 4]{martinez2018skew}.
Furthermore, the generalized Moore matrix $\opMoore{k}{\vecbeta}{\vecxi}$ is a generator matrix of $\linRS{\vecbeta}{\vecxi}{\nVec}{k}$~\cite[Sec.~3.3]{martinez2018skew}.
The {dual code} of $\linRSWithAut{\aut}{\vecbeta}{\vecxi}{\nVec}{k}$ is $\linRSWithAut{\autinv}{\h}{\autInvXiVec}{\nVec}{n-k}$, where $\h \in \Fqm^{n}$ satisfies $\h \cdot \opfullexp{\vecxi}{\vecbeta}{l-1}^{\top} = \0$ for all $l=1,\dots,n-1$ and has sum-rank weight $\SumRankWeight(\h)=n$~\cite{martinez2019reliable,CarusoDurand2022DualsLinearizedReed}.
In other words, the matrix
\begin{equation}\label{eq:parity_check_mat}
    \H \defeq \opMooreInv{n-k}{\h}{\autInvXiVec}
\end{equation}
is a parity-check matrix of $\linRSWithAut{\aut}{\vecbeta}{\vecxi}{\nVec}{k}$.

%% file: vilrs-codes.tex
\section{Vertically Interleaved Linearized Reed--Solomon (VILRS) Codes and Their Decoding\label{sec:vilrs}}

The vertical interleaving of \ac{LRS} codes was first studied in~\cite{bartz2022fast}, where the authors derived a Loidreau--Overbeck-like and an interpolation-based decoder.
They are tailored to an error-only channel and are the only known decoding schemes for \ac{VILRS} codes so far.
Both allow for decoding beyond the unique-decoding radius and can be used as probabilistic unique decoders, which return either the correct solution or a decoding failure.
Further, the interpolation-based approach can be adapted to the list-decoding setting such that the decoder outputs a list of all codewords within a sum-rank ball of a certain radius around the received word.

We introduce a syndrome-based decoding algorithm for \ac{VILRS} codes, which we first present in the error-only setting that was considered for the previously known decoders.
Additionally, we generalize our approach to an error-erasure channel model that allows to correct full errors as well as row and column erasures jointly.
This is an extension of the \acs{ELP}-based error-erasure decoder for non-interleaved \ac{LRS} codes from~\cite{hoermann2022error_erasure}.
Observe that the presented approach has an advantage compared to a potential generalization of the faster interpolation-based error-erasure decoder for vertically interleaved Gabidulin codes from~\cite{Wachter-ZehZeh2014ListUniqueError}.
Namely, the interpolation-based approach requires the code length to equal the extension degree of the ambient finite field and this restriction will most likely carry over to the sum-rank metric in a blockwise manner.
Our syndrome-based decoder, however, does not have these parameter restrictions and can thus decode a larger family of codes.

\subsection{Vertical Interleaving in the Sum-Rank Metric}

Consider an arbitrary $\Fqm$-linear sum-rank-metric code $\mycode{C} \subseteq \Fqm^{n}$ with respect to a length partition $\n = (n_1, \dots, n_{\shots})$.
We define the \emph{vertically interleaved code} $\VInt(\mycode{C}, \intOrder)$ as
\begin{equation}
    \VInt(\mycode{C}, \intOrder) \defeq \left\{
    \C =
    \begin{pmatrix}
        \c_1 \\
        \vdots \\
        \c_{\intOrder}
    \end{pmatrix}
    : \c_j \in \mycode{C} \text{ for all } j = 1, \dots, \intOrder
    \right\} \subseteq \Fqm^{\intOrder \times n}
\end{equation}
and call $\intOrder \in \NN^{\ast}$ its \emph{interleaving order}.

\begin{remark}\label{rem:vert-int-heterogeneous}
    This form of interleaving is called \emph{homogeneous}, as all component codewords $\c_j$ belong to the same code $\mycode{C}$.
    In contrast, \emph{heterogeneous} interleaving allows the component codewords $\c_1, \dots, \c_{\intOrder}$ to belong to possibly different subcodes $\mycode{C}_1, \dots, \mycode{C}_{\intOrder}$ of $\mycode{C}$.
\end{remark}

The length partition $\n$ naturally induces a block structure on every codeword $\C$ of $\VInt(\mycode{C}, \intOrder)$.
Namely,
\begin{equation}
    \C =
    \begin{pmatrix}
        \c_1 \\
        \vdots \\
        \c_{\intOrder}
    \end{pmatrix}
    =
    \left(
    \begin{array}{c|c|c|c}
        \subShot{\c}{1}{1} & \subShot{\c}{1}{2} & \dots & \subShot{\c}{1}{\shots} \\
        \vdots & & & \vdots \\
        \subShot{\c}{\intOrder}{1} & \subShot{\c}{\intOrder}{2} & \dots & \subShot{\c}{\intOrder}{\shots}
    \end{array}
    \right)
    =:
    \bigl( \shot{\C}{1} \mid \dots \mid \shot{\C}{\shots} \bigr)
\end{equation}
with $\shot{\C}{i} \in \Fqm^{\intOrder \times n_i}$ for all $i = 1, \dots, \shots$.
In other words, the columns of $\C$ are divided into the blocks $\shot{\C}{1}, \dots, \shot{\C}{\shots}$ according to the same length partition $\n$.
This suggests a straightforward generalization of the sum-rank weight to codewords of vertically interleaved sum-rank-metric codes and, more generally, to arbitrary matrices $\X \in \Fqm^{\intOrder \times n}$.

We define the \emph{sum-rank weight} of a vertically interleaved matrix $\X = (\shot{\X}{1} \mid \dots \mid \shot{\X}{\shots}) \in \Fqm^{\intOrder \times n}$ as
\begin{equation}\label{eq:sum-rank-for-vertical-interleaving}
    \SumRankWeightWPartition{\n}(\X) \defeq \sum_{i=1}^{\shots} \rkq(\shot{\X}{i}),
\end{equation}
where the $\Fq$-rank of an $\Fqm$-matrix is the maximum number of $\Fq$-linearly independent columns.
Naturally, the corresponding metric is obtained as $\SumRankDistWPartition{\n}(\X, \Y) \defeq \SumRankWeight(\X - \Y)$ for any $\X, \Y \in \Fqm^{\intOrder \times n}$.
We simply write $\SumRankWeight(\cdot)$ and $\SumRankDist(\cdot, \cdot)$ when the length partition $\n$ is clear from the context.

An example of vertically interleaved codes in the sum-rank metric are \emph{\acf{VILRS}} codes, which were first considered in~\cite{bartz2022fast}.
Their component code is an \ac{LRS} code and we use the notation
\begin{equation}
    \vertIntLinRS{\vecbeta}{\vecxi}{\intOrder}{\nVec}{k}
    \defeq \VInt(\linRS{\vecbeta}{\vecxi}{\nVec}{k}, \intOrder)
\end{equation}
to highlight the corresponding code parameters, which have to fulfill the restrictions given in~\autoref{def:LRS_codes}.
\Ac{VILRS} codes have minimum sum-rank distance $d=n-k+1$ and are thus~\ac{MSRD} codes~\cite{bartz2022fast}.

In the following, we discuss channel models that naturally arise in a vertical interleaving setting.
We then propose a probabilistic unique syndrome-based decoder for \ac{VILRS} codes under the error-only model.
We further extend the decoder to the error-erasure channel model and thus describe the first error-erasure decoder for \ac{VILRS} codes.

\subsection{Channel and Error Models}\label{sec:vilrs-channel-error-model}

We first describe the error-only channel for vertical interleaving in the sum-rank metric.
Then, we explain how to generalize the model to the error-erasure setting.

Consider the transmission of a codeword $\C\in\vertIntLinRS{\vecbeta}{\vecxi}{\intOrder}{\nVec}{k}$ over an additive sum-rank error channel such that the received word $\Y \in \Fqm^{\intOrder \times n}$ can be described as
\begin{equation}\label{eq:vint_sum-rank_channel}
    \Y=\C+\E
\end{equation}
for an error matrix $\E \in \Fqm^{\intOrder \times n}$ of sum-rank weight $\SumRankWeightWPartition{\n}(\E) = \numbErrors$.
We assume that $\E$ is chosen uniformly at random from the set of all matrices having sum-rank weight $\numbErrors \in \NN$, i.e., from
\begin{equation}
    \label{eq:mats_given_sr_weight}
    \matGivenSRWeight{q^m}{\intOrder}{\n}{\numbErrors} \defeq \{ \M \in \Fqm^{\intOrder \times n} : \SumRankWeightWPartition{\n}(\M) = \numbErrors \}.
\end{equation}

\begin{remark}\label{rem:fixed-weight-sampling}
    Sampling uniformly at random from $\matGivenSRWeight{q^m}{\intOrder}{\n}{\numbErrors}$ is not straightforward, as the sum-rank weight depends on the rank partition and different rank partitions are not equally likely.
    This topic was studied for vectors of a given sum-rank weight in~\cite{puchinger2020generic,puchinger2022generic} and an algorithm for uniform sampling from $\matGivenSRWeight{q^m}{1}{\n}{\numbErrors}$ based on enumerative coding can be found in~\cite[Alg.~9]{puchinger2022generic}.
    It can be adapted to our case by using $\Fqm^{\intOrder} \cong \Fqms$ and thus $\matGivenSRWeight{q^m}{\intOrder}{\n}{\numbErrors} \cong \matGivenSRWeight{q^{m \intOrder}}{1}{\n}{\numbErrors}$.
\end{remark}

If we focus on the interleaved structure, we can express the channel observation in~\eqref{eq:vint_sum-rank_channel} as
\begin{equation}
    \label{eq:def_vint_sum-rank_channel}
    \begin{pmatrix}
        \y_1   \\
        \vdots \\
        \y_\intOrder
    \end{pmatrix}
    =
    \begin{pmatrix}
        \c_1   \\
        \vdots \\
        \c_\intOrder
    \end{pmatrix}
    +
    \begin{pmatrix}
        \e_1   \\
        \vdots \\
        \e_\intOrder
    \end{pmatrix}
    \quad \text{with }
    \y_j, \c_j, \e_j \in \Fqm^{n}
    \text{ for all } j = 1, \dots, \intOrder.
\end{equation}
We call $\y_1, \dots, \y_{\intOrder} \in \Fqm^{n}$ received component words, $\e_1, \dots, \e_{\intOrder} \in \Fqm^{n}$ component errors, and $\c_1, \dots, \c_{\intOrder} \in \Fqm^{n}$ component codewords, respectively.
We will switch between the perspectives given in~\eqref{eq:vint_sum-rank_channel} and~\eqref{eq:def_vint_sum-rank_channel} depending on the context.

Let $\numbErrorsVec=(\numbErrorsInBlock{1},\dots,\numbErrorsInBlock{\shots})\in\NN^\shots$ denote the rank partition of $\E$ with respect to $\n$.
In other words, we write $\numbErrorsInBlock{i} \defeq \rkq(\E^{(i)})$ for all $i = 1, \dots, \shots$.
We can apply a full-rank decomposition~\cite[Thm.~1]{matsaglia1974equalitiesInequalities} to each block of the error matrix $\E$ and thus obtain a decomposition
\begin{equation}\label{eq:err_mat_decomp}
    \E = (\shot{\E}{1} \mid \dots \mid \shot{\E}{\shots}) =
    \underbrace{(\A^{(1)} \mid \dots \mid \A^{(\shots)})}_{=: \A}
    \cdot
    \underbrace{
    \begin{pmatrix}
     \shot{\B}{1} & \\
     & \ddots & \\
     & & \shot{\B}{\shots}
    \end{pmatrix}
    }_{=:\B}
\end{equation}
with matrices $\A \in \Fqm^{\intOrder \times \numbErrors}$ and $\B \in \Fq^{\numbErrors \times n}$ satisfying $\SumRankWeightWPartition{\numbErrorsVec}(\A) = \numbErrors$ and $\SumRankWeightWPartition{\n}(\B) = \numbErrors$ as in~\cite[Lem.~5]{puchinger2020generic}.
The decomposition naturally induces a block structure on $\A$ and a block-diagonal structure on $\B$, which is shown in~\eqref{eq:err_mat_decomp}.
Namely, the columns of $\A$ and $\B$ are divided into blocks according to the length partitions $\numbErrorsVec$ and $\n$, respectively.

\begin{remark}
    The sum-rank conditions $\SumRankWeightWPartition{\numbErrorsVec}(\A) = \numbErrors$ and $\SumRankWeightWPartition{\n}(\B) = \numbErrors$ on the matrices $\A$ and $\B$ in the above error decomposition are equivalent to $\rkq(\A^{(i)})=\numbErrorsInBlock{i}$ and $\rkq(\shot{\B}{i})=\numbErrorsInBlock{i}$ for all $i = 1, \dots, \shots$.
\end{remark}

Note that the decomposition in~\eqref{eq:err_mat_decomp} is not unique.
In fact, every block-diagonal matrix $\M = \diag(\shot{\M}{1}, \dots, \shot{\M}{\shots}) \in \Fq^{\numbErrors \times \numbErrors}$ with full-rank blocks $\shot{\M}{i} \in \Fq^{\numbErrorsInBlock{i} \times \numbErrorsInBlock{i}}$ for $i = 1, \dots, \shots$ gives rise to a decomposition
\begin{equation}\label{eq:equiv-decomp-vert}
    \E = \A' \cdot \B' = \underbrace{\A \M^{-1}}_{=: \A'} \cdot \underbrace{\M \B}_{=: \B'},
\end{equation}
which satisfies the same sum-rank conditions as~\eqref{eq:err_mat_decomp}.
Even though different decompositions might yield different matrices $\A$ and $\B$, they are always directly linked to the error $\E$.
Namely, the columns of $\A^{(i)}$ span the $\Fq$-column space of $\E^{(i)}$ and are called \emph{error values}.
Similarly, the rows of $\B^{(i)}$ span the $\Fq$-row space of $\E^{(i)}$ and are called \emph{error locations}.
Observe that the error decomposition in~\eqref{eq:err_mat_decomp} shows that all component errors $\e_1, \dots, \e_{\intOrder}$ share the same row space due to the vertically interleaved structure.
This is beneficial for decoding, as it allows to employ the synergies between the error components by working with only \emph{one} set of error locations.

We now move towards the error-erasure channel model.
The main idea behind generalizing the error-only model to the error-erasure case is that the channel provides the receiver with partial knowledge about the error.
For the adapted error model, we divide the $\tau = \SumRankWeight(\E)$ occurred sum-rank errors into three categories according to the available information.
We distinguish between
\begin{itemize}
    \item a \emph{(full) error} or an error of type $\indFullErrors$, for which neither the row nor the column space is known,
    \item a \emph{row erasure} or an error of type $\indRowErasures$, for which the column space is known,
    \item and a \emph{column erasure} or an error of type $\indColErasures$, for which the row space is known.
\end{itemize}
By grouping the errors according to their type $\indErrorType \in \{\indFullErrors, \indRowErasures, \indColErasures\}$, we obtain the decomposition $\E = \E_{\indFullErrors} + \E_{\indRowErasures} + \E_{\indColErasures}$ with $\E_{\indErrorType} \in \Fqm^{\intOrder \times n}$ and $\SumRankWeight(\E) = \sum_{\indErrorType \in \{\indFullErrors, \indRowErasures, \indColErasures\}} \SumRankWeight(\E_{\indErrorType})$.
In the following, we use the notation $\numbErrorType \defeq \SumRankWeight(\E_{\indErrorType})$ for every $\indErrorType \in \{\indFullErrors, \indRowErasures, \indColErasures\}$ and denote the rank partition of $\E_{\indErrorType}$ by $\numbErrorTypeVec = (\numbErrorTypeInBlock{1}, \dots, \numbErrorTypeInBlock{\shots})$.

The application of~\eqref{eq:err_mat_decomp} to the components $\E_{\indFullErrors}$, $\E_{\indRowErasures}$, and $\E_{\indColErasures}$ yields the error decomposition
\begin{equation}
    \label{eq:err_eras_mat_decomp}
    \E = \sum_{\indErrorType \in \{\indFullErrors, \indRowErasures, \indColErasures\}}
    \underbrace{(\A_\indErrorType^{(1)}\mid\dots\mid\A_\indErrorType^{(\shots)})}_{\eqdef \A_\indErrorType \in \Fqm^{\intOrder \times \numbErrorType}}
    \cdot
    \underbrace{
    \setlength{\arraycolsep}{1pt}
    \renewcommand{\arraystretch}{0.6}
    \begin{pmatrix}
     \B_\indErrorType^{(1)} \\[-5pt]
     & \ddots \\
     & & \B_\indErrorType^{(\shots)}
    \end{pmatrix}
    }_{\eqdef \B_\indErrorType \in \Fq^{\numbErrorType \times n}}
\end{equation}
with $\SumRankWeightWPartition{\numbErrorTypeVec}(\A_{\indErrorType}) = \numbErrorType$ and $\SumRankWeightWPartition{\n}(\B_{\indErrorType}) = \numbErrorType$ for each $\indErrorType \in \{\indFullErrors, \indRowErasures, \indColErasures\}$.
As in the error-only setting, the columns of $\A_\indErrorType^{(i)}$ form a basis of the $\Fq$-column space of $\E_\indErrorType^{(i)}$ and the rows of $\B_\indErrorType^{(i)}$ are a basis of its $\Fq$-row space for all $i = 1, \dots, \shots$ and each $\indErrorType \in \{ \indFullErrors, \indRowErasures, \indColErasures \}$.
We thus call the columns of $\A_{\indFullErrors}$, $\A_{\indRowErasures}$, and $\A_{\indColErasures}$ \emph{error values} and the rows of $\B_{\indFullErrors}$, $\B_{\indRowErasures}$, and $\B_{\indColErasures}$ \emph{error locations}.
Again, the error's row space is shared by all component errors due to the structure induced by vertical interleaving.
Observe further that the additional information that is available for the different error types can be translated directly into the knowledge of $\A_{\indRowErasures}$ and $\B_{\indColErasures}$.
This is visualized in~\autoref{fig:error_decomposition} for the non-interleaved setting.

\subsection{Error-Only Decoding}

Assume that we have received a word $\Y \in \Fqm^{\intOrder \times n}$ after a codeword $\C \in \vertIntLinRS{\vecbeta}{\vecxi}{\intOrder}{\nVec}{k}$ was transmitted over the error-only channel described in~\eqref{eq:vint_sum-rank_channel}.
The error $\E = \Y - \C$ is thus assumed to have a fixed sum-rank weight $\numbErrors$.
Let $\H\in\Fqm^{(n-k) \times n}$ be a parity-check matrix of the component code $\linRS{\vecbeta}{\vecxi}{\nVec}{k}$.
Then we can compute the \emph{component syndromes}
\begin{equation}\label{eq:syndromes_vilrs}
    \s_j
    \defeq\y_j\H^\top
    =\e_j\H^\top
    \overset{\eqref{eq:err_mat_decomp}}{=}\a_j\B\H^\top
    \quad\text{for all } j=1,\dots,\intOrder
\end{equation}
with $\a_j$ denoting the $j$-th row of the matrix $\A$ from~\eqref{eq:err_mat_decomp}.
Without loss of generality, we can assume that $\H$ is a generalized Moore matrix with respect to a vector $\h \in \Fqm^{n}$ as described in~\eqref{eq:parity_check_mat}.
We can thus express the $l$-th entry of the syndrome $\s_j$ for $l=1,\dots n-k$ as
\begin{equation}
    \label{eq:vilrs_syndrome_short}
    s_{j, l} = \a_j \B \opfullexpinv{\autInvXiVec}{\h}{l-1}^{\top}
    = \a_j \opfullexpinv{\autInvXiVec}{\underbrace{\h \B^{\top}}_{=: \x}}{l-1}^{\top}
    = \a_j \opfullexpinv{\autInvXiVec}{\x}{l-1}^{\top}
\end{equation}
with $\autInvXiVec \defeq \autinv(\vecxi)$ defined in~\eqref{eq:def_autInvXiVec}.
Observe that the vector $\x \defeq \h \B^{\top} \in \Fqm^{\numbErrors}$ does not depend on the chosen component index $j = 1, \dots, \intOrder$.
We call its entries \emph{error locators} and divide them into blocks according to the rank partition $\numbErrorsVec$ of the error $\e$ and the matrix $\B^{\top}$.

We define the \emph{\acf{ELP}} $\ELP\in\SkewPolyringZeroDerInv$ as the minimal skew polynomial satisfying $\opev{\ELP}{\x}{\autInvXiVec} = \0$.
In other words, $\ELP$ vanishes on the error locators with respect to generalized operator evaluation with evaluation parameters $\autInvXiVec$.
Note that $\ELP$ has degree $\numbErrors$ because the blocks $\shot{\x}{1}, \dots, \shot{\x}{\shots}$ of the error locators have $\Fq$-linearly independent entries and the evaluation parameters $\autInvXiWIndex{1}, \dots, \autInvXiWIndex{\shots}$ belong to pairwise distinct nontrivial conjugacy classes of $\Fqm$.
Next, we define the reversed component-syndrome polynomials which we will shortly relate to the \ac{ELP} by means of a key equation.
For $j = 1, \dots, \intOrder$, the $j$-th \emph{component-syndrome polynomial} $s_j \in\SkewPolyringZeroDerInv$ is the skew polynomial whose coefficients are the entries of the component syndrome $\s_j \in \Fqm^{n-k}$, that is, $s_j(x) \defeq \sum_{l=1}^{n-k}s_{j,l}x^{l-1}$.
We call its skew $\autinv$-reverse with respect to $n-k-1$ the \emph{reversed component-syndrome polynomial} $\skewrev{s}_j \in\SkewPolyringZeroDerInv$.
It is given by
\begin{equation}\label{eq:vilrs_rev_syndromes}
    \skewrev{s}_j(x)=\sum_{l=1}^{n-k}\skewrev{s}_{j,l}x^{l-1} \quad \text{with } \skewrev{s}_{j,l}=\aut^{-(l-n+k)}(s_{j,n-k+1-l}) \quad \text{for } l =1, \dots, n-k \quad \text{and } j = 1, \dots, \intOrder.
\end{equation}
We can now move on to the \ac{ELP} key equation which will allow to jointly recover the error locations from the component syndromes in our decoder.

\begin{theorem}[\ac{ELP} Key Equation]\label{thm:vilrs_elp_key_equation}
    For each $j = 1, \dots, \intOrder$, there is a skew polynomial $\erasureEvalPoly_j\in\SkewPolyringZeroDerInv$ with $\deg(\erasureEvalPoly_j)< \numbErrors$ that satisfies
    \begin{equation}\label{eq:vilrs_key_equation_elp}
        \ELP(x) \cdot \skewrev{s}_j(x) \equiv \erasureEvalPoly_j(x) \modr x^{n-k}.
    \end{equation}
\end{theorem}

\begin{IEEEproof}
    We prove~\eqref{eq:vilrs_key_equation_elp} by showing that $\erasureEvalPoly_{j,l}=0$ holds for all $l = \numbErrors + 1, \dots, n - k$ and all $j = 1, \dots, \intOrder$.
    Namely,
    \begin{align*}
        \erasureEvalPoly_{j,l} &= (\ELP \cdot \skewrev{s}_j)_l
        \overset{\eqref{eq:skew_product_coeffs}}{=}\sum_{\nu=1}^{\numbErrors+1}\ELP_{\nu}\aut^{-(\nu-1)}(\skewrev{s}_{j,l-\nu+1})
        \underset{\eqref{eq:vilrs_syndrome_short}}{\overset{\eqref{eq:vilrs_rev_syndromes}}{=}}\sum_{\nu=1}^{\numbErrors+1}\ELP_{\nu}\aut^{-(l-n+k)}(\a_j \opfullexpinv{\autInvXiVec}{\x}{n-k-l+\nu-1}^{\top})
        \\
        &=\sum_{i=1}^{\shots}\sum_{r=1}^{\numbErrorsInBlock{i}}\aut^{-(l-n+k)}(a_{j,r}^{(i)})\sum_{\nu=1}^{\numbErrors+1}\ELP_{\nu}\aut^{-(\nu-1)}(x_r^{(i)}) \underbrace{\aut^{-(l-n+k)}\bigl(\genNormInv{n-k-l+\nu-1}{\autInvXiWIndex{i}}\bigr)}_{\overset{\eqref{eq:normlemma-1}}{=} \genNormInv{\nu-1}{\autInvXiWIndex{i}} \cdot \genNorm{n-k-l}{\xi_i}}
        \\
        &=\sum_{i=1}^{\shots}\sum_{r=1}^{\numbErrorsInBlock{i}}\aut^{-(l-n+k)}(a_{j,r}^{(i)})\genNorm{n-k-l}{\xi_i}\underbrace{\sum_{\nu=1}^{\numbErrors+1}\ELP_{\nu}\opfullexpinv{\autInvXiWIndex{i}}{x_r^{(i)}}{\nu-1}}_{=\opev{\ELP}{x_r^{(i)}}{\autInvXiWIndex{i}}=0}
        =0.
    \end{align*}
\end{IEEEproof}

The above proof shows that the key equation~\eqref{eq:vilrs_key_equation_elp} can be reformulated as an inhomogeneous system of $\Fqm$-linear equations.
This is useful for the analysis of the decoding radius and the failure probability, and we obtain
\begin{equation}\label{eq:vilrs_key_equation_elp_v1_norm}
   \sum_{\nu=2}^{\numbErrors+1}\ELP_{\nu}\aut^{-(\nu-1)}(\skewrev{s}_{j,l-\nu+1})=-\skewrev{s}_{j,l} \quad \text{for all } l=\numbErrors+1,\dots,n-k \quad \text{and all } j=1,\dots,\intOrder
\end{equation}
by normalizing $\ELP_1=1$ without loss of generality.
The systems for the component indices $j = 1, \dots, \intOrder$ can be joined into the larger system
\begin{equation}\label{eq:vilrs_key_equation_system}
    \begin{pmatrix}
     \skewrev{\S}_1
     \\
     \vdots
     \\
     \skewrev{\S}_\intOrder
    \end{pmatrix}
    \cdot
    \veclambda^\top
    = -
    \begin{pmatrix}
     \skewrev{\s}_1^\top
     \\
     \vdots
     \\
     \skewrev{\s}_\intOrder^\top
    \end{pmatrix},
\end{equation}
where $\skewrev{\S}_j \in \Fqm^{(n-k-\numbErrors) \times \numbErrors}$ is the coefficient matrix
\begin{equation}\label{eq:revSj}
    \skewrev{\S}_j
    =
    \begin{pmatrix}
     \aut^{-1}(\skewrev{s}_{j,\numbErrors}) & \aut^{-2}(\skewrev{s}_{j,\numbErrors-1}) & \dots & \aut^{-\numbErrors}(\skewrev{s}_{j,1})
     \\
     \aut^{-1}(\skewrev{s}_{j,\numbErrors+1}) & \aut^{-2}(\skewrev{s}_{j,\numbErrors}) & \dots & \aut^{-\numbErrors}(\skewrev{s}_{j,2})
     \\
     \vdots & \vdots & \ddots & \vdots
     \\
     \aut^{-1}(\skewrev{s}_{j,n-k-1}) & \aut^{-2}(\skewrev{s}_{j,n-k-2}) & \dots & \aut^{-\numbErrors}(\skewrev{s}_{j,n-k-\numbErrors})
    \end{pmatrix}
\end{equation}
and $\skewrev{\s}_j=(\skewrev{s}_{j, \numbErrors+1},\dots,\skewrev{s}_{j,n-k})$ is the right-hand side of the respective system~\eqref{eq:vilrs_key_equation_elp_v1_norm} for $j=1,\dots,\intOrder$.
The vector $\vec{\ELP}=(\ELP_2,\dots,\ELP_{\numbErrors+1})\in\Fqm^\numbErrors$ contains the unknown coefficients of the \ac{ELP} $\ELP$.
We use the shorthand notation $\skewrev{\S} \cdot \veclambda^{\top} = \skewrev{\s}^{\top}$ to refer to the whole system~\eqref{eq:vilrs_key_equation_system} without focusing on the stacked structure.

\begin{remark}
    While it is possible to solve the key equation in its equivalent formulation~\eqref{eq:vilrs_key_equation_system} by e.g.\ Gaussian elimination, it is beneficial to exploit the particular structure of the system.
    In fact, multisequence skew-feedback shift-register synthesis~\cite{Sidorenko2011SkewFeedback} can solve it in at most $O(\intOrder (n-k)^2)$ operations in $\Fqm$.
\end{remark}

The proof of~\autoref{thm:vilrs_elp_key_equation} showed that the \ac{ELP} is \emph{one} valid solution of the key equation~\eqref{eq:vilrs_key_equation_elp}.
But since we want to be sure to recover the actual error in the decoding process, we require that~\eqref{eq:vilrs_key_equation_elp} is uniquely solvable up to $\Fqm$-multiples.
We can use the equivalent formulation~\eqref{eq:vilrs_key_equation_system} of the key equation to characterize for which parameters this is possible.
The inhomogeneous $\Fqm$-linear system~\eqref{eq:vilrs_key_equation_system} has $\intOrder(n-k-\numbErrors)$ equations in $\numbErrors$ unknowns and can hence only have a one-dimensional solution space if the number of equations is at least the number of unknowns.
Therefore, we directly obtain the necessary condition $\numbErrors \leq \intOrder(n-k-\numbErrors)$ for decoding success.
In other words, the error weight $\numbErrors$ must adhere to
\begin{equation}\label{eq:dec_radius_vilrs}
    \numbErrors \leq \numbErrorsMax \defeq \tfrac{\intOrder}{\intOrder+1}(n-k)
\end{equation}
and $\numbErrorsMax$ is the maximum decoding radius.
Note that this bound is necessary but not sufficient for the decoder's success, that is, successful decoding is \emph{possible} but not guaranteed.

A decoding failure occurs if the key equation~\eqref{eq:vilrs_key_equation_elp} has a solution space of dimension at least two.
This corresponds to the case where the coefficient matrix $\skewrev{\S}$ of~\eqref{eq:vilrs_key_equation_system} has $\Fqm$-rank less than $\numbErrors$.
The next lemma investigates the probability for this event and derives an upper bound on the decoding-failure probability.
Moreover, it shows that correct decoding can be guaranteed as long as the error lies within the unique-decoding radius of the \ac{VILRS} code, i.e., if $\tau \leq \tfrac{1}{2}(d-1) = \tfrac{1}{2}(n-k)$ holds for $d = n - k + 1$ being the minimum distance of the code.
\begin{lemma}\label{lem:failure_prob_vils}
    Let $\skewrev{\S}$ be the coefficient matrix of the system~\eqref{eq:vilrs_key_equation_system}, which arose from a \ac{VILRS} decoding instance $\Y = \C + \E$ with error weight $\SumRankWeight(\E) = \numbErrors \leq \numbErrorsMax$.
    Then, the bound
    \begin{align*}
        \Pr\left\{\rkqm(\skewrev{\S})
            < \numbErrors
            \right\}
            \leq \kappa_q^{\shots+1}q^{-m((\intOrder+1)(\numbErrorsMax-\numbErrors)+1)}
    \end{align*}
    applies for $\kappa_q < 3.5$ being defined in~\eqref{eq:def_kappa_q}.
    Moreover, $\rkqm(\skewrev{\S}) = \numbErrors$ is guaranteed for any error of weight $\numbErrors \leq \tfrac{1}{2}(n - k)$.
\end{lemma}

\begin{IEEEproof}
    Observe from~\eqref{eq:revSj} that the entry in column $\nu - 1$ and row $l + \nu - \numbErrors - 2$ of $\skewrev{\S}_j$ is $\aut^{-(\nu-1)}\left(\skewrev{s}_{j,l-\nu+1}\right)$ for $\nu = 2, \dots, \numbErrors + 1$, $l = \numbErrors + 1, \dots, n - k$, and $j = 1, \dots, \intOrder$.
    We can thus use the equality
    \begin{align}
        \aut^{-(\nu-1)}\left(\skewrev{s}_{j, l-\nu+1}\right)
        &\underset{\eqref{eq:vilrs_syndrome_short}}{\overset{\eqref{eq:vilrs_rev_syndromes}}{=}} \aut^{n-k-l}\bigl(
        \a_j \opfullexpinv{\autInvXiVec}{\x}{n-k-l+\nu-1}^{\top}
        \bigr)
        = \sum_{i=1}^{\shots} \sum_{r=1}^{\numbErrorsInBlock{i}} \aut^{n-k-l}(a_{j, r}^{(i)}) \aut^{-(\nu-1)}(x_r^{(i)})
        \underbrace{\aut^{n-k-l}\bigl(\genNormInv{n-k-l+\nu-1}{\autInvXiWIndex{i}}\bigr)}_{\overset{\eqref{eq:normlemma-1}}{=} \genNormInv{\nu-1}{\autInvXiWIndex{i}} \cdot \genNorm{n-k-l}{\xi_i}}
        \\
        &= \sum_{i=1}^{\shots} \sum_{r=1}^{\numbErrorsInBlock{i}} \opfullexp{\xi_i}{a_{j, r}^{(i)}}{n-k-l} \opfullexpinv{\autInvXiWIndex{i}}{x_r^{(i)}}{\nu-1} \label{eq:vilrs_elp_key_equation_decomp}
        = \opfullexp{\vecxi}{\a_j}{n-k-l} \cdot \opfullexpinv{\autInvXiVec}{\x}{\nu-1}^{\top}
    \end{align}
    for all $\nu=2, \dots, \numbErrors+1$, all $l=\numbErrors+1, \dots, n-k$, and all $j = 1, \dots, \intOrder$ to decompose the matrix $\skewrev{\S}_j$ into a product.
    In particular, we obtain
    \begin{equation*}
        \skewrev{\S}_j =
        \underbrace{
        \begin{pmatrix}
            \opfullexp{\vecxi}{\a_j}{n-k-\numbErrors-1}
            \\
            \vdots
            \\
            \opfull{\vecxi}{\a_j}
            \\
            \a_j
        \end{pmatrix}
        }_{=: \hat{\A}_j \in \Fqm^{(n - k - \numbErrors) \times \numbErrors}}
        \cdot
        \underbrace{
        \begin{pmatrix}
            \opfullinv{\autInvXiVec}{\x}
            \\
            \vdots
            \\
            \opfullexpinv{\autInvXiVec}{\x}{\numbErrors}
        \end{pmatrix}^{\top}
        }_{=: \hat{\X} \in\Fqm^{\numbErrors \times \numbErrors}}
        \quad \text{for all } j = 1, \dots, \intOrder.
    \end{equation*}
    Note that $\hat{\X} = {\opMooreInv{\numbErrors}{\opfullinv{\autInvXiVec}{\x}}{\autInvXiVec}}^{\top}$ does not depend on $j$ and has full $\Fqm$-rank $\numbErrors$ because $\SumRankWeight(\x) = \numbErrors$ holds according to the definition of the code locators.
    Since the matrix $\hat{\A}_j$ contains precisely the rows of $\opMoore{n-k-\numbErrors}{\a_j}{\vecxi}$ in reverse order, $\rkqm(\hat{\A}_j) = \rkqm(\opMoore{n-k-\numbErrors}{\a_j}{\vecxi})$ holds for all $j=1,\dots,\intOrder$.
    Overall, we obtain
    \begin{equation}
    \skewrev{\S} =
    \begin{pmatrix}
         \skewrev{\S}_1
         \\
         \vdots
         \\
         \skewrev{\S}_\intOrder
        \end{pmatrix}
        =
        \underbrace{
        \begin{pmatrix}
         \hat{\A}_1
         \\
         \vdots
         \\
         \hat{\A}_\intOrder
        \end{pmatrix}
        }_{=: \hat{\A}}
        \cdot \hat{\X}
    \end{equation}
    and $\rkqm(\skewrev{\S}) = \rkqm(\hat{\A}) = \rkqm(\opMoore{n-k-\numbErrors}{\A}{\vecxi})$ with $\A$ being the matrix from the error decomposition~\eqref{eq:err_mat_decomp} with the rows $\a_1, \dots, \a_{\intOrder}$.
    Since the error matrix $\E$ is chosen uniformly at random from the set $\matGivenSRWeight{q^m}{\intOrder}{\n}{\numbErrors}$, it follows that $\A$ is distributed uniformly over $\matGivenSRWeight{q^m}{\intOrder}{\numbErrorsVec}{\numbErrors}$.
    In this setting, the proof of~\cite[Lem.~7]{bartz2022fast} yields the stated inequality
    \begin{align*}
        \Pr\left\{\rkqm(\skewrev{\S})
            < \numbErrors
            \right\}
            &=
            \Pr\left\{
            \rkqm\left(\opMoore{n-k-\numbErrors}{\A}{\vecxi}\right)
            < \numbErrors
            \right\}
            \overset{\text{\cite{bartz2022fast}}}{\leq} \kappa_q^{\shots+1}q^{-m((\intOrder+1)(\numbErrorsMax-\numbErrors)+1)}.
    \end{align*}
    
    Let us now consider an error of bounded sum-rank weight $\SumRankWeight(\E) = \numbErrors \leq \tfrac{1}{2}(n-k)$.
    As shown above, it holds that $\rkqm(\skewrev{\S}) = \rkqm(\opMoore{n-k-\numbErrors}{\A}{\vecxi})$.
    The latter matrix has full $\Fqm$-rank $\numbErrors$ according to~\cite[Lem.~6]{bartz2022fast} because no vector in $\Fqm^{\numbErrors}$ can have a sum-rank weight larger than $n-k-\numbErrors \geq \tfrac{1}{2}(n-k) \geq \numbErrors$.
    This concludes the proof.
\end{IEEEproof}

Let us now focus on the steps of the decoder, which starts by setting up the reversed component-syndrome polynomials $\skewrev{s}_1, \dots, \skewrev{s}_{\intOrder}$, i.e., the ingredients for the key equation.
Then, it solves the key equation~\eqref{eq:vilrs_key_equation_elp} for the \ac{ELP} $\ELP$ with multisequence skew-feedback shift-register synthesis~\cite{Sidorenko2011SkewFeedback}, which is briefly summarized in~\autoref{sec:shift_register_synthesis}.
If the solution of~\eqref{eq:vilrs_key_equation_elp} is not unique up to $\Fqm$-multiples, the decoder returns a decoding failure.
Otherwise, it proceeds as follows:
According to the definition of the \ac{ELP}, the equality $\opev{\ELP}{\x}{\autInvXiVec} = \0$ holds and the error locations $\x \in \Fqm^{\numbErrors}$ can thus be recovered by finding the roots of $\ELP$ with respect to generalized operator evaluation and the evaluation parameters $\autInvXiVec \defeq \autinv(\vecxi)$ from~\eqref{eq:def_autInvXiVec}.
This can be done with the Skachek--Roth-like algorithm, whose main idea is described in~\autoref{sec:skachek_roth}.

For the recovery of the error values $\a_1, \dots, \a_{\intOrder} \in \Fqm^{\numbErrors}$, we first note that the application of $\aut^{l-1}$ to $s_{j,l}$ yields
\begin{equation}\label{eq:vilrs-syndrome-theta-l-1}
    \aut^{l-1}(s_{j,l})
    \overset{\eqref{eq:vilrs_syndrome_short}}{=}
    \aut^{l-1}\bigl( \a_j \opfullexpinv{\autInvXiVec}{\x}{l-1}^{\top} \bigr)
    = \sum_{i=1}^{\shots}\sum_{r=1}^{\numbErrorsInBlock{i}} \aut^{l-1}(a_{j, r}^{(i)}) x_{r}^{(i)} \underbrace{\aut^{l-1}\bigl(\genNormInv{l-1}{\autInvXiWIndex{i}}\bigr)}_{\overset{\eqref{eq:normlemma-1}}{=} \genNorm{l-1}{\xi_i}}
    = \opfullexp{\vecxi}{\a_j}{l-1} \cdot \x^{\top}
\end{equation}
for all $l = 1, \dots, n - k$ and all $j = 1, \dots, \intOrder$.
Thus, we can recover $\a_j$ by solving the linear system
\begin{equation}\label{eq:vilrs_gab-like}
    \opMoore{n-k}{\a_j}{\vecxi} \cdot \x^{\top} = \widetilde{\s}_j^{\top}
\end{equation}
with $\widetilde{\s}_j=(s_{j,1},\aut(s_{j,2}),\dots,\aut^{n-k-1}(s_{j,n-k}))\in\Fqm^{n-k}$ for each $j = 1, \dots, \intOrder$.
Since the systems in~\eqref{eq:vilrs_gab-like} have a particular shape, the Gabidulin-like algorithm from~\autoref{sec:efficient_gabidulin-like} is applicable.

We now retrieve the error locations, i.e., the matrix $\B \in \Fq^{\numbErrors \times n}$, from the error locators $\x \in \Fqm^{\numbErrors}$ by applying the methods from~\cite{silva2009error} in a blockwise manner.
We consider the first row $\h \in \Fqm^{n}$ of the parity-check matrix $\H = \opMooreInv{n-k}{\h}{\autInvXiVec}$ of the component \ac{LRS} code from~\eqref{eq:parity_check_mat} and represent it over $\Fq$ as $\H_q \defeq \coeffq{\h} \in \Fq^{m \times n}$.
Next we compute for each block $\H_q^{(i)} \in \Fq^{m \times n_i}$ of $\H_q$ a left inverse $\widetilde{\H}_q^{(i)} \in \Fq^{n_i \times m}$ which satisfies $\widetilde{\H}_q^{(i)} \cdot \H_q^{(i)} = \I_{n_i}$ for the identity matrix $\I_{n_i} \in \Fq^{n_i \times n_i}$ and every $i=1,\dots,\shots$.
We recover the error locations blockwise as
\begin{equation}\label{eq:vilrs-left-inverses}
    \B^{(i)\top} = \widetilde{\H}_q^{(i)}\X_q^{(i)} = \widetilde{\H}_q^{(i)}\H_q^{(i)}\B^{(i)\top} \quad \text{for all } i=1,\dots, \shots
\end{equation}
with $\X_q \defeq \coeffq{\x} \in \Fq^{m \times \numbErrors}$ being the $\Fq$-representation of the code locators.

As we found all rows $\a_1, \dots, \a_{\intOrder}$ of the error-value matrix $\A \in \Fqm^{\intOrder \times \numbErrors}$ and all blocks of the error-location matrix $\B = \diag(\B^{(1)},\dots,\B^{(\shots)}) \in \Fq^{\numbErrors \times n}$, we can compute the error $\E$ as $\E = \A \cdot \B$ according to~\eqref{eq:err_mat_decomp} and return the codeword $\C = \Y - \E$.

\begin{remark}\label{rem:left-inv}
    Note that the computation of the left inverses $\widetilde{\H}_q^{(1)}, \dots, \widetilde{\H}_q^{(\shots)}$ used in~\eqref{eq:vilrs-left-inverses} has a worst-case complexity of $O(\max(m, n)^3)$ and is thus computationally expensive.
    However, since the inverses only depend on the code and not on the particular decoding instance, they can be precomputed and saved in a lookup table.
\end{remark}

\autoref{alg:dec_vilrs} and~\autoref{thm:dec_vilrs} summarize the steps and the properties of the syndrome-based decoder for~\ac{VILRS} codes, respectively.

\begin{algorithm}[ht]
  \caption{\algoname{Error-Only Decoding of \ac{VILRS} Codes}}\label{alg:dec_vilrs}
  \SetKwInOut{Input}{Input}\SetKwInOut{Output}{Output}

  \Input{A channel output $\Y=\C+\E\in\Fqm^{\intOrder\times n}$ with $\C \in \vertIntLinRS{\vecbeta}{\vecxi}{\intOrder}{\nVec}{k}$ and $\SumRankWeight(\E)=\numbErrors \leq \numbErrorsMax$,
  \\
  a parity-check matrix $\H \in \Fqm^{(n-k) \times n}$ of the form $\opMooreInv{n-k}{\h}{\autInvXiVec}$ of $\linRS{\vecbeta}{\vecxi}{\nVec}{k}$,
  \\
  and a left inverse $\widetilde{\H}_q^{(i)} \in \Fq^{n_i \times m}$ of $\coeffq{\h^{(i)}}$ for each $i = 1, \dots, \shots$.}

  \Output{The transmitted codeword $\C \in \vertIntLinRS{\vecbeta}{\vecxi}{\intOrder}{\nVec}{k}$ or \emph{``decoding failure''}.}

  \BlankLine

  \For{$j=1,\dots,\intOrder$}
  {
    Compute the component syndrome $\s_j \gets \y_j\H^\top$ with $\y_j$ being the $j$-th row of $\Y$.
    \\
    Set up the reversed component-syndrome polynomial $\skewrev{s}_j\in\SkewPolyringZeroDerInv$ according to~\eqref{eq:vilrs_rev_syndromes}.
  }
  Solve the key equation~\eqref{eq:vilrs_key_equation_elp} to obtain the \ac{ELP} $\ELP\in\SkewPolyringZeroDerInv$. %
  \\
  \If{the key equation~\eqref{eq:vilrs_key_equation_elp} has a unique solution up to $\Fqm$-multiples}
  {
      Find a basis $\shot{\x}{i}$ of the root space of $\opev{\ELP}{\cdot}{\autInvXiWIndex{i}}$ for each $i = 1, \dots, \shots$ and set $\x \gets (\shot{\x}{1} \mid \dots \mid \shot{\x}{\shots})$.
      \label{alg:vilrs_find_err_loc}
      \\
      Set up $\opMoore{n-k}{\a_j}{\vecxi} \cdot \x^{\top} = \widetilde{\s}_j^{\top}$ from~\eqref{eq:vilrs_gab-like} for each $j = 1, \dots, \intOrder$ and solve it for $\a_j$.
      \\
      Recover the error locations $\B^{(i)\top} \gets \widetilde{\H}_q^{(i)} \cdot \coeffq{\shot{\x}{i}}$ for all $i=1,\dots,\shots$ as in~\eqref{eq:vilrs-left-inverses}.
      \\
      Let $\A$ be the matrix with rows $\a_1, \dots \a_{\intOrder}$ and set $\B \gets \diag(\B^{(1)},\dots,\B^{(\shots)})$.
      \\
      Recover the error matrix $\E \gets \A \B$ as in~\eqref{eq:err_mat_decomp}. \label{alg:vilrs_recover_error}
      \\
      \Return{$\C \gets \Y-\E$.}
    }

    \Return{``decoding failure''.}
\end{algorithm}

\begin{theorem}[Error-Only Decoding of \ac{VILRS} Codes]\label{thm:dec_vilrs}
    Consider the transmission of a codeword $\C \in \vertIntLinRS{\vecbeta}{\vecxi}{\intOrder}{\nVec}{k}$ over the additive error-only channel~\eqref{eq:vint_sum-rank_channel}.
    The error $\E \in \Fqm^{\intOrder \times n}$ of sum-rank weight $\SumRankWeight(\E) = \numbErrors$ is chosen uniformly at random from the set $\matGivenSRWeight{q^m}{\intOrder}{\n}{\numbErrors}$ defined in~\eqref{eq:mats_given_sr_weight} and determines the received word $\Y=\C+\E\in\Fqm^{\intOrder\times n}$.
    The presented syndrome-based decoder can always recover $\C$ from $\Y$ if $\numbErrors \leq \tfrac{1}{2}(n - k)$ holds.
    Moreover, the decoder can be used probabilistically for larger error weights and decoding succeeds with a probability of at least
    \begin{equation}\label{eq:lower_bound_Psucc_vilrs}
    1 - \gammaq^{\ell} q^{-m((s+1)(\numbErrorsMax-\numbErrors)+1)}
    \end{equation}
    as long as the error weight satisfies
    \begin{equation}
        \numbErrors\leq\numbErrorsMax\defeq\tfrac{\intOrder}{\intOrder+1}(n-k).
    \end{equation}
    The decoder requires on average $\oh{\intOrder n^2}$ operations in $\Fqm$ if $m \in O(\intOrder)$ applies.
\end{theorem}

\begin{IEEEproof}
    The correctness of~\autoref{alg:dec_vilrs} follows from the arguments given in this section and in particular from the proof of the key equation in~\autoref{thm:vilrs_elp_key_equation}.
    The only potential point of failure is the key equation having a solution space of dimension at least two, and~\autoref{alg:dec_vilrs} correctly returns \emph{``decoding failure''} in this case.
    
    The stated decoding radius and the bound on the success probability directly follow from the reasoning above, where we transformed the key equation into an equivalent system of linear equations.
    The corresponding results are stated in equation~\eqref{eq:dec_radius_vilrs} and~\autoref{lem:failure_prob_vils}, respectively.
    Note that~\autoref{lem:failure_prob_vils} also shows that successful decoding can be guaranteed as long as the error lies within the unique-decoding radius, that is, as long as $\tau \leq \tfrac{1}{2}(n-k)$ holds.

    Let us now consider the computational complexity of the decoding algorithm with respect to operations in $\Fqm$.
    Note that the fast subroutines we use, i.e., multisequence skew-feedback shift-register synthesis, a Skachek--Roth-like algorithm, and a Gabidulin-like algorithm, are discussed in~\autoref{sec:fast-dec}.
    The computation of the reversed component-syndrome polynomials in lines 1--3 can be achieved in at most $O(\intOrder n^2)$ operations in $\Fqm$.
    Lines 4 and 5 solve the key equation and check the uniqueness of the solution up to $\Fqm$-multiples.
    When multisequence skew-feedback shift-register synthesis is applied, this takes at most $O(\intOrder (n-k)^2)$ operations.
    Next, line 6 computes the roots of $\ELP$ and a Skachek--Roth-like approach has an average complexity in $O(\shots m \deg(\ELP)) = O(\shots m (n-k))$.
    If $m \in O(\intOrder)$ applies, this is in $O(\intOrder n^2)$.
    The system of linear equations in line 7 has a particular form and can be solved with a Gabidulin-like algorithm in at most $O(\intOrder n^2)$ operations in $\Fqm$.
    Since the left inverses in line 8 were precomputed, the remaining basic operations in lines 8 and 9 can be done in $O(n^2)$.
    Overall, this yields an average decoding complexity in $O(\intOrder n^2)$.
    Note that the Skachek--Roth-like algorithm is the only probabilistic ingredient of the decoder and therefore the only part where we consider the average complexity and not the worst-case complexity.
\end{IEEEproof}

\subsection{Error-Erasure Decoding}
\label{sec:vilrs-error-erasure}

Let us now move forward and consider the error-erasure channel for which the sent codeword $\C \in \vertIntLinRS{\vecbeta}{\vecxi}{\intOrder}{\nVec}{k}$ is not only corrupted by $\numbFullErrors$ full errors but also by $\numbRowErasures$ row erasures and $\numbColErasures$ column erasures.
More precisely, the additive error matrix $\E \in \Fqm^{\intOrder \times n}$ has sum-rank weight $\numbErrors = \numbFullErrors + \numbRowErasures + \numbColErasures$ and~\eqref{eq:err_eras_mat_decomp} shows the error decomposition per error type.

Our overall decoding strategy is similar to the error-only case but we incorporate and exploit the information the receiver has about the erasures.
We start by computing the \emph{component syndromes}
\begin{align}
    \s_j
    &\defeq \y_j \H^{\top}
    = (\e_{\indFullErrors, j} + \e_{\indRowErasures, j} + \e_{\indColErasures, j}) \H^{\top}
    \overset{\eqref{eq:err_eras_mat_decomp}}{=} \a_{\indFullErrors, j} \B_{\indFullErrors} \H^{\top} + \a_{\indRowErasures, j} \B_{\indRowErasures} \H^{\top} + \a_{\indColErasures, j} \B_{\indColErasures} \H^{\top}
    \quad \text{for all } j = 1, \dots, \intOrder.
\end{align}
Note that the vector $\a_{\indErrorType, j}$ with $\indErrorType \in \{\indFullErrors, \indRowErasures, \indColErasures\}$ and $j = 1, \dots, \intOrder$ denotes the $j$-th row of the matrix $\A_{\indErrorType}$ from the error decomposition~\eqref{eq:err_eras_mat_decomp}.
When we follow~\eqref{eq:vilrs_syndrome_short} for every error type separately, we can write the $l$-th coefficient of $\s_j$ with $l = 1, \dots, n - k$ and $j = 1, \dots, \intOrder$ as
\begin{equation}\label{eq:component-syndromes_vilrs_error-erasure}
    s_{j, l} = \sum_{\indErrorType \in \{\indFullErrors, \indRowErasures, \indColErasures\}} \a_{\indErrorType, j} \opfullexpinv{\autInvXiVec}{\underbrace{\h \B_{\indErrorType}^{\top}}_{=: \x_{\indErrorType}}}{l-1}^{\top}
    = \sum_{\indErrorType \in \{\indFullErrors, \indRowErasures, \indColErasures\}} \a_{\indErrorType, j} \opfullexpinv{\autInvXiVec}{\x_{\indErrorType}}{l-1}^{\top},
\end{equation}
where $\h \in \Fqm^{n}$ denotes the first row of the parity-check matrix $\H = \opMooreInv{n-k}{\h}{\autInvXiVec}$ of the component \ac{LRS} code and $\autInvXiVec = \aut^{-1}(\vecxi)$ was defined in~\eqref{eq:def_autInvXiVec}.
This allows us to define the \emph{error locators} $\x_{\indErrorType} \defeq \h \B_{\indErrorType}^{\top} \in \Fqm^{\numbErrorType}$ with respect to each error type $\indErrorType \in \{\indFullErrors, \indRowErasures, \indColErasures\}$.
Note that the vector $\x_{\indErrorType}$ with $\indErrorType \in \{\indFullErrors, \indRowErasures, \indColErasures\}$ has a block structure with respect to the rank partition $\numbErrorTypeVec \in \NN^{\shots}$ of $\B_{\indErrorType}^{\top}$ and that $\numbFullErrorsVec + \numbRowErasuresVec + \numbColErasuresVec = \vectau$ holds by definition.

Recall that $\ELP \in \SkewPolyringZeroDerInv$ was defined as the minimal skew polynomial of the error locators in the error-only setting.
We keep the notion of the \emph{\acf{ELP}} and let $\ELP$ be the minimal skew polynomial satisfying $\opev{\ELP}{\x_{\indErrorType}}{\autInvXiVec} = \0$ for all $\indErrorType \in \{\indFullErrors, \indRowErasures, \indColErasures\}$.
It will prove beneficial to express $\ELP$ as a product of skew polynomials related to the different error types.
This allows to incorporate the knowledge about the column erasures into our decoder because their row space translates to the respective error locators and the resulting partial \ac{ELP}.
We write
\begin{align}\label{eq:vilrs_overall_elp}
    \ELP(x) &= \ELProw(x) \cdot \ELPfull(x) \cdot \ELPcol(x),
\end{align}
where the \emph{partial \acp{ELP}} $\ELPfull, \ELProw, \ELPcol \in \SkewPolyringZeroDerInv$
are the minimal skew polynomials satisfying
\begin{align}\label{eq:def-partial-elps}
    \opev{(\ELProw \cdot \ELPfull \cdot \ELPcol)}{\x_{\indRowErasures}}{\autInvXiVec} = \0,
    \quad %
    \opev{(\ELPfull \cdot \ELPcol)}{\x_{\indFullErrors}}{\autInvXiVec} = \0,
    \quad %
    \text{and} \quad \opev{\ELPcol}{\x_{\indColErasures}}{\autInvXiVec} = \0,
\end{align}
respectively.
Observe that the degree of $\ELP_{\indErrorType}$ is $\numbErrorType$ and $\deg(\ELP) = \numbFullErrors + \numbRowErasures + \numbColErasures = \numbErrors$ holds.

Since the column spaces of the row erasures are known for every component error, we capture them in the \emph{partial component \acp{ESP}} $\ESProwWIndex{1}, \dots, \ESProwWIndex{\intOrder}$.
They are defined as the minimal skew polynomials satisfying
\begin{equation}\label{eq:def-component-esps}
    \opev{\ESProwWIndex{j}}{\a_{\indRowErasures, j}}{\autInvXiInvVec} = \0
    \quad \text{for all } j = 1, \dots, \intOrder.
\end{equation}
Here, $\a_{\indRowErasures, j}$ with $j = 1, \dots, \intOrder$ denotes the $j$-th row of the matrix $\A_{\indRowErasures}$ from the error decomposition~\eqref{eq:err_eras_mat_decomp} and we denote its sum-rank weight by $\numbRowErasuresWIndex{j} \defeq \SumRankWeight(\a_{\indRowErasures, j}) = \SumRankWeight(\e_{\indRowErasures, j})$ in the following.
Observe that $\numbRowErasuresWIndex{j} \leq \numbRowErasures$ holds for each $j = 1, \dots, \intOrder$ because the error decomposition imposes $\SumRankWeight(\A_{\indRowErasures}) = \numbRowErasures$.
It is also worth noting that $\sum_{j=1}^{\intOrder} \numbRowErasuresWIndex{j} \geq \numbRowErasures$ applies, while there is no need for equality.

Recall that minimal skew polynomials can be computed efficiently, e.g.\ via the equality given in~\eqref{eq:min_poly}.
Thus, the knowledge of $\B_\indColErasures$ and $\a_{\indRowErasures,1}, \dots, \a_{\indRowErasures, \intOrder}$ directly translates to the knowledge of $\ELPcol$ and $\ESProwWIndex{1}, \dots, \ESProwWIndex{1}$.
We define the \emph{auxiliary component-syndrome polynomials} $\ELPcomponentSyndrome{j} \in \SkewPolyringZeroDerInv$ as
\begin{equation}
    \ELPcomponentSyndrome{j}(x) = \ELPcol(x) \cdot \skewrev{s}_j(x) \cdot \coeffpower{\ESProwRevWIndex{j}(x)}{n-k-1},
\end{equation}
where $\coeffpower{\ESProwRevWIndex{j}(x)}{n-k-1} \defeq \aut^{n-k-1}(\ESProwRevWIndex{j}(\aut^{-(n-k-1)}(x)))$ denotes the skew polynomial obtained from $\ESProwRevWIndex{j}$ by applying $\aut^{n-k-1}$ to all its coefficients.
More precisely, it holds
\begin{equation}\label{eq:coeffpower-esprow-rev}
    \coeffpower{\ESProwRevWIndex{j}(x)}{n-k-1} = \sum_{l=1}^{\numbRowErasures + 1} (\coeffpower{\ESProwRevWIndex{j}}{n-k-1})_l x^{l-1} \quad \text{with } (\coeffpower{\ESProwRevWIndex{j}}{n-k-1})_l = \aut^{n - k + \numbRowErasures - l}(\ESProwWIndex{j, l}) \quad \text{for } l =1, \dots, \numbRowErasures + 1.
\end{equation}
Further, $\skewrev{s}_j$ denotes the $\autinv$-reverse of the component-syndrome polynomial $s_j$ with respect to $n-k-1$ as given in~\eqref{eq:vilrs_rev_syndromes} for each $j = 1, \dots, \intOrder$.

We can now derive an \ac{ELP} key equation that embodies the knowledge about the row and column erasures.
The following theorem relates the \ac{ELP} corresponding to the full errors with the auxiliary component-syndrome polynomials and generalizes the \ac{ELP} key equation from~\autoref{thm:vilrs_elp_key_equation} in the error-only setting.

\begin{theorem}[\ac{ELP} Key Equation for Errors and Erasures]\label{thm:vilrs_error-erasure_key_equation}
    For each $j =1, \dots, \intOrder$, there is a skew polynomial $\erasureEvalPoly_j \in \SkewPolyringZeroDerInv$ with $\deg(\erasureEvalPoly_j) < \numbFullErrors + \numbRowErasuresWIndex{j} + \numbColErasures$ that satisfies
    \begin{equation}\label{eq:vilrs_error-erasure_key_equation}
        \ELPfull(x) \cdot \ELPcomponentSyndrome{j}(x) \equiv \erasureEvalPoly_j(x) \modr x^{n-k}.
    \end{equation}
\end{theorem}

\begin{IEEEproof}
    We prove the statement by showing that $\erasureEvalPoly_{j,l} = 0$ holds for all $l = \numbFullErrors + \numbRowErasuresWIndex{j} + \numbColErasures+1,\dots, n-k$ and all $j = 1, \dots, \intOrder$.
    Therefore, we compute the respective coefficients $\erasureEvalPoly_{j,l}$ for each $j = 1, \dots, \intOrder$ by using the equality
    \begin{equation}
        \erasureEvalPoly_j(x) = \underbrace{\ELPfull(x) \cdot \ELPcol(x)}_{=: \ELPfullcol(x)} \cdot \skewrev{s}_j(x) \cdot \coeffpower{\ESProwRevWIndex{j}(x)}{n-k-1}
    \end{equation}
    and applying~\eqref{eq:skew_product_coeffs} first to the product $\ELPfullcol(x) \cdot \skewrev{s}_j(x)$ and then again to the product of the result and $\coeffpower{\ESProwRevWIndex{j}(x)}{n-k-1}$.
    In the first step, we obtain
    \begin{align}
        (\ELPfullcol \cdot \skewrev{s}_j)_l
        &\overset{\eqref{eq:skew_product_coeffs}}{=} \sum_{\nu=1}^{\numbFullErrors+\numbColErasures+1} \ELPfullcolWIndex{\nu} \aut^{-(\nu-1)}(\revComponentSyndromeWIndex{j}{l-\nu+1})
        \overset{\eqref{eq:vilrs_rev_syndromes}}{\underset{\eqref{eq:component-syndromes_vilrs_error-erasure}}{=}}
        \sum_{\nu=1}^{\numbFullErrors+\numbColErasures+1} \ELPfullcolWIndex{\nu} \sum_{\indErrorType \in \{\indFullErrors, \indRowErasures, \indColErasures\}} \aut^{n-k-l} \bigl( \a_{\indErrorType, j} \opfullexpinv{\autInvXiVec}{\x_{\indErrorType}}{n-k-l+\nu-1}^{\top} \bigr) \\
        &= \sum_{\indErrorType \in \{\indFullErrors, \indRowErasures, \indColErasures\}} \sum_{i=1}^{\shots} \sum_{r=1}^{\numbErrorTypeInBlock{i}} \aut^{n-k-l}(a_{\indErrorType, j, r}^{(i)}) \sum_{\nu=1}^{\numbFullErrors+\numbColErasures+1} \ELPfullcolWIndex{\nu} \aut^{-(\nu-1)}(x_{\indErrorType, r}^{(i)}) \underbrace{\aut^{n-k-l}\bigl(\genNormInv{n-k-l+\nu-1}{\autInvXiWIndex{i}}\bigr)}_{\overset{\eqref{eq:normlemma-1}}{=} \genNormInv{\nu-1}{\autInvXiWIndex{i}} \cdot \genNorm{n-k-l}{\xi_i}} \\
        &= \sum_{\indErrorType \in \{\indFullErrors, \indRowErasures, \indColErasures\}} \sum_{i=1}^{\shots} \sum_{r=1}^{\numbErrorTypeInBlock{i}} \aut^{n-k-l}(a_{\indErrorType, j, r}^{(i)}) \genNorm{n-k-l}{\xi_i} \underbrace{\sum_{\nu=1}^{\numbFullErrors+\numbColErasures+1} \ELPfullcolWIndex{\nu} \aut^{-(\nu-1)}(x_{\indErrorType, r}^{(i)}) \genNormInv{\nu-1}{\autInvXiWIndex{i}}}_{= \opev{\ELPfullcol}{x_{\indErrorType, r}^{(i)}}{\autInvXiWIndex{i}}}  \\
        &= \opfullexp{\vecxi}{\a_{\indRowErasures, j}}{n-k-l} \cdot \underbrace{\opev{\ELPfullcol}{\x_{\indRowErasures}}{\autInvXiVec}^{\top}}_{=: \hat{\x}_{\indRowErasures}^{\top}}
        \label{eq:coeffs-elpfullcol-skewsyndrome}
    \end{align}
    for all $l = \numbFullErrors + \numbColErasures + 1, \dots, n-k$ and all $j = 1, \dots, \intOrder$.
    Note that the last equality follows because $\ELPfullcol$ satisfies $\opev{\ELPfullcol}{\x_{\indFullErrors}}{\autInvXiVec} = \0$ and $\opev{\ELPfullcol}{\x_{\indColErasures}}{\autInvXiVec} = \0$ according to~\eqref{eq:def-partial-elps} and the product rule of the generalized operator evaluation.

    In the second step, we get
    \begin{align}
        \erasureEvalPoly_{j,l}
        &\overset{\eqref{eq:skew_product_coeffs}}{=} \sum_{\nu=1}^{\numbRowErasuresWIndex{j}+1} (\ELPfullcol \cdot \skewrev{s}_j)_{l-\nu+1} \aut^{-(l-\nu)}\left( (\coeffpower{\ESProwRevWIndex{j}}{n-k-1})_{\nu} \right)
        \overset{\eqref{eq:coeffs-elpfullcol-skewsyndrome}}{=} \sum_{\nu=1}^{\numbRowErasuresWIndex{j}+1} \opfullexp{\vecxi}{\a_{\indRowErasures, j}}{n-k-l+\nu-1} \cdot \hat{\x}_{\indRowErasures}^{\top} \aut^{n-k-l+\numbRowErasuresWIndex{j}}(\ESProwWIndex{j,\numbRowErasuresWIndex{j}-\nu+2}) \\
        &= \sum_{i=1}^{\shots} \sum_{r=1}^{\numbRowErasuresInBlock{i}} \hat{x}_{\indRowErasures, r}^{(i)} \aut^{n - k - l + \numbRowErasuresWIndex{j}} \Biggl( \sum_{\nu=1}^{\numbRowErasuresWIndex{j}+1} \aut^{-(\numbRowErasuresWIndex{j} - \nu + 1)}(a_{\indRowErasures, j, r}^{(i)}) \underbrace{\aut^{-(n - k - l + \numbRowErasuresWIndex{j})}\bigl(\genNorm{n-k-l+\nu-1}{\xi_i}\bigr)}_{\overset{\eqref{eq:normlemma-3}}{=} \genNormInv{n-k-l+\numbRowErasuresWIndex{j}}{\autInvXiWIndex{i}} \cdot \genNormInv{\numbRowErasuresWIndex{j}-\nu+1}{\autInvXiInvWIndex{i}}} \ESProwWIndex{j,\numbRowErasuresWIndex{j}-\nu+2} \Biggr)
        \\
        &= \sum_{i=1}^{\shots} \sum_{r=1}^{\numbRowErasuresInBlock{i}} \hat{x}_{\indRowErasures, r}^{(i)} \aut^{n-k-l+\numbRowErasuresWIndex{j}} \biggl( \genNormInv{n-k-l+\numbRowErasuresWIndex{j}}{\autInvXiWIndex{i}} \underbrace{\sum_{\nu=1}^{\numbRowErasuresWIndex{j}+1} \ESProwWIndex{j,\numbRowErasuresWIndex{j}-\nu+2} \aut^{-(\numbRowErasuresWIndex{j}-\nu+1)}(a_{\indRowErasures, j, r}^{(i)}) \genNormInv{\numbRowErasuresWIndex{j}-\nu+1}{\autInvXiInvWIndex{i}}}_{= \opev{\ESProwWIndex{j}}{a_{\indRowErasures, j,r}^{(i)}}{\autInvXiInvWIndex{i}} = 0} \biggr)
        = 0
    \end{align}
    for all $l = \numbFullErrors + \numbColErasures + \numbRowErasuresWIndex{j} + 1, \dots, n-k$ and all $j = 1, \dots, \intOrder$.
    The last step above follows from the definition of the partial component \acp{ESP} $\ESProwWIndex{1}, \dots, \ESProwWIndex{\intOrder}$ in~\eqref{eq:def-component-esps} and this concludes the proof.
\end{IEEEproof}

As the decoder needs to recover the correct \ac{ELP} $\ELPfull$ up to $\Fqm$-multiples, successful decoding requires that the error-erasure key equation~\eqref{eq:vilrs_error-erasure_key_equation} has a one-dimensional solution space.
Let us fix $\ELP_1 = 1$ without loss of generality and express~\eqref{eq:vilrs_error-erasure_key_equation} equivalently as
\begin{equation}\label{eq:vilrs_error-erasure_key_equation_elp_v1_norm}
   \sum_{\nu=2}^{\numbFullErrors+1}\ELPfullWIndex{\nu}\aut^{-(\nu-1)}(\ELPcomponentSyndrome{j,l-\nu+1})=-\ELPcomponentSyndrome{j,l}
    \quad \text{for all } l=\numbFullErrors + \numbRowErasuresWIndex{j} + \numbColErasures+1,\dots,n-k
    \quad \text{and all } j=1,\dots,\intOrder.
\end{equation}
This is an inhomogeneous $\Fqm$-linear system with $\intOrder(n-k-\numbFullErrors-\numbColErasures-\frac{1}{\intOrder} \sum_{j=1}^{\intOrder}\numbRowErasuresWIndex{j})$ equations in $\numbFullErrors$ unknowns and the system can only have a one-dimensional solution space if the number of equations is at least the number of unknowns, i.e., if
\begin{equation}\label{eq:vilrs_error_erasure_dec_radius}
    \numbFullErrors
    \leq \frac{\intOrder}{\intOrder+1} \biggl( n-k-\numbColErasures- \underbrace{\frac{1}{\intOrder} \sum_{j=1}^{\intOrder}\numbRowErasuresWIndex{j}}_{=: \numbRowErasuresBar} \biggr).
\end{equation}
This characterizes the maximal decoding region of the error-erasure decoder.
It is worth noting that~\eqref{eq:vilrs_error_erasure_dec_radius} depends on the \emph{average} number of row erasures per component error, which we denote by $\numbRowErasuresBar$.
Similar to the error-only setting, errors satisfying~\eqref{eq:vilrs_error_erasure_dec_radius} do not always yield a decoding success but the condition is necessary to render it possible.

Let us now summarize how our syndrome-based decoder recovers all errors and erasures step by step.
We first set up all ingredients for the key equation~\eqref{eq:vilrs_error-erasure_key_equation} and then solve~\eqref{eq:vilrs_error-erasure_key_equation} via multisequence skew-feedback shift-register synthesis~\cite{Sidorenko2011SkewFeedback} as discussed in~\autoref{sec:shift_register_synthesis}.
If the key equation has multiple $\Fqm$-linearly independent solutions, a decoding failure is recorded.
Otherwise, we can use the obtained partial \ac{ELP} $\ELPfull$ to determine the product $\ELPfull(x) \cdot \ELPcol(x) \cdot \skewrev{s}_j(x) = \ELPfullcol(x) \cdot \skewrev{s}_j(x)$ for each $j = 1, \dots, \intOrder$ and set up the systems
\begin{equation}\label{eq:vilrs-gabidulin-like-xhat}
    \opMooreInv{n-k-\numbFullErrors-\numbColErasures}{\hat{\x}_{\indRowErasures}}{\autInvXiVec} \cdot \a_{\indRowErasures, j}^{\top} = \v_j^{\top}
\end{equation}
with $\v_{j} \defeq \bigl( (\ELPfullcol \cdot \skewrev{s}_j)_{n-k}, \autinv\bigl((\ELPfullcol \cdot \skewrev{s}_j)_{n-k-1}\bigr), \dots, \aut^{-(n - k - \numbFullErrors - \numbColErasures - 1)}\bigl((\ELPfullcol \cdot \skewrev{s}_j)_{\numbFullErrors + \numbColErasures + 1}\bigr) \bigr)$ for all $j = 1, \dots, \intOrder$.
This is equivalent to the system~\eqref{eq:coeffs-elpfullcol-skewsyndrome} from the proof of~\autoref{thm:vilrs_elp_key_equation}, which follows from applying $\aut^{-(n - k - l)}$ to the $l$-th equation to obtain
\begin{align}
    \aut^{-(n - k - l)}(\ELPfullcol \cdot \skewrev{s}_j)_l
    &= \aut^{-(n - k - l)} \bigl( \opfullexp{\vecxi}{\a_{\indRowErasures, j}}{n-k-l} \cdot \hat{\x}_{\indRowErasures}^{\top} \bigr)
    = \sum_{i=1}^{\shots} \sum_{r=1}^{\numbRowErasuresInBlock{i}} a_{\indRowErasures, j, r}^{(i)}
    \underbrace{\aut^{-(n - k - l)}\bigl( \genNorm{n - k - l}{\xi_i} \bigr)}_{\overset{\eqref{eq:normlemma-3}}{=} \genNormInv{n - k - l}{\autInvXiWIndex{i}}}
    \aut^{-(n - k - l)}(\hat{x}_{\indRowErasures, r}^{(i)})
    \\
    &= \opfullexpinv{\autInvXiVec}{\hat{\x}_{\indRowErasures}}{n - k - l} \cdot \a_{\indRowErasures, j}^{\top}
\end{align}
for all $l = \numbFullErrors + \numbColErasures + 1, \dots, n-k$ and all $j = 1, \dots, \intOrder$ and reversing the order of the equations.
Now we merge the $\intOrder$ systems from~\eqref{eq:vilrs-gabidulin-like-xhat} into one system over the extension field $\Fqms$ of $\Fqm$.
We fix an ordered $\Fqm$-basis $\vecgamma = (\gamma_1, \dots, \gamma_{\intOrder}) \in \Fqms^{\intOrder}$ of $\Fqms$, define the vectors
\begin{equation}
    \a_{\indRowErasures} \defeq \vecgamma \cdot
    \begin{pmatrix}
        \a_{\indRowErasures, 1} \\
        \vdots \\
        \a_{\indRowErasures, \intOrder}
    \end{pmatrix}
    \in \Fqms^{\numbRowErasures}
    \quad \text{and} \quad
    \v \defeq \vecgamma \cdot
    \begin{pmatrix}
        \v_{1} \\
        \vdots \\
        \v_{\intOrder}
    \end{pmatrix}
    \in \Fqms^{n - k - \numbFullErrors - \numbColErasures},
\end{equation}
and set up the $\Fqms$-linear system
\begin{equation}\label{eq:vilrs-Fqms-Gabidulin}
    \opMooreInv{n-k-\numbFullErrors-\numbColErasures}{\hat{\x}_{\indRowErasures}}{\autInvXiVec} \cdot \a_{\indRowErasures}^{\top} = \v^{\top}
\end{equation}
which indeed combines all systems from~\eqref{eq:vilrs-gabidulin-like-xhat}.
We can apply the Gabidulin-like algorithm from~\autoref{sec:efficient_gabidulin-like} over $\Fqms$ to solve~\eqref{eq:vilrs-Fqms-Gabidulin} and obtain a solution $\hat{\x}_{\indRowErasures} \in \Fqms^{\numbRowErasures}$ with sum-rank weight $\numbRowErasures$.
In fact, we can guarantee that $\hat{\x}_{\indRowErasures}$ has only entries from $\Fqm$ because we know that there \emph{is} a solution in $\Fqm^{\numbRowErasures}$ and the solution is unique.
The latter follows since the fact $\coeffq{\a_{\indRowErasures}} = \A_{\indRowErasures}$ implies $\SumRankWeight(\a_{\indRowErasures}) = \numbRowErasures$.

After we have recovered $\hat{\x}_{\indRowErasures}$, we can reconstruct the partial \ac{ELP} $\ELProw$ for the row erasures and set up the overall \ac{ELP} $\ELP(x) = \ELProw(x) \cdot \ELPfull(x) \cdot \ELPcol(x)$ as defined in~\eqref{eq:vilrs_overall_elp}.
Now the missing code locators $\x_{\indRowErasures}$ and $\x_{\indFullErrors}$ can be recovered via the Skachek--Roth-like algorithm from~\autoref{sec:skachek_roth} according to~\eqref{eq:def-partial-elps}.
More precisely, $\x_{\indFullErrors}$ is recovered first by initializing the algorithm with $\x_{\indColErasures}$ and applying it to $\ELPfull(x) \cdot \ELPcol(x)$.
Then, the Skachek--Roth-like algorithm is initialized with $\x_{\indColErasures}$ and $\x_{\indFullErrors}$ and run on $\ELP(x)$.
This yields the missing vector $\x_{\indRowErasures}$.

A system similar to~\eqref{eq:vilrs_gab-like} allows to recover $\A_{\indFullErrors}$ and $\A_{\indColErasures}$.
Namely, equation~\eqref{eq:vilrs-syndrome-theta-l-1} can be applied to~\eqref{eq:component-syndromes_vilrs_error-erasure}, i.e., to each error type separately, and we obtain
\begin{equation}
    \aut^{l-1}(s_{j,l}) - \opfullexp{\vecxi}{\a_{\indRowErasures,j}}{l-1} \cdot \x_{\indRowErasures}^{\top}
    = \sum_{\indErrorType \in \{\indFullErrors, \indColErasures\}} \opfullexp{\vecxi}{\a_{\indErrorType,j}}{l-1} \cdot \x_{\indErrorType}^{\top}
    = \opfullexp{\vecxi}{\a_{\indFullErrors,j} \mid \a_{\indColErasures,j}}{l-1} \cdot (\x_{\indFullErrors} \mid \x_{\indColErasures})^{\top}
    \quad \text{for all } l = 1, \dots, n - k.
\end{equation}
This yields the systems
\begin{equation}\label{eq:vilrs_error-erasure_gab-like_values}
    \opMoore{n-k}{\a_{\indFullErrors, j} \mid \a_{\indColErasures, j}}{\vecxi} \cdot (\x_{\indFullErrors} \mid \x_{\indColErasures})^{\top} = \widetilde{\s}_j^{\top}
\end{equation}
with $\widetilde{\s}_j=\bigl(s_{j,1} - \a_{\indRowErasures,j} \cdot \x_{\indRowErasures}^{\top},\aut(s_{j,2}) - \opfull{\vecxi}{\a_{\indRowErasures,j}} \cdot \x_{\indRowErasures}^{\top},\dots,\aut^{n-k-1}(s_{j,n-k}) - \opfullexp{\vecxi}{\a_{\indRowErasures,j}}{n-k-1} \cdot \x_{\indRowErasures}^{\top}\bigr)\in\Fqm^{n-k}$ for $j = 1, \dots, \intOrder$ and we can solve them efficiently with the Gabidulin-like algorithm.

We can follow the approach~\eqref{eq:vilrs-left-inverses} in the error-only case for every error type to recover the error locations $\B_{\indFullErrors}$ from $\x_{\indFullErrors}$, $\B_{\indRowErasures}$ from $\x_{\indRowErasures}$, and $\B_{\indColErasures}$ from $\x_{\indColErasures}$, respectively.
Finally, we assemble the error matrix $\E$ as $\E = \A_{\indFullErrors} \B_{\indFullErrors} + \A_{\indRowErasures} \B_{\indRowErasures} + \A_{\indColErasures} \B_{\indColErasures}$ according to~\eqref{eq:err_eras_mat_decomp} and recover the codeword $\C = \Y - \E$.

\autoref{alg:error_erasure_dec_vilrs} and~\autoref{thm:error_erasure_dec_vilrs} summarize the syndrome-based error-erasure decoder for~\ac{VILRS} codes and its properties.

\begin{algorithm}[ht]
  \caption{\algoname{Error-Erasure Decoding of \ac{VILRS} Codes}}\label{alg:error_erasure_dec_vilrs}
  \SetKwInOut{Input}{Input}\SetKwInOut{Output}{Output}
  
  \Input{A channel output $\Y=\C+\E\in\Fqm^{\intOrder\times n}$ with $\C \in \vertIntLinRS{\vecbeta}{\vecxi}{\intOrder}{\nVec}{k}$, $\E = \E_{\indFullErrors} + \E_{\indRowErasures} + \E_{\indColErasures}$, and $\SumRankWeight(\E) = \numbFullErrors + \numbRowErasures + \numbColErasures$ satisfying $\numbErrorsWeightedVertical \leq \numbErrorsMax$,
  \\
  a matrix $\A_\indRowErasures \in \Fqm^{\intOrder \times \numbRowErasures}$ of the form in~\eqref{eq:err_eras_mat_decomp} such that $\subShot{\A}{\indRowErasures}{i}$ has the same column space as $\subShot{\E}{\indRowErasures}{i}$ for $i = 1, \dots, \shots$,
  \\
  a matrix $\B_\indColErasures \in \Fq^{\numbColErasures \times n}$ of the form in~\eqref{eq:err_eras_mat_decomp} such that $\subShot{\B}{\indColErasures}{i}$ has the same row space as $\subShot{\E}{\indColErasures}{i}$ for $i = 1, \dots, \shots$,
  \\
  a parity-check matrix $\H \in \Fqm^{(n-k) \times n}$ of the form $\opMooreInv{n-k}{\h}{\autInvXiVec}$ of $\linRS{\vecbeta}{\vecxi}{\nVec}{k}$,
  \\
  and a left inverse $\widetilde{\H}_q^{(i)} \in \Fq^{n_i \times m}$ of $\coeffq{\h^{(i)}}$ for each $i = 1, \dots, \shots$.
  }

  \Output{The transmitted codeword $\C \in \vertIntLinRS{\vecbeta}{\vecxi}{\intOrder}{\nVec}{k}$ or \emph{``decoding failure''}.
  }
  
  \BlankLine

  Set $\x_{\indColErasures} \gets \h \B_\indColErasures^{\top}$ and compute the partial \ac{ELP} $\ELPcol(x) \gets \mpolArgs{\x_{\indColErasures}}{\autInvXiVec}(x)$.

  \For{$j=1,\dots, \intOrder$}
  {
    Compute the component syndrome $\s_j \gets \y_j\H^\top$ with $\y_j$ being the $j$-th row of $\Y$.
    \\
    Set up the reversed component-syndrome polynomial $\skewrev{s}_j\in\SkewPolyringZeroDerInv$ according to~\eqref{eq:vilrs_rev_syndromes}.
    \\
    Compute the partial component \ac{ESP} $\ESProwWIndex{j}(x) \gets \mpolArgs{\a_{\indRowErasures, j}}{\autInvXiInvVec}(x)$ with $\a_{\indRowErasures, j}$ being the $j$-th row of $\A_{\indRowErasures}$.
    \\
    Set up $\coeffpower{\ESProwRevWIndex{j}}{n-k-1}$ according to~\eqref{eq:coeffpower-esprow-rev}.
    \\
    Set up the auxiliary component-syndrome polynomial $\ELPcomponentSyndrome{j}(x) \gets \ELPcol(x) \cdot \skewrev{s}_j(x) \cdot \coeffpower{\ESProwRevWIndex{j}}{n-k-1}(x)$.
  }
  Solve the key equation~\eqref{eq:vilrs_error-erasure_key_equation} to obtain the partial \ac{ELP} $\ELPfull \in \SkewPolyringZeroDerInv$.
  \\
  \If{the key equation~\eqref{eq:vilrs_error-erasure_key_equation} has a unique solution up to $\Fqm$-multiples}
  {
      Set up $(\ELPfullcol \cdot \skewrev{s}_{j})(x) \gets \ELPfull(x) \cdot \ELPcol(x) \cdot \skewrev{s}_j(x)$ for all $j=1,\dots,\intOrder$.
      \\
  \BlankLine
      Set up $\opMooreInv{n-k-\numbFullErrors-\numbColErasures}{\hat{\x}_{\indRowErasures}}{\autInvXiVec} \cdot \a_{\indRowErasures}^{\top} = \v^{\top}$ from~\eqref{eq:vilrs-Fqms-Gabidulin} and solve it for $\hat{\x}_{\indRowErasures}$.
      \\
      Compute $\ELProw(x) \gets \mpolArgs{\hat{\x}_{\indRowErasures}}{\autInvXiVec}(x)$.
      \\
      Set up $\ELPfullcol(x) \gets \ELPfull(x) \cdot \ELPcol(x)$.
      \\
      Find $\subShot{\x}{\indFullErrors}{i}$ whose entries extend $\subShot{\x}{\indColErasures}{i}$ to a basis of the root space of $\opev{\ELPfullcol}{\cdot}{\autInvXiWIndex{i}}$ for each $i = 1, \dots, \shots$.
      \\
      Set up $\ELP(x) \gets \ELProw(x) \cdot \ELPfullcol(x)$.
      \\
      Find $\subShot{\x}{\indRowErasures}{i}$ whose entries extend $(\subShot{\x}{\indColErasures}{i} \mid \subShot{\x}{\indFullErrors}{i})$ to a basis of the root space of $\opev{\ELP}{\cdot}{\autInvXiWIndex{i}}$ for each $i = 1, \dots, \shots$.
      \\
      Set $\x_{\indFullErrors} \gets (\subShot{\x}{\indFullErrors}{1} \mid \dots \mid  \subShot{\x}{\indFullErrors}{\shots})$. %
      \\
      Set up $\opMoore{n-k}{\a_{\indFullErrors, j} \mid \a_{\indColErasures, j}}{\vecxi} \cdot (\x_{\indFullErrors} \mid \x_{\indColErasures})^{\top} = \widetilde{\s}_j^{\top}$ from~\eqref{eq:vilrs_error-erasure_gab-like_values} and solve it for $(\a_{\indFullErrors, j} \mid \a_{\indColErasures, j})$ for all $j=1,\dots,\intOrder$.
      \\
      Recover the error locations $\B_{\indErrorType}^{(i)\top} \gets \widetilde{\H}_q^{(i)} \cdot \coeffq{\subShot{\x}{\indErrorType}{i}}$ for all $i=1,\dots,\shots$ and $\indErrorType \in \{\indFullErrors, \indRowErasures\}$ as in~\eqref{eq:vilrs-left-inverses}.
      \\
      Let $\A_{\indFullErrors}$ and $\A_{\indColErasures}$ be the matrices with rows $\a_{\indFullErrors, 1}, \dots \a_{\indFullErrors, \intOrder}$ and $\a_{\indColErasures, 1}, \dots \a_{\indColErasures, \intOrder}$, respectively.
      \\
      Set $\B_{\indFullErrors} \gets \diag(\B_{\indFullErrors}^{(1)},\dots,\B_{\indFullErrors}^{(\shots)})$ and $\B_{\indRowErasures} \gets \diag(\B_{\indRowErasures}^{(1)},\dots,\B_{\indRowErasures}^{(\shots)})$.
      \\
      Recover the error matrix $\E \gets \A_{\indFullErrors} \B_{\indFullErrors} + \A_{\indRowErasures} \B_{\indRowErasures} + \A_{\indColErasures} \B_{\indColErasures}$ as in~\eqref{eq:err_eras_mat_decomp}.
      \\
      \Return{$\C \gets \Y-\E$.}
    }

    \Return{``decoding failure''.}
\end{algorithm}

\begin{theorem}[Error-Erasure Decoding of \ac{VILRS} Codes]\label{thm:error_erasure_dec_vilrs}
    Consider the transmission of a codeword $\C \in \vertIntLinRS{\vecbeta}{\vecxi}{\intOrder}{\nVec}{k}$ over the additive error-erasure channel~\eqref{eq:vint_sum-rank_channel}.
    The error $\E \in \Fqm^{\intOrder \times n}$ of sum-rank weight $\SumRankWeight(\E) = \numbErrors$ is chosen uniformly at random from the set $\matGivenSRWeight{q^m}{\intOrder}{\n}{\numbErrors}$ defined in~\eqref{eq:mats_given_sr_weight} and determines the received word $\Y=\C+\E\in\Fqm^{\intOrder\times n}$.
    The channel provides partial knowledge of the error which gives rise to a decomposition into $\numbFullErrors$ full errors, $\numbRowErasures$ row erasures, and $\numbColErasures$ column erasures such that $\numbErrors=\numbFullErrors + \numbRowErasures + \numbColErasures$ holds, as explained in~\autoref{sec:vilrs-channel-error-model}.
    The presented syndrome-based decoder can always recover $\C$ from $\Y$ if $\numbFullErrors \leq \tfrac{1}{2}(n - k - \max_{j}\{\numbRowErasuresWIndex{j}\} - \numbColErasures)$ holds with $\numbRowErasuresWIndex{j} \defeq \SumRankWeight(\e_{\indRowErasures, j}) \leq \numbRowErasures$ for $j = 1, \dots, \intOrder$.
    Moreover, the decoder can be used probabilistically for larger error weights and decoding succeeds with a probability of at least
    \begin{equation}\label{eq:lower_bound_Psucc_vilrs_erasure}
        1 - \gammaq^{\ell+1} q^{-m((s+1)(\numbErrorsMax-\numbErrorsWeightedVertical)+1)}
    \end{equation}
    as long as the error weight satisfies
    \begin{equation}
        \numbErrorsWeightedVertical
        \defeq \numbFullErrors + \frac{\intOrder}{\intOrder+1} (\numbColErasures + \numbRowErasuresBar)
        \leq \numbErrorsMax
        \defeq \frac{\intOrder}{\intOrder+1}\left(n-k\right)
    \quad\text{with}\quad \numbRowErasuresBar \defeq \frac{1}{\intOrder}\sum_{j=1}^{\intOrder} \numbRowErasuresWIndex{j}.
    \end{equation}
    The decoder requires on average $\softoh{\intOrder n^2}$ operations in $\Fqm$ if $m \in O(\intOrder)$ applies.
\end{theorem}

\begin{IEEEproof}
    The reasoning in this section and especially the proof of the key equation in~\autoref{thm:vilrs_error-erasure_key_equation} already showed that~\autoref{alg:error_erasure_dec_vilrs} is correct.
    In particular, a decoding failure is returned correctly in case the key equation~\eqref{eq:vilrs_error-erasure_key_equation} admits multiple $\Fqm$-linearly independent solutions.
    This is the only potential problem during decoding since all other steps are guaranteed to succeed.
    The claimed maximum decoding radius was derived in~\eqref{eq:vilrs_error_erasure_dec_radius}.

    The failure probability and the decoding radius for guaranteed success are closely connected to the key equation~\eqref{eq:vilrs_error-erasure_key_equation} and, in particular, to the dimension of its solution space.
    We observed that the coefficient vectors of the auxiliary component-syndrome polynomials $\ELPcomponentSyndrome{1}, \dots, \ELPcomponentSyndrome{\intOrder}$ in the key equation can be interpreted as modified component syndromes corresponding to an error-only transmission.
    The precise relation can be seen when the ideas of the error-erasure decoder for Gabidulin codes from~\cite{gabidulin2008errorAndErasure} are suitably generalized.
    However, this is a nontrivial endeavor and we will present the details in follow-up work.
    The stated observation justifies to apply the upper bound $\failureProb \leq \gammaq^{\ell+1} q^{-m((s+1)(\numbErrorsMax-\numbErrorsWeightedVertical)+1)}$ on the failure probability $\failureProb$, which we derived for the error-only case in~\autoref{lem:failure_prob_vils}.
    Further, a resulting representation of the coefficients of the auxiliary component-syndrome polynomials gives rise to a simple proof that errors satisfying $\numbFullErrors \leq \tfrac{1}{2}(n - k - \numbColErasures - \max_{j}\{\numbRowErasuresWIndex{j}\})$ ensure a successful decoding.
    
    Let us now move on to the complexity analysis of the decoder and focus on the different types of tasks in the algorithm.
    First observe that operations like taking skew reverses and setting up vectors and matrices according to simple rules are essentially for free and we thus do not mention all of them explicitly.
    The vector-matrix and matrix-matrix products in lines 1, 3, 19, and 22 take at most $O(\intOrder n^2)$ operations in $\Fqm$.
    Equation~\eqref{eq:min_poly} shows how a minimal skew polynomial of degree at most $n$ is computed in at most $O(n^2)$ operations in $\Fqm$ and this yields an upper bound of $O(\intOrder n^2)$ for the respective steps in lines 1, 5, and 12.
    Two skew polynomials of degree at most $n$ can be multiplied in $O(n^2)$ operations in $\Fqm$, which lets us summarize the respective tasks in lines 7, 10, 13, and 15 in $O(\intOrder n^2)$.
    
    We deal with the remaining nontrivial tasks with fast subroutines that we describe in~\autoref{sec:fast-dec}.
    For example, multisequence skew-feedback shift-register synthesis solves the key equation~\eqref{eq:vilrs_error-erasure_key_equation} in line 8 in at most $O(\intOrder (n-k)^2)$ operations in $\Fqm$.
    The linear systems in lines 11 and 18 have a special structure and can be tackled with the Gabidulin-like algorithm.
    While the instance in line 18 is over $\Fqm$ and can be solved in at most $O(\intOrder n^2)$ operations in $\Fqm$, the one in line 11 is over $\Fqms$ and requires $O(\intOrder n^2)$ operations in the extension field $\Fqms$.
    When we choose a suitable $\Fqm$-basis of $\Fqms$, this complexity can be upper-bounded by $\softoh{\intOrder n^2}$ operations in $\Fqm$.
    The Skachek--Roth-like algorithm allows to find a basis of the root space of a skew polynomial of degree at most $n$ with respect to generalized operator evaluation with $\shots$ distinct evaluation parameters and takes on average $O(\shots m n)$ operations in $\Fqm$ to do so.
    If we assume $m \in O(\intOrder)$, the occurring instances of this problem in lines 14 and 16 need on average $O(\intOrder n^2)$ operations in $\Fqm$.
    It is worth noting that the Skachek--Roth-like algorithm is the only piece in this complexity analysis where we deal with an average and not a worst-case complexity.
    All in all, we obtain an asymptotic average complexity of $\softoh{\intOrder n^2}$ operations in $\Fqm$ for the syndrome-based error-erasure decoder.
\end{IEEEproof}

\begin{remark}\label{rem:o-tilde-vilrs}
    As the observant reader might have noticed, the stated asymptotic complexity of the error-erasure decoder is $\softO(\intOrder n^2)$, whereas the error-only decoder is in $O(\intOrder n^2)$ over $\Fqm$.
    Note that this only stems from solving~\eqref{eq:vilrs-Fqms-Gabidulin} via the Gabidulin-like algorithm in $O(\numbRowErasures^2)$ operations in $\Fqms$ and assuming a suitable $\Fqm$-basis of $\Fqms$ which allows to upper-bound the necessary $\Fqms$-operations by $\softoh{\intOrder \numbRowErasures^2}$ operations in $\Fqm$.
    
    In practice, it is quite likely that at least one of the vectors $\a_{\indRowErasures, 1}, \dots, \a_{\indRowErasures, \intOrder}$ has full sum-rank weight $\numbRowErasures$.
    In this case, one can restrict to the corresponding $\Fqm$-linear system in~\eqref{eq:vilrs-gabidulin-like-xhat} and solve it over $\Fqm$ to recover $\hat{\x}_{\indRowErasures}$.
    Then, the Gabidulin-like algorithm takes at most $O(\numbRowErasures^2)$ operations in $\Fqm$ and this reduces the overall asymptotic complexity of the decoder to $O(\intOrder n^2)$ in these cases.
\end{remark}

%% file: hilrs-codes.tex
\section{Horizontally Interleaved Linearized Reed--Solomon (HILRS) Codes and Their Decoding\label{sec:hilrs}}

We now introduce {horizontally} interleaved codes in the sum-rank metric and then give a syndrome-based decoder for the family of \acf{HILRS} codes.
Horizontal interleaving has been studied in the rank metric with a focus on the case of Gabidulin codes~\cite{li2014transform,SidorenkoBossert2010DecodingInterleavedGabidulin,Sidorenko2011SkewFeedback,PuchingerRosenkildeneNielsenEtAl2016Rowreductionapplied,PuchingerMueelichEtAl2017DecodingInterleavedGabidulin} and \ac{LRPC} codes~\cite{RennerJerkovitsEtAl2019EfficientDecodingInterleaved}.
Moreover, a Gao-like decoder for the decoding of \ac{HILRS} codes in the sum-rank metric is available~\cite{HoermannBartz2023FastGaoDecoding}.
The Gao-like approach was only investigated in the error-only setting so far and allows probabilistic unique decoding for large-weight errors.

This section has an analog structure as~\autoref{sec:vilrs} to make the two different interleaving approaches easily comparable.
More precisely, we first introduce the concept of horizontal interleaving for the sum-rank metric, we then give the channel and error models of interest, and we finally present syndrome-based decoders for the error-only setting and for the error-erasure setting.
Note that the latter decoder extends the \acs{ESP}-based error-erasure decoder for non-interleaved \ac{LRS} codes from~\cite{hoermann2022error_erasure}.

\subsection{Horizontal Interleaving in the Sum-Rank Metric}

Let $\mycode{C} \subseteq \Fqm^n$ be an $\Fqm$-linear sum-rank-metric code for the length partition $\n = (n_1, \dots, n_\shots)$ and choose an interleaving order $\intOrder \in \NN^{\ast}$.
In the same spirit as for vertical interleaving, we define the \emph{horizontally interleaved code}
\begin{equation}
    \HInt(\mycode{C}, \intOrder) \defeq \left\{ \c = \left( \c_1 \mid \dots \mid \c_\intOrder \right) : \c_j \in \mycode{C} \text{ for all } j = 1, \dots, \intOrder \right\}
    \subseteq \Fqm^{\intOrder n}
\end{equation}
whose codewords are the concatenation of $\intOrder$ codewords of the component code $\mycode{C}$.

\begin{remark}
    Our above definition of horizontal interleaving is called \emph{homogeneous}, as we only allow one component code $\mycode{C}$ to which all component codewords need to belong.
    It is also possible to consider \emph{heterogeneous} interleaving with $\intOrder$ different component codes $\mycode{C}_1, \dots, \mycode{C}_{\intOrder} \subseteq \mycode{C}$, similar to~\autoref{rem:vert-int-heterogeneous} for the vertical setting.
    In fact, we can generalize horizontal interleaving even further and choose component codes $\mycode{C}_j \subseteq \Fqm^{n_j}$ with different lengths $n_j$ and length partitions $\n_j \in \NN^{\shots}$ for $j = 1, \dots, \intOrder$.
    The component codes only need to share the number of blocks $\shots$ to not interfere with the sum-rank weight for horizontally interleaved vectors which we will introduce shortly.
\end{remark}

While vertical interleaving still preserves the block structure of codewords and thus suggests a straightforward definition of the sum-rank weight with respect to the length partition of $\mycode{C}$, the horizontal case needs a more careful treatment.
Observe that every codeword $\c \in \HInt(\mycode{C}, \intOrder)$ has the form
\begin{equation}\label{eq:hor-int-form}
    \c = \left( \c_1 \mid \dots \mid \c_{\intOrder} \right)
    = \left( \subShot{\c}{1}{1} \mid \dots \mid \subShot{\c}{1}{\shots} \Bigm\vert \dots \Bigm\vert \subShot{\c}{\intOrder}{1} \mid \dots \mid \subShot{\c}{\intOrder}{\shots} \right)
\end{equation}
and therefore follows the naturally induced length partition $\vecntilde \defeq (\n, \dots, \n) \in \NN^{\intOrder \shots}$ consisting of $\intOrder$ copies of the length partition $\n$ of $\mycode{C}$.
However, if the sum-rank weight was defined with respect to $\vecntilde$, it would add up the $\Fq$-ranks of the $\intOrder \shots$ blocks and thus equal the sum $\SumRankWeightWPartition{\n}(\c_1) + \dots + \SumRankWeightWPartition{\n}(\c_\intOrder)$ of the sum-rank weights of the component codewords.
This corresponds to independently sending the $\intOrder$ component codewords of $\mycode{C}$ over a non-interleaved sum-rank channel and does not capture horizontal interleaving in the rank metric as a special case.

We regroup the blocks of the interleaved codewords and define the sum-rank weight with respect to the reordering.
This incorporates the rank-metric case and lets us obtain a gain for the joint decoding of the $\intOrder$ concatenated codewords.
We divide $\c$ into $\shots$ blocks $\shot{\c}{1}, \dots, \shot{\c}{\shots}$ and choose the $i$-th block to be the concatenation $( \subShot{\c}{1}{i} \mid \dots \mid \subShot{\c}{\intOrder}{i} )$ of the $i$-th block of every component codeword.
This reordering is illustrated in~\autoref{fig:sum-rank-for-interleaved-vectors} and leads to the length partition $\intOrder \n \defeq (\intOrder n_1, \dots, \intOrder n_{\shots})$ which we will use for the definition of the sum-rank weight.
Note that this reordering process is not limited to codewords but can be applied to arbitrary vectors in $\Fqm^{\intOrder n}$.

We define the \emph{sum-rank weight} of a horizontally interleaved vector $\x \in \Fqm^{\intOrder n}$ of the form~\eqref{eq:hor-int-form} as
\begin{equation}\label{eq:def-hor-sr-weight}
    \SumRankWeightWPartition{\intOrder \n}(\x) \defeq \sum_{i=1}^{\shots} \rkq\bigl((\x_1^{(i)} \mid \dots \mid \x_\intOrder^{(i)})\bigr)
\end{equation}
using the reordered length partition $\intOrder \n \defeq (\intOrder n_1, \dots, \intOrder n_{\shots})$ of $\intOrder n$.
The corresponding sum-rank metric is given by $\SumRankDistWPartition{\intOrder \n}(\x, \y) \defeq \SumRankWeightWPartition{\intOrder \n}(\x - \y)$ for arbitrary $\x, \y \in \Fqm^{\intOrder n}$.
If the length partition $\intOrder \n$ is clear from the context, we will omit it for readability and use the notations $\SumRankWeight(\cdot)$ and $\SumRankDist(\cdot, \cdot)$.
We further use the shorthand $\shot{\x}{i} \defeq ( \subShot{\x}{1}{i} \mid \dots \mid \subShot{\x}{\intOrder}{i} ) \in \Fqm^{\intOrder n_i}$ to denote the $i$-th block of $\x$ with respect to $\intOrder \n$ for all $i = 1, \dots, \shots$ in the following.
Remark that this does not collide with the notation $\subShot{\x}{j}{i}$ for the blocks arising from the length partition $\vecntilde$ with $i = 1, \dots, \shots$ and $j = 1, \dots, \intOrder$.

If we choose an \ac{LRS} code as defined in~\autoref{def:LRS_codes} as the component code $\mycode{C}$ in the construction of a horizontally interleaved code, we arrive at a \emph{\acf{HILRS}} code
\begin{equation}
    \horIntLinRS{\vecbeta}{\vecxi}{\intOrder}{\vecntilde}{k} \defeq \HInt\left(\linRS{\vecbeta}{\vecxi}{\nVec}{k}, \intOrder \right).
\end{equation}
This code has interleaving order $\intOrder \in \NN^{\ast}$, length $\intOrder n$, induced length partition $\vecntilde=(\n,\dots,\n)\in\NN^{\shots\intOrder}$, and dimension $\intOrder k$.
However, recall that we measure the sum-rank weight of its codewords with respect to the length partition $\intOrder \n$, i.e., after reordering the blocks as displayed in~\autoref{fig:sum-rank-for-interleaved-vectors}.
We can explicitly compute the minimum sum-rank distance of \ac{HILRS} codes in the next lemma:

\begin{lemma}[Minimum Distance of HILRS Codes] \label{lem:min-dist-hilrs}
    The \ac{HILRS} code $\horIntLinRS{\vecbeta}{\vecxi}{\intOrder}{\vecntilde}{k}$ has minimum sum-rank distance
    \begin{equation}
        d = n-k+1.
    \end{equation}
\end{lemma}

\begin{IEEEproof}
    Since \ac{HILRS} codes are $\Fqm$-linear, their minimum sum-rank distance equals the minimum weight of a nonzero codeword.
    As the weight of any codeword $\c \in \horIntLinRS{\vecbeta}{\vecxi}{\intOrder}{\vecntilde}{k}$ is computed according to~\eqref{eq:def-hor-sr-weight}, the $i$-th summand is bounded by
    \begin{equation}
        0 \leq \rkq\bigl((\x_1^{(i)} \mid \dots \mid \x_\intOrder^{(i)})\bigr) \leq \sum_{j=1}^{\intOrder} \rkq(\subShot{\x}{j}{i})
        \quad \text{for all } i = 1, \dots, \shots.
    \end{equation}
    Hence, the lowest sum-rank weight can be achieved by choosing the all-zero codeword for all component codewords except for one.
    The optimal choice for the latter is a minimum-weight codeword in the component code $\mathcal{C} = \linRS{\vecbeta}{\vecxi}{\nVec}{k}$ and thus has sum-rank weight $n - k + 1$.
    This shows that an interleaved codeword of minimum weight has weight $d \defeq n - k + 1$ and thus $\horIntLinRS{\vecbeta}{\vecxi}{\intOrder}{\vecntilde}{k}$ has minimum sum-rank distance $d$.
\end{IEEEproof}

Note that \ac{HILRS} codes are \emph{not} \ac{MSRD} for $\intOrder > 1$, as the Singleton-like bound reads $d \leq \intOrder(n-k) + 1$ in this setting.
This behavior is different from \ac{VILRS} codes which are \ac{MSRD} codes for \emph{all} interleaving orders.

\subsection{Channel and Error Models}\label{sec:hilrs-channel-error-model}

Let us first describe the error-only scenario and then move to the error-erasure setting, which incorporates three different error types.
In any case, we consider the transmission of a codeword $\c\in\horIntLinRS{\vecbeta}{\vecxi}{\intOrder}{\vecntilde}{k}$ over a sum-rank channel that returns the received word
\begin{equation}\label{eq:sum_rank_channel_hor_int}
    \y = \c + \e,
\end{equation}
where $\e \in \Fqm^{\intOrder n}$ is an additive error of sum-rank weight $\SumRankWeightWPartition{\intOrder \n}(\e)=\numbErrors$.
Recall once more that the weight of $\e$ is measured according to the reordered length partition $\intOrder \n$ as described in~\eqref{eq:def-hor-sr-weight}.
We assume that the error $\e$ is drawn uniformly at random from the set
\begin{equation}\label{eq:vectors-fixed-horizontal-weight}
    \vecGivenSRWeight{q^m}{\intOrder \n}{\numbErrors} \defeq \left\{ \v \in \Fqm^{\intOrder n} : \SumRankWeightWPartition{\intOrder \n}(\v) = \numbErrors \right\}
\end{equation}
and denote its rank partition by $\numbErrorsVec=(\numbErrorsInBlock{1},\dots,\numbErrorsInBlock{\shots}) \in \NN^\shots$ with $\numbErrorsInBlock{i} \defeq \rkq(\shot{\e}{i}) = \rkq\bigl((\subShot{\e}{1}{i} \mid \dots \mid \subShot{\e}{\intOrder}{i})\bigr)$ for all $i=1,\dots,\shots$.

\begin{remark}
    Note that the error sets $\matGivenSRWeight{q^m}{\intOrder}{\n}{\numbErrors}$ and $\vecGivenSRWeight{q^m}{\intOrder \n}{\numbErrors}$ for vertical and horizontal interleaving are not isomorphic even though $\Fqm^{s \times n} \simeq \Fqm^{sn}$ holds.
    The definitions of the sets look similar but the respective notions of sum-rank weight given in~\eqref{eq:sum-rank-for-vertical-interleaving} and~\eqref{eq:def-hor-sr-weight} make the difference.
    Since a horizontally interleaved error $\e$ can be reordered into $\e = (\shot{\e}{1} \mid \dots \mid \shot{\e}{\shots})$ and the sum-rank condition translates to $\rkq(\shot{\e}{i}) = \numbErrorsInBlock{i}$ for all $i = 1, \dots, \shots$, the uniform sampling of errors from $\vecGivenSRWeight{q^m}{\intOrder \n}{\numbErrors}$ boils down to sampling vectors from $\Fqm^{\intOrder n}$ with $\SumRankWeightWPartition{(\intOrder n_1, \dots, \intOrder n_{\shots})}(\e) = \numbErrors$ uniformly at random.
\end{remark}

Another perspective on the channel in~\eqref{eq:sum_rank_channel_hor_int} is
\begin{equation}
    \left( \y_1 \mid \dots \mid \y_{\intOrder} \right) = \left( \c_1 \mid \dots \mid \c_{\intOrder} \right) + \left( \e_1 \mid \dots \mid \e_{\intOrder} \right)
    \quad \text{with }
    \y_j, \c_j, \e_j \in \Fqm^{n}
    \quad \text{for all } j = 1, \dots, \intOrder.
\end{equation}
It focuses on the components induced by the interleaving structure.
In the following, we switch between this representation and the one given in~\eqref{eq:sum_rank_channel_hor_int} depending on the context.

The condition $\SumRankWeight(\e) = \numbErrors$ on the sum-rank weight of the error is equivalent to requiring $\rkq(\shot{\e}{i}) = \numbErrorsInBlock{i}$ for all $i = 1, \dots, \shots$.
Thus, a full-rank decomposition~\cite[Thm.~1]{matsaglia1974equalitiesInequalities} of each block $\shot{\e}{i}$ yields a vector $\a^{(i)}\in\Fqm^{\numbErrorsInBlock{i}}$ with $\rkq(\a^{(i)})=\numbErrorsInBlock{i}$ and matrices $\B_j^{(i)}\in \Fq^{\numbErrorsInBlock{i} \times n_i}$ with $\rkq\bigl((\B_1^{(i)} \mid \dots \mid \B_\intOrder^{(i)})\bigr)=\numbErrorsInBlock{i}$ such that
\begin{equation}\label{eq:hilrs_error_decomp_per_block}
    \shot{\e}{i} = (\e_1^{(i)} \mid \dots \mid \e_\intOrder^{(i)}) = \shot{\a}{i} \cdot \bigl( \shot{\B}{i}_{1} \mid \dots \mid \shot{\B}{i}_{\intOrder} \bigr)
\end{equation}
holds for every $i = 1, \dots, \shots$.
We will use the shorthand notation $\shot{\B}{i} \defeq (\B_1^{(i)} \mid \dots \mid \B_\intOrder^{(i)})$ for all $i = 1, \dots, \shots$ in the following to highlight the correspondence to the error block $\shot{\e}{i}$.
When we now reorder the blocks of $\e$ according to its natural length partition $\vecntilde$, we obtain the error decomposition
\begin{equation}\label{eq:err_int_vec_decomp}
    \e
    = (\e_1 \mid \dots \mid \e_{\intOrder})
    = \underbrace{(\shot{\a}{1} \mid \dots \mid \shot{\a}{\shots})}_{=: \a}
    \cdot
    \underbrace{\left(
    \begin{array}{ccc|c|ccc}
     \B_1^{(1)} & & & & \B_\intOrder^{(1)} &
     \\
     & \ddots & & \dots & & \ddots &
     \\
    & & \B_1^{(\shots)} & &  & & \B_\intOrder^{(\shots)}
    \end{array}
    \right)}_{=: (\B_1 \mid \dots \mid \B_{\intOrder}) =: \B}
\end{equation}
with $\a \in \Fqm^{\numbErrors}$ having sum-rank weight $\SumRankWeightWPartition{\intOrder \n}(\a) = \tau$.
Further, $\SumRankWeightWPartition{\intOrder \n}(\B) = \numbErrors$ holds and is equivalent to the equality $\rkq(\shot{\B}{i}) = \numbErrorsInBlock{i}$ for every $i = 1, \dots, \shots$.
Note that the decomposition in~\eqref{eq:err_int_vec_decomp} is only unique up to a change of $\Fq$-basis for each block $\subShot{\B}{j}{i}$ for $i = 1, \dots, \shots$ and $j = 1, \dots, \intOrder$.
Namely, any matrix $\M = (\M_1 \mid \dots \mid \M_{\intOrder}) \in \Fq^{\numbErrors \times \intOrder \numbErrors}$ with block-diagonal blocks $\M_j = \diag\bigr( \subShot{\M}{j}{1}, \dots, \subShot{\M}{j}{\shots} \bigl)$ and full-rank $\subShot{\M}{j}{i} \in \Fq^{\numbErrorsInBlock{i} \times \numbErrorsInBlock{i}}$ for all $i =1, \dots, \shots$ and all $j = 1, \dots, \intOrder$ yields another valid error decomposition
\begin{equation}
    \e = \a' \cdot \B' = \underbrace{\a \M^{-1}}_{:= \a'} \cdot \underbrace{\M \B}_{=: \B'}.
\end{equation}

Since the entries of $\a^{(i)}$ span the $\Fq$-column space of the error block $\shot{\e}{i}$ for each $i = 1, \dots, \shots$, we call them \emph{error values}.
Similarly, the rows of $\shot{\B}{i}$ form an $\Fq$-basis of the row space of $\shot{\e}{i}$ for every $i = 1, \dots, \shots$ and are referred to as \emph{error locations}.
The above decomposition~\eqref{eq:err_int_vec_decomp} shows that the error values represented by the vector $\a$ are common for all received component errors $\e_1, \dots, \e_{\intOrder}$.
In contrast, the error locations differ for the components and we use the notation $\B_j \defeq \diag\bigl( \subShot{\B}{j}{1}, \dots, \subShot{\B}{j}{\shots} \bigr) \in \Fq^{\numbErrors \times n}$ to denote the part of the matrix $\B$ corresponding to the $j$-th component error $\e_j$ for $j = 1, \dots, \intOrder$.

\begin{remark}
    The error decomposition in~\eqref{eq:err_int_vec_decomp} corresponds to the one in~\eqref{eq:err_mat_decomp} for vertical interleaving.
    They are both obtained by applying full-rank decompositions to the blocks for which the $\Fq$-rank is known.
    However, vertical and horizontal interleaving establish different structures in the error and result in a somewhat complementary representation.
    Namely, the component errors share their row space for vertical interleaving and their column space for horizontal interleaving, respectively.
    The component errors thus share the matrix $\B$ in~\eqref{eq:err_mat_decomp}, whereas they share the vector $\a$ in~\eqref{eq:err_int_vec_decomp}.
\end{remark}

Let us now discuss the error-erasure setting, in which we incorporate partial knowledge about the error.
The three considered error types are the same as in the case of vertical interleaving and they are illustrated in~\autoref{fig:error_decomposition}.
Namely, we categorize each of the $\numbErrors = \SumRankWeight(\e)$ occurred sum-rank errors
\begin{itemize}
    \item as a \emph{(full) error} or an error of type $\indFullErrors$, if neither the row space nor the column space is known,
    \item as a \emph{row erasure} or an error of type $\indRowErasures$, if its column space is known,
    \item or as a \emph{column erasure} or an error of type $\indColErasures$, if its row space is known.
\end{itemize}
We denote the number of full errors, row erasures, and column erasures by $\numbFullErrors$, $\numbRowErasures$, and $\numbColErasures$, respectively.
The equality $\numbErrors = \numbFullErrors + \numbRowErasures + \numbColErasures$ holds, as every sum-rank error belongs to precisely one error type.
We write $\numbErrorsInBlock{i} = \numbFullErrorsInBlock{i} + \numbRowErasuresInBlock{i} + \numbColErasuresInBlock{i}$ for all $i = 1, \dots, \shots$ to highlight which errors occurred in the $i$-th error block $\shot{\e}{i}$.

The error $\e$ has an additive decomposition $\e = \eFullErrors + \eRowErasures + \eColErasures$
into vectors $\e_\indErrorType \in \Fqm^{\intOrder n}$ that contain the errors of the respective error type $\indErrorType \in \{\indFullErrors, \indRowErasures, \indColErasures\}$ and satisfy $\SumRankWeight(\e_\indErrorType) = \numbErrorType$.
We can decompose all three parts as described in~\eqref{eq:err_int_vec_decomp} and obtain
\begin{equation}\label{eq:hilrs_error-erasure_decomposition}
    \e = \a_{\indFullErrors} \cdot \B_{\indFullErrors} + \a_{\indRowErasures} \cdot \B_{\indRowErasures} + \a_{\indColErasures} \cdot \B_{\indColErasures}
\end{equation}
with $\a_{\indErrorType} \in \Fqm^{\numbErrorType}$ having sum-rank weight $\SumRankWeight(\a_{\indErrorType}) = \numbErrorType$ and $\B_{\indErrorType}$ of the form given in~\eqref{eq:err_int_vec_decomp} with $\rkq(\subShot{\B}{\indErrorType}{i}) = \numbErrorTypeInBlock{i}$ for all $i = 1, \dots, \shots$ and every $\indErrorType \in \{ \indFullErrors, \indRowErasures, \indColErasures \}$.
As in the error-only setting above, the entries of $\subShot{\a}{\indErrorType}{i}$ are a basis of the column space of $\subShot{\e}{\indErrorType}{i}$ and the rows of $\subShot{\B}{\indErrorType}{i}$ span its row space for all $i = 1, \dots, \shots$ and every error type $\indErrorType \in \{\indFullErrors, \indRowErasures, \indColErasures\}$.
Therefore, we call the entries of $\a_{\indFullErrors}$, $\a_{\indRowErasures}$, and $\a_{\indColErasures}$ \emph{error values} and the rows of $\B_{\indFullErrors}$, $\B_{\indRowErasures}$, and $\B_{\indColErasures}$ \emph{error locations}.
\autoref{fig:error_decomposition} depicts the decomposition in~\eqref{eq:hilrs_error-erasure_decomposition} for the non-interleaved setup and highlights the parts that the receiver knows before decoding.
Recall that $\a_\indRowErasures$ and $\B_\indColErasures$ are known for horizontal interleaving according to the definition of row and column erasures.

\subsection{Error-Only Decoding}

We now present a syndrome-based decoder for the error-only setting and the channel~\eqref{eq:sum_rank_channel_hor_int} over which a codeword $\c \in \horIntLinRS{\vecbeta}{\vecxi}{\intOrder}{\vecntilde}{k}$ was transmitted.
Let $\y \in \Fqm^{\intOrder n}$ denote the received word and assume that the sum-rank weight of the error $\e = \y - \c$ is $\numbErrors$.
Further, let $\H\in\Fqm^{(n-k) \times n}$ be a parity-check matrix of the component code $\linRS{\vecbeta}{\vecxi}{\nVec}{k}$ and assume without loss of generality that $\H$ is a generalized Moore matrix with respect to the vector $\h \in \Fqm^{n}$ as described in~\eqref{eq:parity_check_mat}.

Compute the \emph{component syndromes} $\s_1, \dots, \s_{\intOrder}$ as
\begin{equation}\label{eq:syndromes_hilrs}
    \s_j
    =\y_j\H^\top
    =\e_j\H^\top
    \overset{\eqref{eq:err_int_vec_decomp}}{=}\a\B_j\H^\top
    \quad \text{for all } j = 1, \dots, \intOrder.
\end{equation}
Now the $l$-th entry $s_{j, l}$ of $\s_j$ for $l = 1, \dots, n-k$ and $j = 1, \dots, \intOrder$ is
\begin{equation}\label{eq:hilrs_syndrome_entries}
    s_{j,l}
    = \a \B_j \opfullexpinv{\autInvXiVec}{\h}{l-1}^{\top}
    = \a \opfullexpinv{\autInvXiVec}{\underbrace{\h \B_j^{\top}}_{=: \x_j}}{l-1}^{\top}
    = \a \opfullexpinv{\autInvXiVec}{\x_j}{l-1}^{\top},
\end{equation}
where $\autInvXiVec = \autinv(\vecxi)$ according to~\eqref{eq:def_autInvXiVec}.
We define the \emph{error locators} of the $j$-th component as $\x_j \defeq \h \B_j^{\top} \in \Fq^{\tau}$ for $j = 1, \dots, \intOrder$ and denote the vector's block structure with respect to the rank partition $\vectau$ of the error and the matrix $\B_j^{\top}$ by $\x_j = (\subShot{\x}{j}{1} \mid \dots \mid \subShot{\x}{j}{\shots})$.
This is in the same spirit as in the vertical case but now the error locators depend on the component index $j = 1, \dots, \intOrder$ and the vector $\a$ containing the error values is independent of it.
Therefore, the \ac{HILRS} decoder starts by recovering the common error values, in contrast to first recovering the error locators in the vertical setting.

We will shortly derive a key equation that is focused on the \acl{ESP} and allows to recover it by exploiting its relations with the component syndromes $\s_1, \dots, \s_{\intOrder}$.
The \emph{\acf{ESP}} is the minimal skew polynomial $\ESP\in\SkewPolyringZeroDerInv$ vanishing on the error values $\a$ with respect to generalized operator evaluation and the evaluation parameters $\autInvXiInvVec = \autinv(\vecxi^{-1})$ defined in~\eqref{eq:def_autInvXiVec}.
In other words, $\ESP$ is the skew polynomial of minimal degree satisfying $\opev{\ESP}{\a}{\autInvXiInvVec} = \0$.
Because the evaluation parameters in $\autInvXiInvVec$ belong to pairwise distinct nontrivial conjugacy classes of $\Fqm$ and the blocks $\shot{\a}{1}, \dots, \shot{\a}{\shots}$ of the error values are $\Fq$-linearly independent, the \ac{ESP} has degree $\numbErrors$.
We further associate to each component syndrome $\s_j \in \Fqm^{n-k}$ the corresponding \emph{component-syndrome polynomial} $s_j\in\SkewPolyringZeroDerInv$ given by $s_j(x)=\sum_{l=1}^{n-k}s_{j,l}x^{l-1}$ for all $j=1,\dots,\intOrder$.
Now we can state the \ac{ESP} key equation which is the key ingredient for the syndrome-based decoding of \ac{HILRS} codes:

\begin{theorem}[\ac{ESP} Key Equation]\label{thm:hilrs_esp_key_equation}
    For each $j = 1, \dots, \intOrder$, there is a skew polynomial $\errEvalPoly_j\in\SkewPolyringZeroDerInv$ with $\deg(\errEvalPoly_j) < \numbErrors$ that satisfies
    \begin{equation}\label{eq:hilrs_key_equation_esp}
        \ESP(x) \cdot s_j(x) \equiv \errEvalPoly_j(x) \modr x^{n-k}.
    \end{equation}
\end{theorem}

\begin{IEEEproof}
    We show~\eqref{eq:hilrs_key_equation_esp} by proving that $\errEvalPoly_{j,l}=0$ holds for all $l = \numbErrors + 1, \dots, n-k$ and all $j = 1, \dots, \intOrder$.
    We obtain
    \begin{align*}
        \errEvalPoly_{j,l}
        &= (\ESP \cdot s_j)_l
        \overset{\eqref{eq:skew_product_coeffs}}{=}\sum_{\nu=1}^{\numbErrors+1} \ESP_\nu \aut^{-(\nu-1)}(s_{j,l-\nu+1})
        \overset{\eqref{eq:hilrs_syndrome_entries}}{=} \sum_{\nu=1}^{\numbErrors+1} \ESP_\nu \aut^{-(\nu-1)}\bigl( \a \opfullexpinv{\autInvXiVec}{\x_j}{l-\nu}^{\top} \bigr)
        \\
        &=\sum_{i=1}^{\shots} \sum_{r=1}^{\numbErrorsInBlock{i}} \aut^{-(l-1)}(x_{j,r}^{(i)}) \sum_{\nu=1}^{\numbErrors+1} \ESP_\nu \aut^{-(\nu-1)}(a_r^{(i)})
        \underbrace{\aut^{-(\nu-1)}\bigl(\genNormInv{l-\nu}{\autInvXiWIndex{i}}\bigr)}_{\overset{\eqref{eq:normlemma-2}}{=} \genNormInv{l-1}{\autInvXiWIndex{i}} \cdot \genNormInv{\nu-1}{\autInvXiInvWIndex{i}}}
        \\
        &=\sum_{i=1}^{\shots} \sum_{r=1}^{\numbErrorsInBlock{i}} \aut^{-(l-1)}(x_{j,r}^{(i)}) \genNormInv{l-1}{\autInvXiWIndex{i}} \underbrace{\sum_{\nu=1}^{\numbErrors+1} \ESP_\nu \aut^{-(\nu-1)}(a_r^{(i)}) \genNormInv{\nu-1}{\autInvXiInvWIndex{i}}}_{= \opev{\ESP}{a_r^{(i)}}{\autInvXiInvWIndex{i}} = 0}
        = 0.
    \end{align*}
\end{IEEEproof}

The above proof shows that the key equation in~\eqref{eq:hilrs_key_equation_esp} is equivalent to a set of $\Fqm$-linear equations.
Namely, we get
\begin{equation}\label{eq:hilrs_key_equation_esp_v1_norm}
    \sum_{\nu=2}^{\numbErrors+1}\ESP_\nu\aut^{-(\nu-1)}\left(s_{j,l-\nu+1}\right)=-s_{j,l}
    \quad \text{for all } l=\numbErrors+1,\dots,n-k \quad \text{and all } j=1,\dots,\intOrder
\end{equation}
when the normalization $\ESP_1=1$ is assumed without loss of generality.
This can be expressed as
\begin{equation}\label{eq:hilrs_key_equation_system}
    \begin{pmatrix}
     \S_1
     \\
     \vdots
     \\
     \S_\intOrder
    \end{pmatrix}
    \cdot
    \vec{\sigma}^\top
    = -
    \begin{pmatrix}
     \s_1^{\prime\top}
     \\
     \vdots
     \\
     \s_\intOrder^{\prime\top}
    \end{pmatrix},
\end{equation}
where the matrix
\begin{equation}\label{eq:def_esp_syndrome_matrix}
    \S_j
    =
    \begin{pmatrix}
     \aut^{-1}(s_{j,\numbErrors}) & \aut^{-2}(s_{j,\numbErrors-1}) & \dots & \aut^{-\numbErrors}(s_{j,1})
     \\
     \aut^{-1}(s_{j,\numbErrors+1}) & \aut^{-2}(s_{j,\numbErrors}) & \dots & \aut^{-\numbErrors}(s_{j,2})
     \\
     \vdots & \vdots & \ddots & \vdots
     \\
     \aut^{-1}(s_{j,n-k-1}) & \aut^{-2}(s_{j,n-k-2}) & \dots & \aut^{-\numbErrors}(s_{j,n-k-\numbErrors})
    \end{pmatrix}
    \in \Fqm^{(n-k-\numbErrors) \times \numbErrors}
\end{equation}
and the vector $\s_j'=(s_{j, \numbErrors+1},\dots,s_{j,n-k})$ are defined for each $j=1,\dots,\intOrder$ and the vector $\vec{\ESP}=(\ESP_2,\dots,\ESP_{\numbErrors+1})\in\Fqm^\numbErrors$ captures the unknown coefficients of the \ac{ESP}.
For brevity, we often write $\S \cdot \vec{\ESP}^{\top} = \s'^{\top}$ to represent the system~\eqref{eq:hilrs_key_equation_system} in the following.

\begin{remark}
    The reformulation~\eqref{eq:hilrs_key_equation_system} can be used to solve the key equation~\eqref{eq:hilrs_key_equation_esp} by Gaussian elimination.
    However, multisequence skew-feedback shift-register synthesis~\cite{Sidorenko2011SkewFeedback} is a tool that is more tailored to this problem and has less computational complexity.
    We discuss it briefly in~\autoref{sec:shift_register_synthesis}.
\end{remark}

The inhomogeneous linear system~\eqref{eq:hilrs_key_equation_system} consists of $\intOrder(n-k-\numbErrors)$ equations in $\numbErrors$ unknowns and can only have a one-dimensional solution space if the number of equations is at least the number of unknowns, i.e., if $\numbErrors \leq \intOrder(n-k-\numbErrors)$ applies.
Since~\eqref{eq:hilrs_key_equation_system} is equivalent to the \ac{ESP} key equation~\eqref{eq:hilrs_key_equation_esp}, this directly yields the necessary condition
\begin{equation}\label{eq:hilrs_dec-radius}
    \numbErrors \leq \numbErrorsMax \defeq \tfrac{\intOrder}{\intOrder+1}(n-k)
\end{equation}
on the error weight $\numbErrors$ for potentially successful decoding.

A decoding failure occurs if the key equation~\eqref{eq:hilrs_key_equation_esp} has multiple solutions, which corresponds to the case that the matrix $\S$ on the left-hand side of~\eqref{eq:hilrs_key_equation_system} has $\Fqm$-rank less than $\numbErrors$.
The following lemma gives an upper bound on the probability for this scenario and also shows when $\S$ is guaranteed to have full rank and unique decoding is possible.

\begin{lemma}\label{lem:failure_prob_hilrs}
    Let $\S$ be the coefficient matrix of the system~\eqref{eq:hilrs_key_equation_system}, which arose from an \ac{HILRS} decoding instance $\y = \c + \e$ with error weight $\SumRankWeight(\e) = \numbErrors \leq \numbErrorsMax$.
    Then, the bound
    \begin{equation}
        \Pr\left\{\rkqm(\S) < \numbErrors \right\}
        \leq \kappa_q^{\shots+1}q^{-m((\intOrder+1)(\numbErrorsMax-\numbErrors)+1)}
    \end{equation}
    applies for $\kappa_q < 3.5$ being defined in~\eqref{eq:def_kappa_q}.
    Moreover, $\rkqm(\S) = \numbErrors$ is guaranteed for any error $\e$ of weight $\numbErrors \leq \tfrac{1}{2}(n-k)$.
\end{lemma}

\begin{IEEEproof}
    The definition of $\S_j$ in~\eqref{eq:def_esp_syndrome_matrix} shows that its entry in column $\nu - 1$ and row $l + \nu - \tau - 2$ is $\aut^{-(\nu-1)}\left(s_{j,l-\nu+1}\right)$ for $\nu=2, \dots, \numbFullErrors+1$, $l=\numbErrors+1, \dots, n-k$, and $j = 1, \dots, \intOrder$.
    Since
    \begin{align}
        \aut^{-(\nu-1)}(s_{j,l-\nu+1})
        &\overset{\eqref{eq:hilrs_syndrome_entries}}{=} \aut^{-(\nu-1)} \bigl( \a \opfullexpinv{\autInvXiVec}{\x_j}{l-\nu}^{\top} \bigr)
        = \sum_{i=1}^{\shots} \sum_{r=1}^{\numbErrors} \aut^{-(\nu-1)}(a_r^{(i)}) \aut^{-(l-1)}(x_{j,r}^{(i)})
        \underbrace{\aut^{-(\nu-1)}\bigl(\genNormInv{l-\nu}{\autInvXiWIndex{i}}\bigr)}
        _{\overset{\eqref{eq:normlemma-2}}{=} \genNormInv{l-1}{\autInvXiWIndex{i}} \cdot \genNormInv{\nu-1}{\autInvXiInvWIndex{i}}}
        \\
        &= \sum_{i=1}^{\shots} \sum_{r=1}^{\numbErrors} \opfullexpinv{\autInvXiWIndex{i}}{x_{j,r}^{(i)}}{l-1} \opfullexpinv{\autInvXiInvWIndex{i}}{a_{r}^{(i)}}{\nu-1} \label{eq:esp_key_equation_decomp}
        = \opfullexpinv{\autInvXiVec}{\x_j}{l-1} \cdot \opfullexpinv{\autInvXiInvVec}{\a}{\nu-1}^{\top}
    \end{align}
    holds for all $\nu=2, \dots, \numbErrors + 1$, all $l=\numbErrors+1, \dots, n-k$, and all $j = 1, \dots, \intOrder$, we obtain the decomposition
    \begin{equation}
        \S_j
        =
        \underbrace{
        \begin{pmatrix}
            \opfullexpinv{\autInvXiVec}{\x_j}{\numbErrors} \\
            \vdots \\
            \opfullexpinv{\autInvXiVec}{\x_j}{n - k - 1}
        \end{pmatrix}
        }_{=: \hat{\X}_j \in \Fqm^{(n - k - \numbErrors) \times \numbErrors}}
        \cdot
        \underbrace{
        \begin{pmatrix}
            \opfullinv{\autInvXiInvVec}{\a} \\
            \vdots \\
            \opfullexpinv{\autInvXiInvVec}{\a}{\numbErrors} \\
        \end{pmatrix}
        ^{\top}
        }_{=: \hat{\A} \in \Fqm^{\numbErrors \times \numbErrors}}
        \quad \text{for all } j = 1, \dots, \intOrder.
    \end{equation}
    Observe that $\hat{\A} = {\opMooreInv{\numbErrors}{\opfullinv{\autInvXiInvVec}{\a}}{\autInvXiInvVec}}^\top$ is independent of $j$ and has full $\Fqm$-rank $\numbErrors$ because the sum-rank weight of $\a$ equals $\numbErrors$ by definition.
    When we now combine $\S_1, \dots, \S_{\intOrder}$ into $\S$, we get
    \begin{equation}
    \S =
    \begin{pmatrix}
         \S_1
         \\
         \vdots
         \\
         \S_\intOrder
        \end{pmatrix}
        =
        \underbrace{
        \begin{pmatrix}
         \hat{\X}_1
         \\
         \vdots
         \\
         \hat{\X}_\intOrder
        \end{pmatrix}
        }_{=: \hat{\X}}
        \cdot \hat{\A}
    \end{equation}
    and hence $\rkqm(\S) = \rkqm(\hat{\X})$.
    Note that $\hat{\X}_j = \opMooreInv{n-k-\numbErrors}{\opfullexpinv{\autInvXiVec}{\x_j}{\numbErrors}}{\autInvXiVec}$ applies for all $j = 1, \dots, \intOrder$.
    This yields $\hat{\X} = \opMooreInv{n-k-\numbErrors}{\opfullexpinv{\autInvXiVec}{\X}{\numbErrors}}{\autInvXiVec}$, where the rows of the matrix $\X$ are precisely $\x_1, \dots, \x_{\intOrder}$.
    In fact, $\X$ contains all error locators and can be computed as $\X = \diag(\h, \dots, \h) \cdot \B^{\top}$ with $\h$ being the first row of the parity-check matrix $\H$ of the component \ac{LRS} code given in~\eqref{eq:parity_check_mat} and $\B$ arising from the error decomposition~\eqref{eq:err_int_vec_decomp}.
    Recall that $\h \in \Fqm^{n}$ only depends on the considered code and not on the particular decoding instance.
    We can thus consider $\h$ as a fixed given part and since $\e$ is chosen uniformly at random from $\vecGivenSRWeight{q^m}{\intOrder \n}{\numbErrors}$, the matrix $\X$ is distributed uniformly over the set $\matGivenSRWeight{q^m}{\intOrder}{\numbErrorsVec}{\numbErrors}$.
    The proof of~\cite[Lem.~7]{bartz2022fast} then yields the desired bound
    \begin{align*}
        \Pr\left\{\rkqm(\S) < \numbErrors\right\}
            &=
            \Pr\bigl\{\rkqm(\opMooreInv{n-k-\numbErrors}{\opfullexpinv{\autInvXiVec}{\X}{\numbErrors}}{\autInvXiVec})
            < \numbErrors
            \bigr\}
            \overset{\text{\cite{bartz2022fast}}}{\leq}
            \kappa_q^{\shots+1}q^{-m((\intOrder+1)(\numbErrorsMax-\numbErrors)+1)}.
    \end{align*}
    
    Let us focus on errors with sum-rank weight at most $\tfrac{1}{2}(n-k)$ and use the equality
    $\rkqm(\S) = \rkqm(\opMooreInv{n-k-\numbErrors}{\opfullexpinv{\autInvXiVec}{\X}{\numbErrors}}{\autInvXiVec})$ from above to see that $\S$ has always full $\Fqm$-rank $\numbErrors$ in this case.
    Since $\opfullexpinv{\autInvXiVec}{\X}{\numbErrors} \in \Fqm^{\intOrder \times \numbErrors}$ has sum-rank weight $\numbErrors$,~\cite[Lem.~6]{bartz2022fast} states that $\opMooreInv{n-k-\numbErrors}{\opfullexpinv{\autInvXiVec}{\X}{\numbErrors}}{\autInvXiVec}$ cannot be rank-deficient.
    This follows because a vector $\b \in \Fqm^{\numbErrors}$ with $\SumRankWeight(\b) > n-k-\numbErrors \geq \tfrac{1}{2}(n-k) \geq \numbErrors$ cannot exist.
\end{IEEEproof}

We now describe step by step how the syndrome-based decoder for \ac{HILRS} codes proceeds.
First, the component-syndrome polynomials $s_1, \dots, s_{\intOrder}$ and the key equation~\eqref{eq:hilrs_key_equation_esp} are set up.
The in~\autoref{sec:shift_register_synthesis} described multisequence skew-feedback shift-register synthesis~\cite{Sidorenko2011SkewFeedback} is used to recover the \ac{ESP} from~\eqref{eq:hilrs_key_equation_esp} and to decide whether the solution space of the key equation has dimension one or more.
In case it has dimension at least two, a decoding failure is returned.
Otherwise, the \ac{ESP} $\ESP$ was successfully recovered and the Skachek--Roth-like algorithm from~\autoref{sec:skachek_roth} allows to retrieve the error values $\a \in \Fqm^{\numbErrors}$ from $\ESP$ by finding bases $\shot{\a}{1}, \dots, \shot{\a}{\shots}$ of the root spaces of $\ESP$ with respect to generalized operator evaluation and evaluation parameter $\autInvXiInvWIndex{i}$ for each $i = 1, \dots, \shots$.

Next, the error locators $\x_1, \dots, \x_{\intOrder} \in \Fqm^{\numbErrors}$ are recovered by making use of the equivalent formulation
\begin{equation}\label{eq:hilrs_gab-like}
    \opMooreInv{n-k}{\x_j}{\autInvXiVec} \cdot \a^\top=\s_j^\top \quad \text{for all } j=1,\dots,\intOrder
\end{equation}
of~\eqref{eq:hilrs_syndrome_entries}.
Since the above systems have a particular form, we can solve them with the generalized version of Gabidulin's algorithm from~\autoref{sec:efficient_gabidulin-like}.
We recover the matrices $\B_1, \dots, \B_{\intOrder} \in \Fq^{\numbErrors \times n}$ containing the error locations by applying the techniques from~\cite{silva2009error} blockwise.
More precisely, we consider the matrix $\H_q \defeq \coeffq{\h} \in \Fq^{m \times n}$ with $\h \in \Fqm^{n}$ being the first row of the parity-check matrix $\H = \opMooreInv{n-k}{\h}{\autInvXiVec}$ of the component code given in~\eqref{eq:parity_check_mat}.
We compute for each block $\H_q^{(i)} \in \Fq^{m \times n_i}$ with $i = 1, \dots, \shots$ a left inverse $\widetilde{\H}_q^{(i)} \in \Fq^{n_i \times m}$ satisfying $\widetilde{\H}_q^{(i)} \cdot \H_q^{(i)} = \I_{n_i}$ for the identity matrix $\I_{n_i} \in \Fq^{n_i \times n_i}$.
Then, we can recover the $i$-th block of $\B_{j}$ as
\begin{equation}\label{eq:left-inverses-hilrs}
    \B_{j}^{(i)\top} = \widetilde{\H}_q^{(i)}\X_{j,q}^{(i)} = \widetilde{\H}_q^{(i)}\H_q^{(i)}\B_{j}^{(i)\top} \quad \text{for all } i=1,\dots, \shots \quad \text{and all } j = 1, \dots, \intOrder.
\end{equation}
Here, $\X_{j,q} \defeq \coeffq{\x_j} \in \Fq^{m \times \numbErrors}$ denotes the $\Fq$-representation of the vector $\x_j \in \Fqm^{\numbErrors}$ for $j = 1, \dots, \intOrder$.
We set $\B_j = \diag(\B_j^{(1)},\dots,\B_j^{(\shots)}) \in \Fq^{\numbErrors \times n}$ for each $j = 1, \dots, \intOrder$ and finally obtain $\B$ as $\B = (\B_1 \mid \dots \mid \B_{\intOrder}) \in \Fq^{\numbErrors \times \intOrder n}$.
Ultimately, we can compute the error $\e$ as $\e = \a \cdot \B$ according to~\eqref{eq:err_int_vec_decomp} and return the correct codeword $\c = \y - \e$.

It is worth noting that the left inverses $\widetilde{\H}_q^{(1)}, \dots, \widetilde{\H}_q^{(\shots)}$ can be precomputed as described in~\autoref{rem:left-inv}.
\autoref{alg:dec_hilrs} and~\autoref{thm:dec_hilrs} summarize the syndrome-based decoder for~\ac{HILRS} codes in the error-only setting.

\begin{algorithm}[ht]
  \caption{\algoname{Error-Only Decoding of \ac{HILRS} Codes}}\label{alg:dec_hilrs}
  \SetKwInOut{Input}{Input}\SetKwInOut{Output}{Output}

  \Input{A channel output $\y=\c+\e \in \Fqm^{\intOrder n}$ with $\c \in \horIntLinRS{\vecbeta}{\vecxi}{\intOrder}{\vecntilde}{k}$ and $\SumRankWeight(\e)=\numbErrors \leq \numbErrorsMax$,
  \\
  a parity-check matrix $\H \in \Fqm^{(n-k) \times n}$ of the form $\opMooreInv{n-k}{\h}{\autInvXiVec}$ of $\linRS{\vecbeta}{\vecxi}{\nVec}{k}$,
  \\
  and a left inverse $\widetilde{\H}_q^{(i)} \in \Fq^{n_i \times m}$ of $\coeffq{\h^{(i)}}$ for each $i = 1, \dots, \shots$.}

  \Output{The transmitted codeword $\c \in \horIntLinRS{\vecbeta}{\vecxi}{\intOrder}{\vecntilde}{k}$ or \emph{``decoding failure''}.}

  \BlankLine

  \For{$j=1,\dots,\intOrder$}
  {
    Compute the component syndrome $\s_j \gets \y_j\H^\top$ with $\y_j$ being the $j$-th component of $\y$.
    \\
    Set up the component-syndrome polynomial $s_j\in\SkewPolyringZeroDerInv$.
  }
  Solve the key equation~\eqref{eq:hilrs_key_equation_esp} to obtain the \ac{ESP} $\ESP\in\SkewPolyringZeroDerInv$.
  \\
  \If{the key equation~\eqref{eq:hilrs_key_equation_esp} has a unique solution up to $\Fqm$-multiples}
  {
      Find a basis $\shot{\a}{i}$ of the root space of $\opev{\ESP}{\cdot}{\autInvXiInvWIndex{i}}$ for each $i = 1, \dots, \shots$ and set $\a \gets (\shot{\a}{1} \mid \dots \mid \shot{\a}{\shots})$.
        \label{alg:hilrs_find_err_val}
      \\
      Set up $\opMooreInv{n-k}{\x_j}{\autInvXiVec} \cdot \a^\top=\s_j^\top$ from~\eqref{eq:hilrs_gab-like} for each $j = 1, \dots, \intOrder$ and solve it for $\x_j$.
      \\
      Recover the error locations $\B_{j}^{(i)\top} = \widetilde{\H}_q^{(i)} \cdot \coeffq{\subShot{\x}{j}{i}}$ for all $i=1,\dots,\shots$ and all $j = 1, \dots, \intOrder$ as in~\eqref{eq:left-inverses-hilrs}.
      \\
      Set up $\B \gets (\B_1 \mid \dots \mid \B_{\intOrder})$ with $\B_j \gets \diag(\subShot{\B}{j}{1}, \dots, \subShot{\B}{j}{\shots})$ for all $j = 1, \dots, \intOrder$.
      \\
      Recover the error vector $\e \gets \a\B$ as in~\eqref{eq:err_int_vec_decomp}. \label{alg:hilrs_recover_error}
      \\
      \Return{$\c \gets \y-\e$.}
    }

    \Return{``decoding failure''.}
\end{algorithm}

\begin{theorem}[Error-Only Decoding of \ac{HILRS} Codes]
    \label{thm:dec_hilrs}
    Consider the transmission of a codeword $\c \in \horIntLinRS{\vecbeta}{\vecxi}{\intOrder}{\vecntilde}{k}$ over the additive error-only channel~\eqref{eq:sum_rank_channel_hor_int}.
    The error $\e \in \Fqm^{\intOrder n}$ of sum-rank weight $\SumRankWeight(\e) = \numbErrors$ is chosen uniformly at random from the set $\vecGivenSRWeight{q^m}{\intOrder \n}{\numbErrors}$ defined in~\eqref{eq:vectors-fixed-horizontal-weight} and determines the received word $\y = \c + \e \in \Fqm^{\intOrder n}$.
    The presented syndrome-based decoder can always recover $\c$ from $\y$ if $\numbErrors \leq \tfrac{1}{2}(n - k)$ holds.
    Moreover, the decoder can be used probabilistically for larger error weights and decoding succeeds with a probability of at least
    \begin{equation}
        \label{eq:lower_bound_Psucc_hilrs}
        1 - \gammaq^{\ell} q^{-m((s+1)(\numbErrorsMax-\numbErrors)+1)}
    \end{equation}
    as long as the error weight satisfies
    \begin{equation}
        \numbErrors \leq \numbErrorsMax\defeq\tfrac{\intOrder}{\intOrder+1}(n-k).
    \end{equation}
    The decoder requires on average $O(\intOrder n^2)$ operations in $\Fqm$ if $m \in O(\intOrder)$ applies.
\end{theorem}

\begin{IEEEproof}
    The reasoning above and the proof of the key equation~\eqref{eq:hilrs_key_equation_esp} ensure the correctness of~\autoref{alg:dec_hilrs}, as a decoding failure is returned if and only if the key equation has a solution space of dimension greater than one and this is the only potential point of failure.
    
    The maximum decoding radius was established in~\eqref{eq:hilrs_dec-radius} and~\autoref{lem:failure_prob_hilrs} showed that successful decoding is guaranteed as long as the error weight is at most $\tfrac{1}{2}(n-k)$.
    For larger error weights that are still within the decoding radius~\eqref{eq:hilrs_dec-radius}, the claimed upper bound on the failure probability was proved in~\autoref{lem:failure_prob_hilrs} as well.
    
    We now focus on the asymptotic complexity of~\autoref{alg:dec_hilrs} and make use of the fast subroutines outlined in~\autoref{sec:fast-dec} for computational gains.
    The computation of the component syndromes $\s_1, \dots, \s_{\intOrder}$ in line 2 and the setup of the corresponding component-syndrome polynomials in line 3 need at most $O(\intOrder n^2)$ operations in $\Fqm$.
    Next, multisequence skew-feedback shift-register synthesis can be applied to lines 4 and 5 to solve the key equation~\eqref{eq:hilrs_key_equation_esp} and test the uniqueness of its solution up to $\Fqm$-multiples in at most $O(\intOrder (n - k)^2)$ operations in $\Fqm$.
    Line 6 derives the error values $\a \in \Fqm^{\numbErrors}$ from the \ac{ESP} $\ESP$ and the Skachek--Roth-like algorithm can achieve this on average in $O(\shots m \deg(\ESP)) = O(\shots m (n - k))$ operations in $\Fqm$.
    The average complexity of this step is in $O(\intOrder n^2)$ for $m \in O(\intOrder)$.
    The system~\eqref{eq:hilrs_gab-like} in line 7 is solved with the Gabidulin-like algorithm in at most $O(\intOrder n^2)$ operations in $\Fqm$.
    Line 8 makes use of the precomputed left inverses $\widetilde{\H}_q^{(i)}$ for $i = 1, \dots, \shots$ which allows to solve \eqref{eq:left-inverses-hilrs} in at most $O(\intOrder n^2)$ operations in $\Fqm$.
    The remaining lines consist of basic operations that require at most $O(n^2)$ operations in $\Fqm$.
    In the end, we obtain an average complexity in $O(\intOrder n^2)$ if $m \in O(\intOrder)$ applies.
    It is worth noting that we accounted for the worst-case complexity in most steps and only the Skachek--Roth-like algorithm is probabilistic and was thus assessed in terms of average complexity.
\end{IEEEproof}

\subsection{Error-Erasure Decoding}

In this section, we generalize the presented syndrome-based decoder to work with the error-erasure channel model.
Hence, we consider an error vector $\e \in \Fqm^{\intOrder n}$ which can be additively decomposed into three parts containing $\numbFullErrors$ full errors, $\numbRowErasures$ row erasures, and $\numbColErasures$ column erasures, respectively, and has sum-rank weight $\numbErrors = \numbFullErrors + \numbRowErasures + \numbColErasures$.
With a parity-check matrix $\H \in \Fqm^{(n-k) \times n}$ of the component code $\linRS{\vecbeta}{\vecxi}{\nVec}{k}$ and the error decomposition shown in~\eqref{eq:hilrs_error-erasure_decomposition}, we can compute the \emph{component syndromes}
\begin{equation}
    \s_j
    \defeq \y_j \H^{\top}
    = (\e_{\indFullErrors, j} + \e_{\indRowErasures, j} + \e_{\indColErasures, j}) \H^{\top}
    \overset{\eqref{eq:hilrs_error-erasure_decomposition}}{=} \a_{\indFullErrors} \B_{\indFullErrors, j} \H^{\top} + \a_{\indRowErasures} \B_{\indRowErasures, j} \H^{\top} + \a_{\indColErasures} \B_{\indColErasures, j} \H^{\top}
    \quad \text{for all } j = 1, \dots, \intOrder,
\end{equation}
where each matrix $\B_{\indErrorType, j}$ with $\indErrorType \in \{\indFullErrors, \indRowErasures, \indColErasures\}$ and $j = 1, \dots, \intOrder$ has a block-diagonal structure as shown in~\eqref{eq:err_int_vec_decomp}.
Since there is a suitable vector $\h \in \Fqm^{n}$ of sum-rank weight $n$ for which the generalized Moore matrix $\opMooreInv{n-k}{\h}{\autInvXiVec} \in \Fqm^{(n-k) \times n}$ is a parity-check matrix of $\linRS{\vecbeta}{\vecxi}{\nVec}{k}$ according to~\eqref{eq:parity_check_mat}, we can express the $l$-th entry of the component syndrome $\s_j$ as
\begin{align}\label{eq:hilrs_syndrome}
    s_{j, l} = \sum_{\indErrorType \in \{\indFullErrors, \indRowErasures, \indColErasures\}} \a_{\indErrorType} \opfullexpinv{\autInvXiVec}{\underbrace{\h \B_{\indErrorType, j}^{\top}}_{=: \x_{\indErrorType, j}}}{l-1}^{\top}
    = \sum_{\indErrorType \in \{\indFullErrors, \indRowErasures, \indColErasures\}} \a_{\indErrorType} \opfullexpinv{\autInvXiVec}{\x_{\indErrorType, j}}{l-1}^{\top}
\end{align}
for all $l=1,\dots n-k$ and every $j = 1, \dots, \intOrder$.
This motivates to define the \emph{error locators} per component and error type, i.e., as $\x_{\indErrorType, j} \defeq \h \B_{\indErrorType, j}^{\top} \in \Fqm^{\numbErrorType}$ for each $\indErrorType \in \{\indFullErrors, \indRowErasures, \indColErasures\}$ and all $j = 1, \dots, \intOrder$.

Since the receiver already knows $\x_{\indColErasures, 1}, \dots, \x_{\indColErasures, \intOrder}$, we encode this knowledge in the \emph{partial component \acfp{ELP}} which we define as the minimal skew polynomials satisfying
\begin{equation}
    \opev{\ELPcolWIndex{j}}{\x_{\indColErasures,j}}{\autInvXiVec} = \0
    \quad \text{for all } j = 1, \dots, \intOrder.
\end{equation}
As the evaluation parameters in $\autInvXiInvVec$ belong to pairwise distinct nontrivial conjugacy classes of $\Fqm$, the skew polynomial $\ELPcolWIndex{j}$ with $j = 1, \dots, \intOrder$ has degree $\numbColErasuresWIndex{j} \defeq \SumRankWeightWPartition{\numbColErasuresVec}(\x_{\indColErasures, j}) = \SumRankWeightWPartition{\n}(\e_{\indColErasures, j})$.
Observe that $\numbColErasuresWIndex{j}$ is upper-bounded by $\numbColErasures$ for each $j = 1, \dots, \intOrder$ since $\numbColErasuresWIndex{j} = \SumRankWeightWPartition{\n}(\B_{\indColErasures, j})$ applies and the error decomposition enforces $\SumRankWeightWPartition{\intOrder \n}(\B_{\indColErasures}) = \sum_{i=1}^{\shots} \rkq\bigl((\B_{\indColErasures, 1}^{(i)} \mid \dots \mid \B_{\indColErasures, \intOrder}^{(i)})\bigr) = \numbColErasures$.
We also get the bound $\sum_{j=1}^{\intOrder} \numbColErasuresWIndex{j} \geq \numbColErasures$.

Further, we keep using the \emph{\acf{ESP}} $\ESP \in \SkewPolyringZeroDerInv$ which is the minimal skew polynomial vanishing on all error values, that is, $\opev{\ESP}{\a_{\indErrorType}}{\autInvXiInvVec} = \0$ holds for all $\indErrorType \in \{ \indFullErrors, \indRowErasures, \indColErasures \}$.
We express $\ESP$ as a product of partial \acp{ESP} related to the different error types to make the knowledge about the row erasures more accessible to the decoder.
Namely, we write
\begin{align}\label{eq:hilrs_overall_esp}
    \ESP(x) &= \ESPcol(x) \cdot \ESPfull(x) \cdot \ESProw(x),
\end{align}
where the \emph{partial \acfp{ESP}} $\ESPfull, \ESProw, \ESPcol \in \SkewPolyringZeroDerInv$ are defined as the minimal skew polynomials satisfying
\begin{align}\label{eq:def_part_esp_w_erasures}
    \opev{(\ESPcol \cdot \ESPfull \cdot \ESProw)}{\a_{\indColErasures}}{\autInvXiInvVec} = \0,
    \quad
    \opev{(\ESPfull \cdot \ESProw)}{\a_{\indFullErrors}}{\autInvXiInvVec} = \0,
    \quad
    \text{and} \quad \opev{\ESProw}{\a_{\indRowErasures}}{\autInvXiInvVec} = \0,
\end{align}
respectively.
Recall again that since $\a_\indRowErasures$ and $\B_{\indColErasures}$ are known, we can compute $\ESProw$ and $\ELPcolWIndex{1},\dots,\ELPcolWIndex{\intOrder}$ efficiently using~\eqref{eq:min_poly}.
We use them to compute the \emph{auxiliary component-syndrome polynomials} which we define as
\begin{equation}
    \ESPcomponentSyndrome{j}(x) \defeq \ESProw(x) \cdot s_j(x) \cdot \ELPcolRevWIndex{j}(x) \in \SkewPolyringZeroDerInv \quad \text{for all } j = 1, \dots, \intOrder.
\end{equation}
Here, $\ELPcolRevWIndex{j}$ denotes the $\autinv$-reverse of $\ELPcolWIndex{j}$ with respect to $\numbColErasuresWIndex{j}$ for each $j = 1, \dots, \intOrder$ and is explicitly given as
\begin{equation}\label{eq:def-ELPcol-reverse}
    \ELPcolRevWIndex{j}(x) = \sum_{\nu = 1}^{\numbColErasuresWIndex{j}+1} \ELPcolRevWIndex{j, \nu} x^{\nu - 1} \quad \text{with } \ELPcolRevWIndex{j, \nu} = \aut^{-(\nu - \numbColErasuresWIndex{j} - 1)}(\ELPcolRevWIndex{j, \numbColErasuresWIndex{j} - \nu + 2})
    \quad \text{for all } \nu = 1, \dots, \numbColErasuresWIndex{j}+1.
\end{equation}
We are now ready to derive a key equation that relates the auxiliary component-syndrome polynomials with the partial \ac{ESP} corresponding to the full errors.

\begin{theorem}[ESP Key Equation for Errors and Erasures]\label{thm:hilrs_error-erasure_key_equation}
    For each $j = 1, \dots, \intOrder$, there is a skew polynomial $\errEvalPoly_j \in \SkewPolyringZeroDerInv$ with $\deg(\errEvalPoly_j) < \numbFullErrors + \numbRowErasures + \numbColErasuresWIndex{j}$ that satisfies
    \begin{equation}\label{eq:hilrs_error-erasure_key_equation}
        \ESPfull(x) \cdot \ESPcomponentSyndrome{j}(x) \equiv \errEvalPoly_j(x) \modr x^{n-k}.
    \end{equation}
\end{theorem}

\begin{IEEEproof}
    Instead of showing~\eqref{eq:hilrs_error-erasure_key_equation} directly, we prove the equivalent statement that $\errEvalPoly_{j,l}=0$ holds for all $l = \numbFullErrors + \numbRowErasures + \numbColErasuresWIndex{j}+1, \dots, n-k$ and all $j = 1, \dots, \intOrder$.
    Therefore, we compute the coefficients of $\errEvalPoly_{j}$ for each $j = 1, \dots, \intOrder$ via the equality
    \begin{equation}
        \errEvalPoly_j(x) = \underbrace{\ESPfull(x) \cdot \ESProw(x)}_{=: \ESPfullrow(x)} \cdot s_j(x) \cdot \ELPcolRevWIndex{j}(x)
    \end{equation}
    by applying~\eqref{eq:skew_product_coeffs} first to $\ESPfullrow(x) \cdot s_j(x)$ and then to the product of the obtained result and $\ELPcolRevWIndex{j}(x)$.
    The first part yields
    \begin{align}
        (\ESPfullrow \cdot s_j)_{l}
        &\overset{\eqref{eq:skew_product_coeffs}}{=} \sum_{\nu=1}^{\numbFullErrors+\numbRowErasures+1} \ESPfullrowWIndex{\nu} \aut^{-(\nu-1)}(\componentSyndromeWIndex{j}{l-\nu+1})
        \overset{\eqref{eq:hilrs_syndrome}}{=} \sum_{\nu=1}^{\numbFullErrors+\numbRowErasures+1} \ESPfullrowWIndex{\nu} \sum_{\indErrorType \in \{\indFullErrors, \indRowErasures, \indColErasures\}} \aut^{-(\nu-1)} \bigl( \a_{\indErrorType} \opfullexpinv{\autInvXiVec}{\x_{\indErrorType, j}}{l-\nu}^{\top} \bigr)
        \\
        &= \sum_{\indErrorType \in \{\indFullErrors, \indRowErasures, \indColErasures\}} \sum_{i=1}^{\shots} \sum_{r=1}^{\numbErrorTypeInBlock{i}} \aut^{-(l-1)}(x_{\indErrorType, j,r}^{(i)})
        \sum_{\nu=1}^{\numbFullErrors+\numbRowErasures+1} \ESPfullrowWIndex{\nu} \aut^{-(\nu-1)}(a_{\indErrorType, r}^{(i)}) \underbrace{\aut^{-(\nu-1)}\bigl(\genNormInv{l-\nu}{\autInvXiWIndex{i}}\bigr)}_{\overset{\eqref{eq:normlemma-2}}{=} \genNormInv{l-1}{\autInvXiWIndex{i}} \cdot \genNormInv{\nu-1}{\autInvXiInvWIndex{i}}}
        \\
        &= \sum_{\indErrorType \in \{\indFullErrors, \indRowErasures, \indColErasures\}} \sum_{i=1}^{\shots} \sum_{r=1}^{\numbErrorTypeInBlock{i}} \aut^{-(l-1)}(x_{\indErrorType, j,r}^{(i)}) \genNormInv{l-1}{\autInvXiWIndex{i}}
        \underbrace{\sum_{\nu=1}^{\numbFullErrors+\numbRowErasures+1} \ESPfullrowWIndex{\nu} \aut^{-(\nu-1)}(a_{\indErrorType, r}^{(i)})  \genNormInv{\nu-1}{\autInvXiInvWIndex{i}}}_{= \opev{\ESPfullrow}{a_{\indErrorType, r}^{(i)}}{\autInvXiInvWIndex{i}}}
        \\
        &= \opfullexpinv{\autInvXiVec}{\x_{\indColErasures, j}}{l-1} \cdot \underbrace{\opev{\ESPfullrow}{\a_{\indColErasures}}{\autInvXiInvVec}^{\top}}_{=: \hat{\a}_{\indColErasures}^{\top}}
        \label{eq:ESP_proof_other_LRS_decoding_problem}
    \end{align}
    for all $l = \numbFullErrors + \numbRowErasures + 1, \dots, n-k$ and all $j = 1, \dots, \intOrder$.
    Here, the last equality follows from the fact that $\ESPfullrow$ vanishes at the error values corresponding to full errors and row erasures, i.e., $\opev{\ESPfullrow}{\a_{\indFullErrors}}{\autInvXiInvVec} = \0$ and $\opev{\ESPfullrow}{\a_{\indRowErasures}}{\autInvXiInvVec} = \0$ apply according to~\eqref{eq:def_part_esp_w_erasures}.
    The second step yields
    \begin{align*}
        \errEvalPoly_{j,l}
        &\overset{\eqref{eq:skew_product_coeffs}}{=} \sum_{\nu=1}^{\numbColErasuresWIndex{j}+1} (\ESPfullrow \cdot \componentSyndrome{j})_{l-\nu+1} \aut^{-(l-\nu)}(\ELPcolRevWIndex{j,\nu})
        \overset{\eqref{eq:def-ELPcol-reverse}}{\underset{\eqref{eq:ESP_proof_other_LRS_decoding_problem}}{=}} \sum_{\nu=1}^{\numbColErasuresWIndex{j}+1} \opfullexpinv{\autInvXiVec}{\x_{\indColErasures,j}}{l-\nu} \cdot \hat{\a}_{\indColErasures}^{\top} \cdot
        \aut^{-(l-\numbColErasuresWIndex{j}-1)}(\ELPcolWIndex{j,\numbColErasuresWIndex{j}-\nu+2})
        \\
        &= \sum_{i=1}^{\shots} \sum_{r=1}^{\numbColErasuresInBlock{i}} \hat{a}_{\indColErasures,r}^{(i)} \aut^{-(l-\numbColErasuresWIndex{j}-1)} \Biggl(
        \sum_{\nu=1}^{\numbColErasuresWIndex{j}+1} \aut^{-(\numbColErasuresWIndex{j} - \nu + 1)}(x_{\indColErasures,j,r}^{(i)}) \underbrace{\aut^{l-\numbColErasuresWIndex{j}-1}\bigl(\genNormInv{l-\nu}{\autInvXiWIndex{i}}\bigr)}_{\overset{\eqref{eq:normlemma-1}}{=} \genNormInv{\numbColErasuresWIndex{j} - \nu + 1}{\autInvXiWIndex{i}} \cdot \genNorm{l - \numbColErasuresWIndex{j} - 1}{\xi_i}} \ELPcolWIndex{j,\numbColErasuresWIndex{j}-\nu+2} \Biggr)
        \\
        &= \sum_{i=1}^{\shots} \sum_{r=1}^{\numbColErasuresInBlock{i}} \hat{a}_{\indColErasures,r}^{(i)}
        \aut^{-(l-\numbColErasuresWIndex{j}-1)} \biggl( \genNorm{l-\numbColErasuresWIndex{j}-1}{\xi_i} \underbrace{\sum_{\nu=1}^{\numbColErasuresWIndex{j}+1} \ELPcolWIndex{j,\numbColErasuresWIndex{j}-\nu+2} \aut^{-(\numbColErasuresWIndex{j} - \nu + 1)}(x_{\indColErasures,j,r}^{(i)}) \genNormInv{\numbColErasuresWIndex{j} - \nu + 1}{\autInvXiWIndex{i}}}_{= \opev{\ELPcol}{x_{\indColErasures,j,r}^{(i)}}{\autInvXiWIndex{i}} = 0} \biggr)
        = 0
    \end{align*}
    for all $l = \numbFullErrors + \numbRowErasures + \numbColErasuresWIndex{j} + 1, \dots, n- k$ and every $j = 1, \dots, \intOrder$.
\end{IEEEproof}

We can formulate the key equation~\eqref{eq:hilrs_error-erasure_key_equation} equivalently as
\begin{equation}\label{eq:hilrs_error-erasure_key_equation_esp_v1_norm}
    \sum_{\nu=2}^{\numbFullErrors+1}\ESPfullWIndex{\nu}\aut^{-(\nu-1)}\left(\ESPcomponentSyndrome{j,l-\nu+1}\right)=-\ESPcomponentSyndrome{j,l}
    \quad \text{for all } l=\numbFullErrors + \numbRowErasures + \numbColErasuresWIndex{j}+1,\dots,n-k
    \quad \text{and all } j=1,\dots,\intOrder
\end{equation}
when we normalize $\ESPfullWIndex{1}=1$ without loss of generality.
Hence, the key equation~\eqref{eq:hilrs_error-erasure_key_equation} corresponds to an inhomogeneous $\Fqm$-linear system of $\intOrder(n-k-\numbFullErrors-\numbRowErasures-\frac{1}{\intOrder} \sum_{j=1}^{\intOrder}\numbColErasuresWIndex{j})$ equations in $\numbFullErrors$ unknowns.
This system can have a unique solution up to $\Fqm$-multiples only if the number of equations is at least the number of unknowns, i.e., if
\begin{equation}\label{eq:hilrs_error_erasure_dec_radius}
 \numbFullErrors \leq \frac{\intOrder}{\intOrder+1}\biggl(n-k-\numbRowErasures-
    \underbrace{\frac{1}{\intOrder} \sum_{j=1}^{\intOrder}\numbColErasuresWIndex{j}}_{=: \numbColErasuresBar}
\biggr).
\end{equation}
However, this condition is not sufficient to ensure a one-dimensional solution space and decoding failures can occur.
Remark that the maximal decoding region defined by~\eqref{eq:hilrs_error_erasure_dec_radius} depends on $\numbColErasuresBar$, that is, on the \emph{average} number of column erasures per component error.

Overall, we can summarize the steps of the syndrome-based error-erasure decoder as follows:
The partial \ac{ESP} $\ESProw$ as well as the the partial component \acp{ELP} $\ELPcolWIndex{1}, \dots, \ELPcolWIndex{\intOrder}$ and the auxiliary component-syndrome polynomials $\ESPcomponentSyndrome{1}, \dots, \ESPcomponentSyndrome{\intOrder}$ are determined to set up the key equation~\eqref{eq:hilrs_error-erasure_key_equation}.
We can solve~\eqref{eq:hilrs_error-erasure_key_equation} by means of multisequence skew-feedback shift-register synthesis~\cite{Sidorenko2011SkewFeedback} which is briefly explained in~\autoref{sec:shift_register_synthesis}.
In case the solution space of the key equation has dimension at least two, the decoder returns a decoding failure.
Otherwise, the obtained partial \ac{ESP} $\ESPfull$ corresponds to the actual error and can be used to set up the skew polynomials $\ESPfull(x) \cdot \ESProw(x) \cdot s_j(x) = \ESPfullrow(x) \cdot s_j(x)$ for all $j=1,\dots,\intOrder$.
We then make use of equality~\eqref{eq:ESP_proof_other_LRS_decoding_problem} from the proof of~\autoref{thm:hilrs_error-erasure_key_equation} and apply $\aut^{l - 1}$ to it to obtain
\begin{align}
    \aut^{l - 1}\bigl((\ESPfullrow \cdot s_j)_{l}\bigr) &\overset{\eqref{eq:ESP_proof_other_LRS_decoding_problem}}{=} \aut^{l - 1}\bigl( \opfullexpinv{\autInvXiVec}{\x_{\indColErasures, j}}{l-1} \cdot \hat{\a}_{\indColErasures}^{\top} \bigr)
    = \sum_{i = 1}^{\shots} \sum_{r = 1}^{\numbFullErrorsInBlock{i}} \subShot{x}{\indColErasures, j, r}{i}
    \underbrace{\aut^{l - 1}\bigl(\genNormInv{l - 1}{\autInvXiWIndex{i}}\bigr)}_{\overset{\eqref{eq:normlemma-1}}{=} \genNorm{l - 1}{\xi_{i}}}
    \aut^{l - 1}(\subShot{\hat{a}}{\indColErasures, r}{i})
    = \opexp{\vecxi}{\hat{\a}_{\indColErasures}}{l - 1} \cdot \x_{\indColErasures, j}^{\top}
\end{align}
for all $l = \numbFullErrors + \numbRowErasures + 1, \dots, n - k$.
Now we can set up the system
\begin{equation}\label{eq:hilrs-gabidulin-like-ahat}
    \opMoore{n - k - \numbFullErrors - \numbRowErasures}{\opfullexp{\vecxi}{\hat{\a}_{\indColErasures}}{\numbFullErrors + \numbRowErasures}}{\vecxi} \cdot \x_{\indColErasures, j}^{\top} = \v_{j}^{\top}
\end{equation}
with $\v_{j} \defeq \bigl( (\ESPfullrow \cdot s_j)_{\numbFullErrors + \numbRowErasures + 1}, \dots, (\ESPfullrow \cdot s_j)_{n - k} \bigr) \in \Fqm^{n - k - \numbFullErrors - \numbRowErasures}$ for each $j = 1, \dots, \intOrder$.
Then we combine the $\intOrder$ systems from~\eqref{eq:hilrs-gabidulin-like-ahat} into one $\Fqms$-linear system.
Therefore, fix an $\Fqm$-basis $\vecgamma = (\gamma_1, \dots, \gamma_{\intOrder}) \in \Fqms^{\intOrder}$ of $\Fqms$ and set
\begin{equation}
    \x_{\indColErasures} \defeq \vecgamma \cdot
    \begin{pmatrix}
        \x_{\indColErasures, 1} \\
        \vdots \\
        \x_{\indColErasures, \intOrder}
    \end{pmatrix}
    \in \Fqms^{\numbColErasures}
    \quad \text{and} \quad
    \v \defeq \vecgamma \cdot
    \begin{pmatrix}
        \v_{1} \\
        \vdots \\
        \v_{\intOrder}
    \end{pmatrix}
    \in \Fqms^{n - k - \numbFullErrors - \numbRowErasures}.
\end{equation}
This yields the combined system
\begin{equation}\label{eq:hilrs-Fqms-Gabidulin}
    \opMoore{n - k - \numbFullErrors - \numbRowErasures}{\opfullexp{\vecxi}{\hat{\a}_{\indColErasures}}{\numbFullErrors + \numbRowErasures}}{\vecxi} \cdot \x_{\indColErasures}^{\top} = \v^{\top}.
\end{equation}
We can use the Gabidulin-like algorithm over $\Fqms$ to solve~\eqref{eq:hilrs-Fqms-Gabidulin} and refer to~\autoref{sec:efficient_gabidulin-like} for a detailed description of the method.
Further note that $\SumRankWeight(\x_{\indColErasures}) = \numbColErasures$ implies the uniqueness of the solution of~\eqref{eq:hilrs-Fqms-Gabidulin} and since the decoding problem makes sure that there \emph{is} a valid solution $\hat{\a}_{\indColErasures}$ over $\Fqm$, we will recover the correct one.

After we have found $\hat{\a}_{\indColErasures}$, we can reconstruct the partial \ac{ESP} $\ESPcol$ for the column erasures as the minimal skew polynomial $\mpolArgs{\hat{\a}_{\indColErasures}}{\autInvXiInvVec}$ of $\hat{\a}_{\indColErasures}$.
We can also finally set up the overall \ac{ESP} $\ESP(x) = \ESPcol(x) \cdot \ESPfull(x) \cdot \ESProw(x)$ as defined in~\eqref{eq:hilrs_overall_esp}.
The characterization of the partial \acp{ESP} in~\eqref{eq:def_part_esp_w_erasures} allows us to recover the missing error values $\a_{\indFullErrors}$ and $\a_{\indColErasures}$.
Namely, we first initialize the Skachek--Roth-like algorithm from~\autoref{sec:skachek_roth} with $\a_{\indRowErasures}$ and run it on $\ESPfull(x) \cdot \ESProw(x)$ to obtain $\a_{\indFullErrors}$.
Then we initialize the algorithm with $\a_{\indRowErasures}$ and $\a_{\indFullErrors}$ to recover $\a_{\indColErasures}$ from the full \ac{ESP} $\ESP$.
The missing code locators $\x_{\indFullErrors}$ and $\x_{\indRowErasures}$ can be recovered in a similar fashion as described in~\eqref{eq:hilrs_gab-like} for the error-only case.
Rephrasing~\eqref{eq:hilrs_syndrome} yields
\begin{equation}
    s_{j,l} - \opfullexpinv{\autInvXiVec}{\x_{\indColErasures, j}}{l-1} \cdot \a_{\indColErasures}^{\top}
    = \sum_{\indErrorType \in \{\indFullErrors, \indRowErasures\}} \opfullexpinv{\autInvXiVec}{\x_{\indErrorType, j}}{l-1} \cdot \a_{\indErrorType}^{\top}
    = \opfullexpinv{\autInvXiVec}{\x_{\indFullErrors, j} \mid \x_{\indRowErasures, j}}{l-1} \cdot (\a_{\indFullErrors} \mid \a_{\indRowErasures})^{\top}
    \quad \text{for all } l = 1, \dots, n - k
\end{equation}
and we obtain the Gabidulin-like systems
\begin{equation}\label{eq:gab-like_hilrs_error-erasure_locators}
    \opMooreInv{n-k}{\x_{\indFullErrors, j} \mid \x_{\indRowErasures, j}}{\autInvXiVec} \cdot (\a_{\indFullErrors} \mid \a_{\indRowErasures})^\top
    = \widetilde{\s}_j^\top
\end{equation}
with $\widetilde{\s}_j=\bigl(s_{j,1} - \x_{\indColErasures, j} \cdot \a_{\indColErasures}^{\top},\aut(s_{j,2}) - \opfullinv{\autInvXiVec}{\x_{\indColErasures, j}} \cdot \a_{\indColErasures}^{\top},\dots,\aut^{n-k-1}(s_{j,n-k}) - \opfullexpinv{\autInvXiVec}{\x_{\indColErasures, j}}{n-k-1} \cdot \a_{\indColErasures}^{\top}
\bigr)\in\Fqm^{n-k}$ for all $j = 1, \dots, \intOrder$.
They can be solved with the Gabidulin-like algorithm which is explained in~\autoref{sec:efficient_gabidulin-like}.

The error locations $\B_{\indFullErrors}$, $\B_{\indRowErasures}$, and $\B_{\indColErasures}$ can be obtained from the respective error locators $\x_{\indFullErrors}$, $\x_{\indRowErasures}$, and $\x_{\indColErasures}$ by applying the approach from the error-only setting given in~\eqref{eq:left-inverses-hilrs} to each error type.
The overall error $\e$ is then the sum of the vectors $\e_{\indErrorType} = \a_{\indErrorType} \cdot \B_{\indErrorType}$ for all $\indErrorType \in \{\indFullErrors, \indRowErasures, \indColErasures\}$ and the decoder returns $\c = \y - \e$.

\autoref{alg:error_erasure_dec_hilrs} summarizes the steps of the decoder compactly and~\autoref{thm:error_erasure_dec_hilrs} states the main attributes of the syndrome-based error-erasure decoder for~\ac{HILRS} codes.

\begin{algorithm}[ht]
  \caption{\algoname{Error-Erasure Decoding of \ac{HILRS} Codes}}\label{alg:error_erasure_dec_hilrs}
  \SetKwInOut{Input}{Input}\SetKwInOut{Output}{Output}
  
  \Input{A channel output $\y=\c+\e \in \Fqm^{\intOrder n}$ with $\c \in \horIntLinRS{\vecbeta}{\vecxi}{\intOrder}{\vecntilde}{k}$, $\e = \e_{\indFullErrors} + \e_{\indRowErasures} + \e_{\indColErasures}$, and $\SumRankWeight(\e)= \numbFullErrors + \numbRowErasures + \numbColErasures$ satisfying $\numbErrorsWeightedHorizontal \leq \numbErrorsMax$,
  \\
  a vector $\a_\indRowErasures \in \Fqm^{\numbRowErasures}$ of the form in~\eqref{eq:err_int_vec_decomp} such that $\subShot{\a}{\indRowErasures}{i}$ has the same column space as $\subShot{\e}{\indRowErasures}{i}$ for $i = 1, \dots, \shots$,
  \\
  and a matrix $\B_\indColErasures \in \Fqm^{\numbColErasures \times \intOrder n}$ of the form in~\eqref{eq:err_int_vec_decomp} such that $\subShot{\B}{\indColErasures}{i}$ has the same row space as $\subShot{\e}{\indColErasures}{i}$ for $i = 1, \dots, \shots$,
  \\
  a parity-check matrix $\H \in \Fqm^{(n-k) \times n}$ of the form $\opMooreInv{n-k}{\h}{\autInvXiVec}$ of $\linRS{\vecbeta}{\vecxi}{\nVec}{k}$,
  \\
  a left inverse $\widetilde{\H}_q^{(i)} \in \Fq^{n_i \times m}$ of $\coeffq{\h^{(i)}}$ for each $i = 1, \dots, \shots$.
  }
  
  \Output{The transmitted codeword $\c \in \horIntLinRS{\vecbeta}{\vecxi}{\intOrder}{\vecntilde}{k}$ or \emph{``decoding failure''}.}
  
  \BlankLine

  Compute the partial \ac{ESP} $\ESProw(x) \gets \mpolArgs{\a_{\indRowErasures}}{\autInvXiInvVec}(x)$.

  \For{$j=1,\dots, \intOrder$}
  {
    Compute the component syndrome $\s_j \gets \y_j\H^\top$ with $\y_j$ being the $j$-th component of $\y$.
    \\
    Set up the component-syndrome polynomial $s_j\in\SkewPolyringZeroDerInv$.
    \\
    Compute $\x_{\indColErasures,j} \gets \h\B_{\indColErasures,j}^{\top}$ and set $\x_{\indColErasures,j} \gets (\x_{\indColErasures,j}^{(1)} \mid \dots \mid \x_{\indColErasures,j}^{(\shots)})$.
    \\
    Compute the partial component \ac{ELP} $\ELPcolWIndex{j}(x) \gets \mpolArgs{\x_{\indColErasures,j}}{\autInvXiVec}(x)$.
    \\
    Set up $\ELPcolRevWIndex{j}$ according to~\eqref{eq:def-ELPcol-reverse}.
    \\
    Set up the auxiliary component-syndrome polynomial $\ESPcomponentSyndrome{j}(x) \gets \ESProw(x) \cdot s_j(x) \cdot \ELPcolRevWIndex{j}(x)$.
  }
  Solve the key equation~\eqref{eq:hilrs_error-erasure_key_equation} to obtain the partial \ac{ESP} $\ESPfull \in \SkewPolyringZeroDerInv$.
  \\
  \If{the key equation~\eqref{eq:hilrs_error-erasure_key_equation} has a unique solution up to $\Fqm$-multiples}
  {
      Set up $(\ESPfullrow \cdot s_j)(x) \gets \ESPfull(x) \cdot \ESProw(x) \cdot s_j(x)$ for all $j=1,\dots,\intOrder$.
      \\
      Set up $\opMoore{n - k - \numbFullErrors - \numbRowErasures}{\opfullexp{\vecxi}{\hat{\a}_{\indColErasures}}{\numbFullErrors + \numbRowErasures}}{\vecxi} \cdot \x_{\indColErasures}^{\top} = \v^{\top}$ from~\eqref{eq:hilrs-Fqms-Gabidulin} and solve it for $\hat{\a}_{\indColErasures}$.
      \\
      Compute $\ESPcol(x) \gets \mpolArgs{\hat{\a}_{\indColErasures}}{\autInvXiInvVec}(x)$.
      \\
      Set up $\ESPfullrow(x) \gets \ESPfull(x) \cdot \ESProw(x)$.
      \\
      Find $\subShot{\a}{\indFullErrors}{i}$ whose entries extend $\subShot{\a}{\indRowErasures}{i}$ to a basis of the root space of $\opev{\ESPfullrow}{\cdot}{\autInvXiInvWIndex{i}}$ for all $i = 1, \dots, \shots$.
      \\
      Set up $\ESP(x) \gets \ESPcol(x) \cdot \ESPfullrow(x)$.
      \\
      Find $\subShot{\a}{\indColErasures}{i}$ whose entries extend $(\subShot{\a}{\indRowErasures}{i} \mid \subShot{\a}{\indFullErrors}{i})$ to a basis of the root space of $\opev{\ESP}{\cdot}{\autInvXiInvWIndex{i}}$ for all $i = 1, \dots, \shots$.
      \\
      Set up $\opMooreInv{n-k}{\x_{\indFullErrors, j} \mid \x_{\indRowErasures, j}}{\autInvXiVec} \cdot (\a_{\indFullErrors} \mid \a_{\indRowErasures})^\top = \widetilde{\s}_j^\top$ from~\eqref{eq:gab-like_hilrs_error-erasure_locators} and solve it for $(\x_{\indFullErrors, j} \mid \x_{\indRowErasures, j})$ for all $j = 1, \dots, \intOrder$.
      \\
      Recover the error locations $\B_{\indErrorType, j}^{(i)\top} \gets \widetilde{\H}_q^{(i)} \cdot \coeffq{\subShot{\x}{\indErrorType, j}{i}}$ for $i=1,\dots,\shots$, $\indErrorType \in \{\indFullErrors, \indRowErasures\}$, and $j = 1, \dots, \intOrder$ as in~\eqref{eq:left-inverses-hilrs}.
      \\
      Set up $\B_{\indErrorType, j} \gets \diag(\subShot{\B}{\indErrorType, j}{1}, \dots, \subShot{\B}{\indErrorType, j}{\intOrder})$ for all $\indErrorType \in \{\indFullErrors, \indRowErasures\}$ and all $j = 1, \dots, \intOrder$.
      \\
      Set up $\B_{\indFullErrors} \gets (\B_{\indFullErrors, 1} \mid \dots \mid \B_{\indFullErrors, \intOrder})$ and $\B_{\indRowErasures} \gets (\B_{\indRowErasures, 1} \mid \dots \mid \B_{\indRowErasures, \intOrder})$.
      \\
      Recover the error vector $\e \gets \a_{\indFullErrors} \B_{\indFullErrors} + \a_{\indRowErasures} \B_{\indRowErasures} + \a_{\indColErasures} \B_{\indColErasures}$ as in~\eqref{eq:hilrs_error-erasure_decomposition}.
      \\
      \Return{$\c \gets \y-\e$.}
    }

    \Return{``decoding failure''.}
\end{algorithm}

\begin{theorem}[Error-Erasure Decoding of \ac{HILRS} Codes]\label{thm:error_erasure_dec_hilrs}
    Consider the transmission of a codeword $\c \in \horIntLinRS{\vecbeta}{\vecxi}{\intOrder}{\vecntilde}{k}$ over the additive error-erasure channel~\eqref{eq:sum_rank_channel_hor_int}.
    The error $\e \in \Fqm^{\intOrder n}$ of sum-rank weight $\SumRankWeight(\e) = \numbErrors$ is chosen uniformly at random from the set $\vecGivenSRWeight{q^m}{\intOrder \n}{\numbErrors}$ defined in~\eqref{eq:vectors-fixed-horizontal-weight} and determines the received word $\y = \c + \e \in \Fqm^{\intOrder n}$.
    The channel provides partial knowledge of the error which gives rise to a decomposition into $\numbFullErrors$ full errors, $\numbRowErasures$ row erasures, and $\numbColErasures$ column erasures such that $\numbErrors=\numbFullErrors + \numbRowErasures + \numbColErasures$ holds, as explained in~\autoref{sec:hilrs-channel-error-model}.
    The presented syndrome-based decoder can always recover $\c$ from $\y$ if $\numbFullErrors \leq \tfrac{1}{2}(n - k - \numbRowErasures - \max_{j}\{\numbColErasuresWIndex{j}\})$ holds with $\numbColErasuresWIndex{j} = \SumRankWeightWPartition{\n}(\e_{\indColErasures, j}) \leq \numbColErasures$ for $j = 1, \dots, \intOrder$.
    Moreover, the decoder can be used probabilistically for larger error weights and decoding succeeds with a probability of at least
    \begin{equation}\label{eq:lower_bound_Psucc_hilrs_erasure}
        1 - \gammaq^{\ell+1} q^{-m((s+1)(\numbErrorsMax-\numbErrorsWeightedHorizontal)+1)}
    \end{equation}
    as long as the error weight satisfies
    \begin{gather}
        \numbErrorsWeightedHorizontal
        \defeq \numbFullErrors + \frac{\intOrder}{\intOrder+1}(\numbRowErasures + \numbColErasuresBar)
        \leq \numbErrorsMax
        \defeq \frac{\intOrder}{\intOrder+1}\left(n-k \right)
        \quad\text{with}\quad
        \numbColErasuresBar \defeq \frac{1}{\intOrder}\sum_{j=1}^{\intOrder} \numbColErasuresWIndex{j}.
    \end{gather}
    The decoder requires on average $\softoh{\intOrder n^2}$ operations in $\Fqm$ if $m \in O(\intOrder)$ applies.
\end{theorem}

\begin{IEEEproof}
    As the argumentation in this section and, specifically, the proof of~\autoref{thm:hilrs_error-erasure_key_equation} have shown, the decoder in~\autoref{alg:error_erasure_dec_hilrs} is correct.
    There is only one step that can potentially lead to a decoding failure.
    Namely, the solution space of the key equation~\eqref{eq:hilrs_error-erasure_key_equation} could have dimension larger than one.
    The decoder handles this case correctly and returns a decoding failure.
    Further, the claimed decoding radius was derived in~\eqref{eq:hilrs_error_erasure_dec_radius}.
    
    The statements about the success probability of the decoder and the condition on the error weight for guaranteed unique decoding both directly depend on properties of the key equation~\eqref{eq:hilrs_error-erasure_key_equation}.
    The coefficient vectors of the auxiliary component-syndrome polynomials $\ESPcomponentSyndrome{1}, \dots, \ESPcomponentSyndrome{\intOrder}$ in the key equation can in fact be expressed as modified component syndromes of an error-only decoding instance.
    This follows from suitable extensions of the error-erasure decoder from~\cite{gabidulin2008errorAndErasure}, which was presented for horizontally interleaved Gabidulin codes in the rank metric.
    As the details overstretch the scope of this paper, we will discuss them in follow-up work.
    In any case, the outlined result justifies that we obtain the upper bound $\failureProb \leq \gammaq^{\ell+1} q^{-m((s+1)(\numbErrorsMax-\numbErrorsWeightedHorizontal)+1)}$ on the failure probability $\failureProb$ by applying~\autoref{lem:failure_prob_hilrs} from the error-only scenario.
    Moreover, the insights about the coefficients of the auxiliary component-syndrome polynomials allow to derive a simple proof that the decoder always decodes correctly when $\numbFullErrors \leq \tfrac{1}{2}(n - k - \numbRowErasures - \max_{j}\{\numbColErasuresWIndex{j}\})$ applies.
    
    Next, we analyze the asymptotic complexity of the presented decoding algorithm by grouping similar tasks.
    We do not mention every step explicitly, as e.g.\ setting up vectors and matrices according to simple rules or determining a skew reverse are essentially for free.
    Lines 3, 5, 19, and 22 contain vector-matrix and matrix-matrix products which can be computed in at most $O(\intOrder n^2)$ operations in $\Fqm$.
    Minimal skew polynomials are given by the formula~\eqref{eq:min_poly} and since the occurrences in lines 1, 6, and 13 concern skew polynomials of degree bounded by $n$, their computation can be done in $O(\intOrder n^2)$.
    The skew-polynomial products in lines 8, 11, 14, and 16 involve factors of degree at most $n$ and thus require at most $O(\intOrder n^2)$ operations in $\Fqm$.
    
    In the following, we will apply fast subroutines to achieve the stated overall decoding complexity.
    We describe the respective methods in~\autoref{sec:fast-dec}.
    Multisequence skew-feedback shift-register synthesis solves the key equation~\eqref{eq:hilrs_error-erasure_key_equation} in line 9 in at most $O(\intOrder (n-k)^2)$ operations in $\Fqm$.
    Lines 12 and 18 contain linear systems of a particular form, which allow for a fast solution via the Gabidulin-like algorithm.
    Observe that the system in line 18 is over $\Fqm$ and thus in $O(\intOrder n^2)$ but the one in line 12 is linear over the extension field $\Fqms$ of $\Fqm$.
    Therefore, the latter requires at most $O(\intOrder n^2)$ operations in $\Fqms$ and we bound this complexity by $\softoh{\intOrder n^2}$ operations in $\Fqm$ by means of a suitable $\Fqm$-basis of $\Fqms$.
    In lines 15 and 17, the decoder needs to find a basis of the root space of a skew polynomial of degree at most $n$ with respect to generalized operator evaluation with $\shots$ distinct evaluation parameters.
    The probabilistic Skachek--Roth-like algorithm achieves this in an average complexity of $O(\shots m n)$ over $\Fqm$.
    Under the assumption $m \in O(\intOrder)$, this is upper-bounded by $O(\intOrder n^2)$ operations in $\Fqm$.
    The Skachek--Roth-like algorithm is the only probabilistic component of the decoder and we measure its complexity in terms of average complexity.
    All other parts were assessed with respect to worst-case complexity.
    In summary, the error-erasure decoder for \ac{HILRS} codes has an asymptotic average complexity of $\softoh{\intOrder n^2}$ over $\Fqm$.
\end{IEEEproof}

\begin{remark}
    Observe that the above theorem states an asymptotic complexity of $\softoh{\intOrder n^2}$ for the error-erasure decoder, while the error-only decoder can be executed in at most $O(\intOrder n^2)$ operations in $\Fqm$.
    This originates from the fact that we solve the $\Fqms$-linear system~\eqref{eq:hilrs-Fqms-Gabidulin} in $O(\numbColErasures^2)$ operations over $\Fqms$ and bound this step's complexity by $\softoh{\intOrder \numbColErasures^2}$ operations in $\Fqm$ for a suitable $\Fqm$-basis of $\Fqms$.
    
    Note however that this is not necessary in many cases:
    it is likely that there is a $j = 1, \dots, \intOrder$ such that $\x_{\indColErasures, j}$ has full sum-rank weight.
    In this case, $\hat{\a}_{\indColErasures}$ can be recovered by solving only the corresponding $\Fqm$-linear system~\eqref{eq:hilrs-gabidulin-like-ahat}.
    As the Gabidulin-like algorithm over $\Fqm$ can achieve this in at most $O(\numbColErasures^2)$ operations in $\Fqm$, the overall decoding complexity is reduced to $O(\intOrder n^2)$ in these cases.
\end{remark}

%% file: simulations.tex
\section{Simulations}\label{sec:simulations}

We now present Monte Carlo simulations to experimentally verify the tightness of the upper bound on the decoding-failure probability of the derived error-only decoders for \ac{VILRS} and \ac{HILRS} codes.
We use SageMath~\cite{stein_sagemath} for our implementations and fix a random component \ac{LRS} code with the chosen parameters in each step of the Monte Carlo simulations.
Then, we decode random codewords that were distorted by uniformly distributed sum-rank errors of the predefined weight and collect 100 decoding failures.
Note that the parameters are selected such that decoding failures are observable within a reasonable time and they are thus far from suitable for practical applications.
Namely, we consider a component \ac{LRS} code with $q = 3$, $m = 4$, $k = 3$, and $n = 8$ with $\n = (4, 4)$ and $\shots = 2$.
We investigate the two interleaving orders $\intOrder \in \{4, 5\}$ for both vertical and horizontal interleaving.
In all cases, successful decoding can be guaranteed for all errors of sum-rank weight $\numbErrors \leq \tfrac{1}{2}(n - k) = 2.5$.
Probabilistic decoding is possible for error weights $\numbErrors \leq \numbErrorsMax = \tfrac{\intOrder}{\intOrder + 1} (n - k)$ and our parameter choices yield $\numbErrorsMax = 4$ for $\intOrder = 4$ and $\numbErrorsMax = 4.167$ for $\intOrder = 5$.
Thus, up to $\numbErrors = 4$ sum-rank errors can be decoded with high probability and the two choices $\numbErrors \in \{3, 4\}$ cover all possible scenarios for probabilistic decoding.

Note that the derived bounds on the failure probability for \ac{VILRS} and \ac{HILRS} codes coincide.
The corresponding results were derived in~\autoref{lem:failure_prob_vils} and in~\autoref{lem:failure_prob_hilrs}, respectively, and the bound reads as follows:
\begin{equation}
    \label{eq:Pf_upper_bound}
    \failureProb \leq \kappa_q^{\shots+1}q^{-m((\intOrder+1)(\numbErrorsMax-\numbErrors)+1)}.
\end{equation}
Here, $\numbErrorsMax \defeq \tfrac{\intOrder}{\intOrder + 1} (n - k)$ is the maximal decoding radius and $\kappa_q < 3.5$ was defined as $\kappa_q = \prod_{i=1}^{\infty}\tfrac{1}{1-q^{-i}}$ for integers $q \geq 2$ in~\eqref{eq:def_kappa_q}.
When we compute the bound explicitly in the following, we use the first $100$ factors to approximate $\kappa_q$ from above, i.e., we compute $\prod_{i=1}^{100}\tfrac{1}{1-q^{-i}}   \geq \kappa_q$.
\autoref{tab:simulation_results} shows the evaluation of the standard bound~\eqref{eq:Pf_upper_bound} for the considered parameters as well as the experimentally observed failure probabilities for \ac{VILRS} and \ac{HILRS} codes.
Note that the bounds for $\numbErrors = 3$ are too small to be experimentally observed with reasonable constraints in time and resources.
We thus only simulated the case $\numbErrors = 4$ and obtained results between $1.3 \cdot 10^{-2}$ and $1.4 \cdot 10^{-2}$ for interleaving order $\intOrder = 4$ and between $1.4 \cdot 10^{-4}$ and $1.6 \cdot 10^{-4}$ for $\intOrder = 5$.
The theoretical bound from~\eqref{eq:Pf_upper_bound} is about $5$ to $6$ times larger than the experimental observations.

\begin{table}[ht]
    \centering
    \caption{Predicted and experimentally observed failure probability in the decoding of vertically and horizontally interleaved \ac{LRS} codes. The parameters of the component \ac{LRS} code are $q=3$, $m=4$, $k=3$, and $\n = (4,4)$.}
    \label{tab:simulation_results}
    \begin{tabular}{|c||c|c||c|c||c|c|}
        \hline
        \multirow{3}{*}{$\failureProb$} & \multicolumn{2}{|c||}{$\numbErrors = 3$} & \multicolumn{4}{|c|}{$\numbErrors = 4$} \\
        \cline{2-7}
        & \multicolumn{2}{|c||}{Upper Bounds} & \multicolumn{2}{|c||}{Upper Bounds} & \multicolumn{2}{|c|}{Simulation Results} \\
        \cline{2-7}
        & standard~\eqref{eq:Pf_upper_bound} & improved~\eqref{eq:Pf_improved_upper_bound} & standard~\eqref{eq:Pf_upper_bound} & improved~\eqref{eq:Pf_improved_upper_bound} & VILRS & HILRS \\
        \hline
        \hline
        {$\intOrder = 4$} & {2.015e-11} & {1.143e-11} & {7.026e-02} & {3.985e-02} & 1.302e-02 & 1.348e-02 \\
        {$\intOrder = 5$} & {3.071e-15} & {1.742e-15} & {8.674e-04} & {4.920e-04} & 1.569e-04 & 1.431e-04 \\
        \hline
    \end{tabular}
\end{table}

The main ingredient for the standard bound in~\eqref{eq:Pf_upper_bound} is a result from~\cite{bartz2022fast}.
However, this result can be improved by having a closer look at the proof of~\cite[Lem.~7]{bartz2022fast} which reveals that one factor $\kappa_q$ can actually be replaced by $\kappa_{q^m}$.
As $\kappa_q$ is decreasing and converges to $1$ for growing $q$, this yields the improved upper bound
\begin{equation}
    \label{eq:Pf_improved_upper_bound}
    \failureProb \leq \kappa_{q^m}^{\phantom{1}} \kappa_q^{\shots}q^{-m((\intOrder+1)(\numbErrorsMax-\numbErrors)+1)}.
\end{equation}

\begin{figure}[ht!]
    \centering
        \input{./simulation_results.tikz}
    \caption{Visualization of the observed decoding-failure probability for \ac{VILRS} and \ac{HILRS} codes and theoretical upper bounds. See~\autoref{tab:simulation_results} for numerical values.}
    \label{fig:simulation_results}
\end{figure}
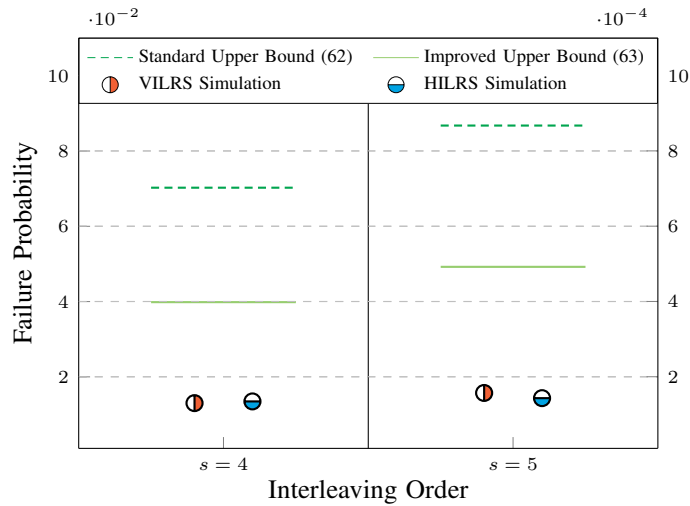

The ratio of $\kappa_{q^m}$ and $\kappa_q$ determines the multiplicative gain obtained from~\eqref{eq:Pf_improved_upper_bound} with respect to~\eqref{eq:Pf_upper_bound}.
But since $\kappa_q$ converges quickly to $1$ for $q \to \infty$, this is mostly attractive for relatively small $q$ or large $m$.
The small parameters we selected for the presented simulations lie in this regime and we obtain $\kappa_3 \approx 1.785$ and $\kappa_{3^4} \approx 1.013$.
Consequently, their ratio $\tfrac{\kappa_{3^4}}{\kappa_{3}} \approx 0.567$ shows that the values of~\eqref{eq:Pf_improved_upper_bound} almost halve the ones obtained from~\eqref{eq:Pf_upper_bound}.
The concrete values are part of~\autoref{tab:simulation_results} and are less than a factor 3.5 larger than the simulation results for all cases.
\autoref{fig:simulation_results} summarizes the results of this section graphically and visualizes the gain we obtained from the improved bound~\eqref{eq:Pf_improved_upper_bound}.

%% file: simulation_results.tikz
\begin{tikzpicture}
    \begin{axis}[
        scale only axis,
        scaled y ticks={base 10:2},
        axis y line*=left,
		height=.3\textwidth,
		width=.42\textwidth,
        ylabel={Failure Probability},
        xlabel={Interleaving Order},
       	xtick={0, 2},
        xticklabels={$s = 4$, $s = 5$},
		ymajorgrids,
		ymin=.001,
		ymax=.11,
        xmin=-1,
        xmax=3,
    ]

		\addplot[color=Green, densely dashed, forget plot] coordinates {
			(-.5, 0.07025182374202299)
			(.5, 0.07025182374202299)
		};
        \addplot[color=YellowGreen, forget plot] coordinates {
			(-.5, 0.0398479013370711)
			(.5, 0.0398479013370711)
		};

        \addplot[only marks, mark color=RedOrange, mark=halfcircle, mark options={rotate=90}] table[row sep=crcr]{%
			-.2 		1.302083e-02 \\
		};
        \addplot[only marks, mark color=Cerulean, mark=halfcircle] table[row sep=crcr]{%
			.2		1.347527e-02 \\
		};
		
		\addplot[black, thin, no marks] table[row sep=crcr]{%
			1		.001 \\
            1       .099 \\
		};
    \end{axis}
	
    \begin{axis}[
        scale only axis,
	    scaled y ticks={base 10:4},
        axis y line*=right,
        axis x line=none,
		height=.3\textwidth,
		width=.42\textwidth,
        xlabel={Interleaving Order},
       	xtick={0, 2},
        xticklabels={$s = 4$, $s = 5$},
		ymajorgrids,
		ymin=.00001,
		ymax=.0011,
        xmin=-1,
        xmax=3,
		legend style={at={(axis cs:-1,.0011)},anchor=north west,legend cell align=left,align=left,line width=.25pt,legend columns=2},
    ]
		\addplot[color=Green, densely dashed] coordinates {
			(1.5, 0.0008673064659508943)
			(2.5, 0.0008673064659508943)
		};
		\addlegendentry{Standard Upper Bound~\eqref{eq:Pf_upper_bound}~\;\,}
    
    	\addplot[color=YellowGreen] coordinates {
			(1.5, 0.0004919493992230962)
			(2.5, 0.0004919493992230962)
		};
		\addlegendentry{Improved Upper Bound~\eqref{eq:Pf_improved_upper_bound}}
		
        \addplot[only marks, mark color=RedOrange, mark=halfcircle, mark options={rotate=90}] table[row sep=crcr]{%
            1.8		1.569031e-04 \\
		};
        \addlegendentry{VILRS Simulation}
        \addplot[only marks, mark color=Cerulean, mark=halfcircle] table[row sep=crcr]{%
            2.2		1.431184e-04 \\
		};
        \addlegendentry{HILRS Simulation}
	\end{axis}
\end{tikzpicture}

%% file: efficient_subroutines.tex
\section{Efficient Subroutines for Fast Decoding\label{sec:fast-dec}}

This section contains fast algorithms that can be applied to certain computationally expensive steps of the presented syndrome-based decoders to speed them up.
Even though the techniques themselves are only remotely connected to the syndrome-based decoding approach, they are important to reach the claimed complexities.

\subsection{Solving Key Equations}\label{sec:shift_register_synthesis}

Each of the presented decoders relies on a key equation which needs to be solved in order to recover the \ac{ELP} or the \ac{ESP} corresponding to the full errors, respectively.
The key equations can be expressed as systems of linear equations as shown in~\eqref{eq:vilrs_key_equation_elp_v1_norm} and~\eqref{eq:vilrs_error-erasure_key_equation_elp_v1_norm} for \ac{VILRS} codes and in~\eqref{eq:hilrs_key_equation_esp_v1_norm} and~\eqref{eq:hilrs_error-erasure_key_equation_esp_v1_norm} for \ac{HILRS} codes.
Since the systems have a particular form, we do not need to rely on classical Gaussian elimination but can solve them faster by applying multisequence skew-feedback shift-register synthesis as proposed in~\cite{Sidorenko2011SkewFeedback}.
This allows us to achieve a complexity in $O(\intOrder (n-k)^2)$ over $\Fqm$ for this decoding step.

Multisequence skew-feedback shift-register synthesis takes $\intOrder$ sequences $\s_j = (s_{j,d_j+1}, \dots, s_{j,n-k}) \in \Fqm^{n-k-d_j}$ for $j = 1, \dots, \intOrder$ of potentially different length as input and finds the shortest connection vector $\vecsigma = (\sigma_2, \dots, \sigma_{\tau+1}) \in \Fqm^{\tau}$ satisfying the shift-register relations
\begin{equation}
    \sum_{\nu=2}^{\numbErrors+1}\ESP_\nu\aut^{-(\nu-1)}\left(s_{j,l-\nu+1}\right)=-s_{j,l} \quad \text{for all } l=d_j + \numbErrors + 1,\dots,n-k \quad \text{and all } j=1,\dots,\intOrder
\end{equation}
with a fixed field automorphism $\aut$ of $\Fqm$~\cite[Prob.~1]{Sidorenko2011SkewFeedback}.
The algorithm~\cite[Alg.~2]{Sidorenko2011SkewFeedback} synthesizes $\vecsigma$ by means of an iterative procedure starting from the trivial connection vector $\vecsigma = (1)$ and trying to adapt and lengthen the shift register as necessary to accommodate each entry of every input sequence step by step.
Note that the algorithm also outputs how many degrees of freedom were involved in every iteration.
This allows to easily verify the uniqueness of the found solution up to $\Fqm$-multiples by making sure that there were no ambiguous choices in any step.
The explicit condition for checking this is given in~\cite[Cor.~7]{Sidorenko2011SkewFeedback}.
A fast variant of skew-feedback shift-register synthesis can be found in~\cite{SidorenkoBossert2012Fastskewfeedback}.
Since solving the key equation is not the computational bottleneck for our syndrome-based decoders, we do not discuss the speedup here.

\subsection{Finding Roots of Skew Polynomials}\label{sec:skachek_roth}

All discussed decoders use \acfp{ELP} or \acfp{ESP} which are skew polynomials of minimal degree that vanish precisely at the error locators or at the error values, respectively.
Therefore, a fast method for finding a basis of the root space of a skew polynomial with respect to generalized operator evaluation is crucial to recover the information about the error from the skew polynomials.
We now present a probabilistic Skachek--Roth-like algorithm that accomplishes the task with an average complexity in $O(\shots m \deg(p))$ over $\Fqm$ for a skew polynomial $p \in \SkewPolyringZeroDer$ and its root spaces with respect to $\shots$ chosen evaluation parameters.
In contrast, adapting the conventional approach for linearized polynomials described in~\cite[Chap.~11.1]{berlekamp2015algebraic} would require $O(\shots m \deg(p)^{2})$ operations in $\Fqm$ in the worst case.

The authors showed in the patent~\cite{patentSkachekRoth2023} that Skachek and Roth's approach for \emph{linearized} polynomials~\cite{skachek2008probabilistic} can be generalized to skew polynomials with respect to generalized operator evaluation.
One of the main observations is that the skew polynomial $n(x) \defeq x^m - \genNorm{m}{\xi} \in \SkewPolyringZeroDer$ is the minimal skew polynomial vanishing at all $\Fqm$-elements with respect to generalized operator evaluation and evaluation parameter $\xi$.
We can factor it into $n(x) = h(x) \cdot g(x)$ with $h(x) = \gcrd(n(x), p(x))$ and $g(x) = \ldivNoArg(n(x), h(x))$.
Then, the root spaces of $h$ and $p$ with respect to the chosen evaluation parameter $\xi$ coincide and further, the root space of $h$ is precisely the image space of $g$.
When we want to find a basis of the zeros of $p \in \SkewPolyringZeroDer$ with respect to $\opev{p}{\cdot}{\xi}$, we can thus instead probabilistically find a basis of the image of $g$.

The resulting procedure is depicted in~\autoref{alg:skachek-roth_like} and multiple evaluation parameters can be incorporated by computing a basis of the root space for one evaluation parameter at a time.
Moreover, the algorithm allows to initialize the basis $\mathcal{B}_{i}$ of the root space of $p$ with respect to $\opev{p}{\cdot}{\xi_i}$ with a basis $\mathcal{S}_{i}$ of a subspace of the respective root space for each $i = 1, \dots, \shots$.
This incorporation of partial knowledge reduces the overall runtime of the algorithm.

\begin{algorithm}[ht]
    \caption{\algoname{Skachek--Roth-like Algorithm}}\label{alg:skachek-roth_like}

    \Input{%
        A skew polynomial $p \in \SkewPolyringZeroDer$,
        \\
  	    a vector $\vecxi = (\xi_1,\dots,\xi_{\shots}) \in \Fqm^{\shots}$ of evaluation parameters,
        \\
        and a basis $\mathcal{S}_{i}$ of a subspace of the root space of $p$ with respect to $\opev{p}{\cdot}{\xi_i}$ (or $\mathcal{S}_{i} = \emptyset$) for each $i = 1, \dots, \shots$.
    }
    \Output{A basis $\mathcal{B}_i$ of the root space of $p$ with respect to $\opev{p}{\cdot}{\xi_i}$ for every $i = 1, \dots, \shots$.}

    \BlankLine

    \For{$i = 1, \dots, \shots$}{
        $n_i(x) \gets x^m - \genNorm{m}{\xi_i}$\;
        $h_i(x) \gets \gcrd(n_i(x), p(x))$\;
        $g_i(x) \gets \ldivNoArg(n_i(x), h_i(x))$\;

        \BlankLine

        $\mathcal{B}_i = \mathcal{S}_i$\;
        \While{$\vert \mathcal{B}_i \vert < \deg(g_i)$}{
            $b \sample \Fqm^{\ast}$\;
            \If{$\opev{h}{b}{\xi_i} \not\in \langle \opev{h}{\mathcal{B}_i}{\xi_i} \rangle_{\Fq}$}{
                $\mathcal{B}_i \gets \mathcal{B}_i \cup \{b\}$\;
            }
        }
    }

    \Return{$\mathcal{B}_1, \dots, \mathcal{B}_{\shots}$}\;
\end{algorithm}

\subsection{Recovering Error Values and Error Locators}\label{sec:efficient_gabidulin-like}

Our syndrome-based decoding schemes obtain the error by recovering the error values and the error locators separately.
Recall that the \ac{VILRS} decoders first recover the error locators and then compute the missing error values, whereas the \ac{HILRS} decoders proceed in the opposite order and recover the error values first and then the error locators.
When one of the parts is known and the other needs to be found, we are confronted with a system of $\Fqm$-linear equations of a particular form.
Such systems arise e.g.\ in~\eqref{eq:vilrs_gab-like} for \ac{VILRS} decoding and in~\eqref{eq:hilrs_gab-like} for \ac{HILRS} decoding and look as follows:

Given $\a =(\shot{\a}{1} \mid \dots \mid \shot{\a}{\shots}) \in \Fqm^t$ with $\SumRankWeight(\a)=t \leq n-k$, $\s \in \Fqm^{n-k}$ and a vector $\vecxi' \in \Fqm^\shots$ with entries belonging to pairwise distinct nontrivial conjugacy classes of $\Fqm$, we want to find the solution $\x = (\shot{\x}{1} \mid \dots \mid \shot{\x}{\shots}) \in \Fqm^t$ with $\SumRankWeight(\x)=t$ of the $\Fqm$-linear system
\begin{equation}\label{eq:gabidulin-like_system}
	\opMoore{n-k}{\x}{\vecxi'} \cdot \a^\top = \s^\top.
\end{equation}
This system is equivalent to
\begin{equation}\label{eq:equiv-gabidulin-like_system}
    \opMooreInv{n-k}{\a}{\autinv(\vecxi')} \cdot \x^\top = \widetilde{\s}^\top
\end{equation}
with $\widetilde{\s} = ( s_1, \autinv(s_2), \dots, \aut^{-(n-k-1)}(s_{n-k}) )$.
This can be verified by applying $\aut^{-(l-1)}$ to the $l$-th equation for $l = 1, \dots, n-k$ and using~\eqref{eq:normlemma-3}.
Recall that the generalized Moore matrix $\opMooreInv{n-k}{\a}{\autinv(\vecxi')}$ has full $\Fqm$-rank $t$ if and only if the entries of $\autinv(\vecxi')$ belong to pairwise distinct nontrivial conjugacy classes of $\Fqm$ and $\SumRankWeight(\a)=t$ applies~\cite[Thm.~4.5]{lam1988vandermonde}.
Thus, the requirements on $\vecxi'$ and $\a$ directly imply that the coefficient matrix of~\eqref{eq:equiv-gabidulin-like_system} has full rank and the solution $\x$ is unique.
Since~\eqref{eq:gabidulin-like_system} and~\eqref{eq:equiv-gabidulin-like_system} are equivalent, the matrix $\opMoore{n-k}{\x}{\vecxi'}$ needs to have full $\Fqm$-rank and hence $\SumRankWeight(\x) = t$ follows for the unique solution $\x \in \Fqm^{t}$ by~\cite[Thm.~4.5]{lam1988vandermonde}.

A special case of problem~\eqref{eq:gabidulin-like_system} is the rank-metric setting, where the vectors $\a$ and $\x$ have only one block, $\vecxi' = (1)$ contains the only evaluation parameter $1$, and $\aut$ is the Frobenius automorphism $\cdot^q$.
It arises in syndrome-based decoding of Gabidulin codes and thus Gabidulin proposed an efficient algorithm to solve it in~\cite{Gabidulin_TheoryOfCodes_1985}.
Gabidulin's algorithm exploits the structure of the underlying Moore matrix which allows for the successive elimination of variables.
It requires at most $\oh{n^2}$ operations in $\Fqm$~\cite{gadouleau2008complexity}.

We now generalize his approach to the sum-rank-metric case that we stated in~\eqref{eq:gabidulin-like_system}.
  In order to simplify the notation, we neglect the blockwise structure of the vectors $\a$ and $\x$, i.e., we write $\a=(a_1,\dots,a_t)$ and $\x=(x_1, \dots, x_t)$, and let the $i$-th entry of the vector $\vecxi = (\xi_1,\dots,\xi_t) \in \Fqm^t$ be the evaluation parameter corresponding to the $i$-th block with respect to the length partition of $\a$ and $\x$ for each $i = 1, \dots, t$.
\autoref{alg:gabidulin_like} describes the procedure and we analyze its complexity in the next lemma.

\begin{algorithm}[ht]
  \caption{\algoname{Gabidulin-like Algorithm}}\label{alg:gabidulin_like}
  \SetKwInOut{Input}{Input}\SetKwInOut{Output}{Output}

  \Input{A vector $\a \in \Fqm^t$ with $\SumRankWeight(\a)=t \leq n-k$, \\
    a vector $\vecxi' \in \Fqm^{\shots}$ with entries from pairwise distinct nontrivial conjugacy classes of $\Fqm$ \\ from which the vector $\vecxi = (\xi_1, \dots, \xi_t) \in \Fqm^{t}$ is constructed as described in the text, \\
    and a vector $\s \in \Fqm^{n-k}$.
  }
  \Output{The solution $\x \in \Fqm^t$ of the $\Fqm$-linear system $\opMoore{n-k}{\x}{\vecxi'} \cdot \a^\top = \s^\top$ from~\eqref{eq:gabidulin-like_system} with $\SumRankWeight(\x)=t$.}

  \BlankLine

  $\A, \Q \in \Fqm^{t \times t} \gets \0$
  \\ 
  \For{$j=1,\dots,t$}{
  	$A_{1,j} \gets a_j$ \hfill \tcc{Initialize first row of $\A$ with $\a$}
  	$Q_{1,j} \gets s_j$ \hfill \tcc{Initialize first row of $\Q$ with $(s_1,\dots,s_t)$}
  }

  \For{$i=1,\dots,t-1$\label{line:outer_for}}
  {
  	$\kappa_{i} \gets {A_{i,i}^{-1}}{\opfull{\xi_{i}^{-1}}{A_{i,i}}}$ \label{line:compute_kappa} \\
  	\For{$j=1,\dots,t$\label{line:inner_for}}
  	{
  		\If{$j > i$}{
  			$A_{i+1,j} \gets A_{i,j} - \autinv\left(\kappa_{i} A_{i,j} \xi_j\right)$
  		}

  		\If{$j \leq t-i$}{
  			$Q_{i+1,j} \gets Q_{i,j} - \autinv\left(\kappa_{i} Q_{i,j+1}\right)$
  		}
  	}
  }

  $x_t \gets A_{t,t}^{-1} {Q_{t,1}}$ \label{line:back_sub_start}

  \For{$i=t-1,\dots,1$}{
  	$x_i \gets A_{i,i}^{-1}\bigl(Q_{i,1} - \sum_{j=i+1}^{t}A_{i,j}x_j\bigr)$
  }\label{line:back_sub_end}

  \Return{$\x \gets (x_1,\dots,x_t)$}
\end{algorithm}

\begin{lemma}
	\autoref{alg:gabidulin_like} solves the problem described in~\eqref{eq:gabidulin-like_system} and requires at most $\oh{t^2}$ operations in $\Fqm$ to do so.
\end{lemma}

\begin{IEEEproof}
 Note that the system~\eqref{eq:gabidulin-like_system} is equivalent to
 \begin{equation}\label{eq:gabidulin-like_system_trunc}
	\sum_{j=1}^{t} A_{1,j} \opfullexp{\xi_j}{x_j}{l-1} = Q_{1,l} \quad \text{for all } l=1,\dots,t
 \end{equation}
 according to how we defined $\xi_1,\dots,\xi_t$ and the matrices $\A$ and $\Q$.
 We can define $\kappa_1 = {A_{1,1}^{-1}}{\opfull{\xi_{1}^{-1}}{A_{1,1}}}$ and use it to eliminate the unknown $x_1$ from all but the first equation as follows:
 Multiply the $l$-th equation for each $l = 2, \dots, t$ with $\kappa_1$, apply $\autinv$ to it, and subtract it from the previous equation with index $l-1$.
 Then, the new $l$-th equation has a zero at the first coefficient and we get the system
 \begin{equation}\label{eq:gabidulin-like_system_trunc_red}
	\sum_{j=2}^{t} A_{2, j} \opfullexp{\xi_j}{x_j}{l-1}  = Q_{2,l} \quad \text{for all } l=1,\dots,t-1.
 \end{equation}
 In particular,~\eqref{eq:gabidulin-like_system_trunc_red} does not contain $x_1$ anymore and when we add the first equation from~\eqref{eq:gabidulin-like_system_trunc} to it, the system is equivalent to~\eqref{eq:gabidulin-like_system_trunc}.
 We then repeat this procedure for the unknowns $x_2, \dots, x_{t-1}$ in a similar fashion and finally obtain an $\Fqm$-linear system with an upper-triangular matrix $\A$ that is equivalent to~\eqref{eq:gabidulin-like_system_trunc}.
 In particular, we get
 \begin{equation}
	\A \x^\top = \q_1
 \end{equation}
 where $\q_1$ denotes the first column of $\Q$.
 Due to the upper-triangular structure of $\A$, the unknowns $x_1, \dots, x_t$ can be recovered via back substitution, i.e., we obtain
 \begin{align*}
  x_t = A_{t,t}^{-1} {Q_{t,1}}
  \quad \text{and} \quad
  x_i = A_{i,i}^{-1}\biggl(Q_{i,1} - \sum_{j=i+1}^{t}A_{i,j}x_j\biggr) \quad \text{for all } i = t-1, \dots, 1.
 \end{align*}

 The complexity analysis proceeds in a similar way as for the original Gabidulin algorithm~\cite{gadouleau2008complexity}.
 The computation of the coefficients $A_{i+1, j}$ and $Q_{i+1, j}$ in the inner loop in lines 7 to 11 requires $\oh{t}$ operations in $\Fqm$ and $\kappa_{i}$ can be computed in $\oh{1}$ in line 6.
 Therefore, the outer loop spanning lines 5 to 11 requires at most $\oh{t^2}$ operations in $\Fqm$.
 Due to the upper-triangular structure of $\A$, the unknowns $x_1,\dots,x_t$ can be recovered via back substitution in lines 12 to 14.
 This requires at most $\oh{t^{2}}$ operations in $\Fqm$ and we obtain an overall worst-case complexity in $O(t^2)$ over $\Fqm$ for the Gabidulin-like algorithm.
\end{IEEEproof}

Let us conclude this section with a short example that illustrates the triangular structure of $\A$ and thus the applicability of back substitution.

\begin{example}
  Consider the finite field $\F_{3^2}$ constructed by the primitive polynomial $p(x)=x^2 + 2x + 2 \in \F_{3}[x]$ corresponding to the primitive element $\pe$ and let $\aut = \cdot^3$ be the Frobenius automorphism.
  Consider the vectors $\a = \bigl( (\pe^7, \pe^{6}) \mid (\pe) \bigr)$ with $\SumRankWeight(\a)=3$ and $\s = (\pe, \pe^{4}, \pe^{3})$.
  Suppose we want to find the solution $\x = \bigl((x_1^{(1)}, x_2^{(1)}) \mid (x_1^{(2)})\bigr)$ of the system
  \begin{equation}\label{eq:gab-example-system}
    \opMoore{3}{\x}{\vecxi'} \cdot \a^\top = \s^\top
  \end{equation}
  with $\vecxi' = (1, \pe)$ and thus $\vecxi = (1, 1, \pe)$.
  \autoref{alg:gabidulin_like} first constructs the two matrices
  \begin{align*}
    \A = 
    \left(
      \begin{array}{ccc}
        \pe^7 & \pe^{6} & \pe \\
        0 & \pe^{5} & \pe^{7} \\
        0 & 0 & \pe^{5}
      \end{array}
    \right)
    \qquad \text{and} \qquad
    \Q =
    \left(
      \begin{array}{ccc}
        \pe & \pe^{4} & \pe^3 \\
        \pe^{3} & \pe^{2} & 0 \\
        \pe^{6} & 0 & 0
      \end{array}
    \right),
  \end{align*}
  where the first rows of $\A$ and $\Q$ correspond to $\a$ and $\s$, respectively.
  In fact, $\A$ is a \acl{REF} of the matrix
  \begin{equation}
    \opMooreInv{3}{\a}{\autinv(\vecxi')} =
    \begin{pmatrix}
        \pe^{7} & \pe^{6} & \pe \\
        \pe^{5} & \pe^{2} & \pe^{6} \\
        \pe^{7} & \pe^{6} & \pe^{5}
    \end{pmatrix}
\end{equation}
  which contains the first $t = 3$ rows of the coefficient matrix of the equivalent formulation
  \begin{equation}
      \opMooreInv{n-k}{\a}{\autinv(\vecxi')} \cdot \x^{\top} = \bigl(\pe, \autinv(\pe^{4}), \aut^{-2}(\pe^{3})\bigr)^{\top}
  \end{equation}
  of the system~\eqref{eq:gab-example-system}.
  This leads to the linear system
  \begin{equation}\label{eq:nicer_system}
    \underbrace{ 
    \left(
      \begin{array}{ccc}
        \pe^7 & \pe^{6} & \pe \\
        0 & \pe^{5} & \pe^{7} \\
        0 & 0 & \pe^{5}
      \end{array}
    \right)
    }_{\A}
    \cdot
    \underbrace{
    \left(
      \begin{array}{c}
        x_1^{(1)} \\
        x_2^{(1)} \\
        x_1^{(2)}
      \end{array}
    \right)
    }_{\x}
    =
    \underbrace{
    \left(
      \begin{array}{c}
        \pe \\
        \pe^{3} \\
        \pe^{6}
      \end{array}
    \right)
    }_{\q_1},
  \end{equation}
  where $\q_1$ is the first column of $\Q$.
  Due to the upper-triangular structure of $\A$ we can solve~\eqref{eq:nicer_system} efficiently for $\x = \bigl((\pe^2, 1) \mid (\pe)\bigr)$ via back substitution.
\end{example}

As we have seen in the previous example, the Gabidulin-like algorithm implicitly transforms the system~\eqref{eq:gabidulin-like_system} into its equivalent formulation~\eqref{eq:equiv-gabidulin-like_system} and simultaneously brings its coefficient matrix $\opMooreInv{n-k}{\a}{\autinv(\vecxi')}$, or rather the first $t$ rows of it, into~\acl{REF}.
Since the algorithm exploits the particular form of the system, this can be achieved much faster than with classical Gaussian elimination.

%% file: conclusion.tex
\section{Conclusion}\label{sec:conclusion}

We showed that both vertically and horizontally interleaved \acl{LRS} (VILRS and HILRS) codes can be decoded with a syndrome-based approach.
More precisely, we gave an error-only decoder in both cases and generalized it to an error-erasure scenario.
The decoders for \acs{VILRS} codes first determine the \ac{ELP} and thus the row space of the full errors via an \ac{ELP} key equation and then recover the missing column space.
In contrast, \acs{HILRS} decoders use a key equation based on the \ac{ESP} to first recover the column space of the full errors and then continue to retrieve their row space.
This duality follows from the interleaving construction, as the components of vertically interleaved errors share the same row space, whereas horizontal interleaving leads to a shared column space for all component errors.
The two presented error-only decoders have an average  complexity in $O(\intOrder n^2)$ and the error-erasure variants need on average $\softoh{\intOrder n^2}$ operations in the ambient field $\Fqm$, where $\intOrder \in O(m)$ is the interleaving order and $n$ the length of the component code.
When $\numbFullErrors$, $\numbRowErasures$, and $\numbColErasures$ denote the number of full errors, of row erasures, and of column erasures, respectively, successful decoding is guaranteed as long as the error weight $\tau = \numbFullErrors + \numbRowErasures + \numbColErasures$ satisfies $\numbFullErrors \leq \tfrac{1}{2}(n-k-\numbRowErasures - \numbColErasures)$.
Moreover, probabilistic unique decoding allows to decode an error of weight $\tau$ with high probability if $\numbFullErrors \leq \tfrac{\intOrder}{\intOrder + 1} (n - k - \numbRowErasures - \numbColErasures)$ applies.
In fact, the error-erasure decoders have a slightly larger decoding region and the gain is with respect to row erasures for vertical interleaving~\eqref{eq:vilrs_error_erasure_dec_radius} and with respect to column erasures for horizontal interleaving~\eqref{eq:hilrs_error_erasure_dec_radius}.
We gave a tight upper bound on the probability of decoding failures and showcased its tightness for the error-only case with Monte Carlo simulations.

A straightforward theoretical generalization of the presented methods is the study of inhomogeneous interleaving.
Moreover, \ac{LRS} codes arising from skew-polynomial rings with nonzero derivation could be considered.
However, it has to be taken into account that the dual of an \ac{LRS} code might be a generalized Goppa code and not an \ac{LRS} code in this setting~\cite{CarusoDurand2022DualsLinearizedReed}.

Further, lifted variants of \acs{VILRS} and \acs{HILRS} codes and their properties in the sum-subspace metric are promising candidates for multishot network coding.
The lifting of Gabidulin codes in the rank metric was studied and applied to single-shot network coding in~\cite{silva_rank_metric_approach} and the usage of multivariate polynomials in the decoder led to gains in the decoding radius~\cite{BartzMeierSid_SubspaceSyndrome_SCC,BartzSidorenko_ISubSyndrome_DCC}.
Generalizations of these techniques should yield similar results for lifted interleaved \ac{LRS} codes and their application to multishot network coding.

Of course, the syndrome-based approach is not the only way to decode \acs{VILRS} and \acs{HILRS} codes.
It stays an interesting question to study known and new decoders for these codes and compare their advantages and limitations.
One open point in this area is the understanding of the error patterns for which probabilistic decoders fail and whether the failures depend on the decoding scheme or purely on the error.